\documentclass[runningheads]{llncs}

\usepackage{amsmath,amssymb,mathtools}
\usepackage{tikz}
\usetikzlibrary{calc,arrows.meta,decorations.pathmorphing,bending,backgrounds,positioning,fit,petri,shapes.misc}
\usepackage{pgfplots}
\pgfplotsset{compat=1.18}
\usepackage[dvipsnames]{xcolor}
\usepackage[labelfont=bf,labelsep=period,font=small]{caption}
\usepackage{subcaption}
\usepackage{csquotes}
\usepackage{hyperref}
\usepackage{xurl}

\DeclareFontEncoding{LS1}{}{}
\DeclareFontSubstitution{LS1}{stix2}{m}{n}
\DeclareSymbolFont{symbols4}{LS1}{stixscr}{m}{n}
\DeclareMathSymbol{\hexagon}{\mathord}{symbols4}{"AE}

\tikzset{
    triangular_slant/.style={
        cm={1,0,-0.5,0.866,(0, 0))}
    },
    robot/.style={
        circle, draw=blue!50, fill=blue!20,
thick, inner sep=1pt, minimum size=4.5mm
    },
    robottiny/.style={
        circle, draw=blue!50, fill=blue!20,
        thick, inner sep=1pt, minimum size=4mm
    },
    moves/.style={
        very thick, -{Stealth[scale length=0.8, scale width=0.95]}
    },
    movestiny/.style={
        very thick, -{Stealth[scale length=0.6, scale width=0.9]}
    },
    cross/.style={cross out, draw,
         minimum size=2*(#1-\pgflinewidth),
         inner sep=0pt, outer sep=0pt},
     cross/.default={1pt},
     mark/.style={
        circle, thick, inner sep=0pt, minimum size=#1, draw=black!90,fill=white!100
     },
     mark/.default={5pt},
     rblack/.style={draw=black!90,fill=black!5},
     rblue/.style={draw=blue!60,fill=blue!10},
     rgreen/.style={draw=green!70,fill=green!10},
     rred/.style={draw=red!70,fill=red!10},
     rorange/.style={draw=Orange!80,fill=YellowOrange!13},
     rviolet/.style={draw=violet!70,fill=violet!10},
}

\newcommand{\tri}[1]{(#1)_\triangle}

\newcommand*\rulescale{0.85}
\newcommand*\adjscale{0.65}

\newcommand{\triangulargrid}[2]{
    \begin{scope}
        \clip (0.25,0.5) -- (#2/2-0.25,#2-0.5) -- (#1-0.25,#2-0.5) -- (#1-#2/2+0.25, 0.5) -- cycle;
        \draw (0,0) grid (#1, #2);
        \foreach \i in {-#2, ..., #1} {
            \draw (\i, 0) -- (\i+#2, #2);
        }
    \end{scope}
}

\newcommand{\triangulargridview}[1]{
    \path (3*#1/4+0.4, -0.4) -- (#1/4-0.4, #1+0.4);
    \begin{scope}
        \clip (0,0) -- (#1/2, 0) -- (#1,#1/2) -- (#1, #1) -- (#1/2, #1) -- (0, #1/2) -- cycle;
        \draw (0,0) grid (#1, #1);
        \foreach \i in {-#1, ..., #1} {
            \draw (\i, 0) -- (\i+#1, #1);
        }
    \end{scope}
    \begin{scope}
    \draw (0,0) -- (#1/2, 0) -- (#1,#1/2) -- (#1, #1) -- (#1/2, #1) -- (0, #1/2) -- cycle;
    \end{scope}
}

\newcommand{\hexgrid}[3][(0,0)]{
    \begin{scope}
        \clip (-0.25,0.5) -- (#3/2-0.75,#3-0.5) -- (#2-0.25,#3-0.5) -- (#2-#3/2+0.25, 0.5) -- cycle;

\begin{scope}[shift={#1}]
        \foreach \x in {0, ..., #2}
            \foreach \y in {0, ..., #3}
            {
                \def\posa{(3*\x+\y,2*\y)};
                \def\posb{(3*\x+\y+2,2*\y+1)};
                \foreach \pos in {\posa, \posb} {
                    \draw \pos -- +(1,1);
                    \draw \pos -- +(-1,0);
                    \draw \pos -- +(0,-1);
                }
            }
        \end{scope}
    \end{scope}
}

\newcommand{\hexviewoneL}{
    \path (-1, 0) -- (1, 0);
    \path (0, 1.4) -- (0, -1.4);
    \def\pos{(0,0)}
    \draw \pos -- +(1,1);
    \draw \pos -- +(-1,0);
    \draw \pos -- +(0,-1);
}

\newcommand{\hexviewtwoL}{
    \path (-2, 0) -- (2, 0);
    \path (0, 2.4) -- (0, -2.4);
    \begin{scope}
        \foreach \pos in {(1, 1), (-1,0), (0,-1)} {
            \draw \pos -- +(-1,-1);
            \draw \pos -- +(1,0);
            \draw \pos -- +(0,1);
        }
    \end{scope}
}

\newcommand{\hexviewtwoR}{
    \path (-2, 0) -- (2, 0);
    \path (0, 2.4) -- (0, -2.4);
    \begin{scope}
        \foreach \pos in {(-1, -1), (1,0), (0,1)} {
            \draw \pos -- +(1,1);
            \draw \pos -- +(-1,0);
            \draw \pos -- +(0,-1);
        }
    \end{scope}
}

\newcommand{\hex}[1]{(#1)_\hexagon}

\begin{document}

\title{Exploration of Infinite Triangular and Hexagonal Grids by
Synchronous Robots without Chirality}
\titlerunning{Exploration of Infinite Triangular and Hexagonal Grids}

\author{Daniel Babický \and Michal Opler}
\authorrunning{D. Babický and M. Opler}

\institute{Faculty of Information Technology, Czech Technical University in Prague,\\
Prague, Czech Republic\\
\email{babicdan@fit.cvut.cz}, \email{michal.opler@fit.cvut.cz}}

\maketitle

\begin{abstract}
We study exploration of infinite triangular and hexagonal grids by a
swarm of anonymous robots equipped with a light that can take one of
finitely many colors. The robots operate synchronously and execute a
shared movement protocol in which every move depends only on a local
view of bounded visibility range, including the colors of nearby robots;
the robots have neither a compass nor a sense of chirality.

For triangular grids, we present three new algorithms, each optimal in
one of the three respective resources: the number of colors, the
visibility range, and the number of robots. One of them strictly
improves the only previously known exploration algorithm for triangular
grids both in visibility range and in the number of colors, while still
using the optimal number of five robots. All three algorithms are
deterministic and never require the robots to change their color.

For hexagonal grids, we provide the first exploration algorithms, again
achieving optimality in the number of colors, the visibility range, and
the number of robots, respectively.

\keywords{infinite exploration \and robot swarms \and triangular grid \and hexagonal grid}
\end{abstract}

\section{Introduction}

\newcommand{\sqalg}[1]{$\mathcal{B}^\square_{#1}$}
\newcommand{\trialgb}[1]{$\mathcal{B}^\triangle_{#1}$}
\newcommand{\trialg}[1]{$\mathcal{A}^\triangle_{#1}$}
\newcommand{\hexalg}[1]{$\mathcal{A}^\hexagon_{#1}$}

Exploring an unknown environment with a swarm of simple mobile agents is an extensively researched problem.
The variants differ in whether the environment is finite or infinite, continuous or discrete, and in the computational power, synchronicity and communication abilities of the agents.
Its roots reach back to the question of which labyrinths, i.e., finite connected subgraphs of the infinite square grid, can be explored by a single finite automaton aided by pebbles or by fellow automata~\cite{BK78,BS77}.
A modern incarnation of the same question asks how many finite automata, able to communicate if and only if they occupy the same cell, are needed to visit every cell of the \emph{infinite} square grid~\cite{EmekLSUW15,EmekLUW14,FeinermanK17}.
There, the answer is by now known exactly, and it depends delicately on the scheduler and on the access to randomness~\cite{BUW20,CELU17}.

The robots we consider follow the \emph{Look-Compute-Move} (LCM) model of Suzuki and Yamashita~\cite{suzuki1999distributedCFORMATION}, in which anonymous robots repeatedly observe their surroundings, compute an action and then execute it.
Robots in this model are considerably weaker than finite automata: in the basic setting they are oblivious, and the only state they carry between cycles is a light that takes one of finitely many colors and is visible to the robots around them~\cite{DAS2016171SYNC}.
Crucially, they are also disoriented, agreeing on no global compass.
In some works, including ours, the robots moreover lack a common chirality, that is, a shared notion of clockwise direction.
This is modeled by letting an adversary map the view of each robot through an arbitrary symmetry of the grid, rotations and reflections alike, independently in each round, before the robot decides on its next move.

In this setting, the benchmark task is the \emph{infinite grid exploration} (IGE) problem: every node of an infinite grid must eventually be visited by some robot.
We work in the \emph{fully synchronous} (FSYNC) model, where all robots execute their LCM cycle simultaneously, and our robots are myopic and opaque, seeing only the nodes within a fixed visibility range~$\Phi$ that are not hidden behind another robot.
Exploration must moreover be collision-free, meaning no two robots ever occupy the same node or swap along an edge.
The quality of a solution is measured in three values that cannot be minimized simultaneously: the number of robots, the number of colors of their lights and the visibility range~$\Phi$.
In the square grid, Bramas et al.~\cite{BRAMAS2021} showed that five robots are necessary no matter how many colors and how large a visibility range are available, and later~\cite{BRAMAS2025193} that no algorithm can succeed with robots of a single color under visibility range one.
The goal is therefore not a single best algorithm, but the best possible trade-offs between these parameters.

\paragraph{State of the art}
For the infinite square grid, this programme is nearly complete.
Bramas et al.~\cite{BRAMAS2021} first considered the model with common chirality: six robots of three fixed colors under visibility range one suffice for IGE when collisions are allowed, and five robots with five modifiable color states under visibility range one suffice for the collision-free variant, matching the lower bound above.
In a follow-up paper~\cite{BRAMAS2025193}, the same authors removed the assumption of chirality and gave three algorithms, each optimal in one of the three resources: eight robots of a single color under visibility range two, six robots of two colors under visibility range one, and five robots of twelve colors under visibility range one.
The same programme has been carried out for finite square grids as well, where perpetual exploration is understood both with~\cite{BLD23} and without~\cite{RBDLL21} chirality, as well as under an asynchronous scheduler~\cite{BDDLL24}.

Beyond the square grid, much less is known.
Das and Sau~\cite{DASRAJA} recently gave the first, and so far only, algorithm for the infinite \emph{triangular} grid, using five robots of four fixed colors under visibility range two, with no matching lower bound.
In the finite triangular setting, perpetual exploration has been studied for grids cut out by a rectangle~\cite{DGMS25,DGS24ssss}; for gathering on the infinite triangular grid, visibility range one is known to be insufficient without axis agreement~\cite{GSGS25}.

For the infinite \emph{hexagonal} grid, finally, we are not aware of any exploration result at all.
It has been considered for arbitrary pattern formation, which has been solved uniformly for all three regular tessellation graphs~\cite{CDDN23}, albeit with robots of unlimited visibility.

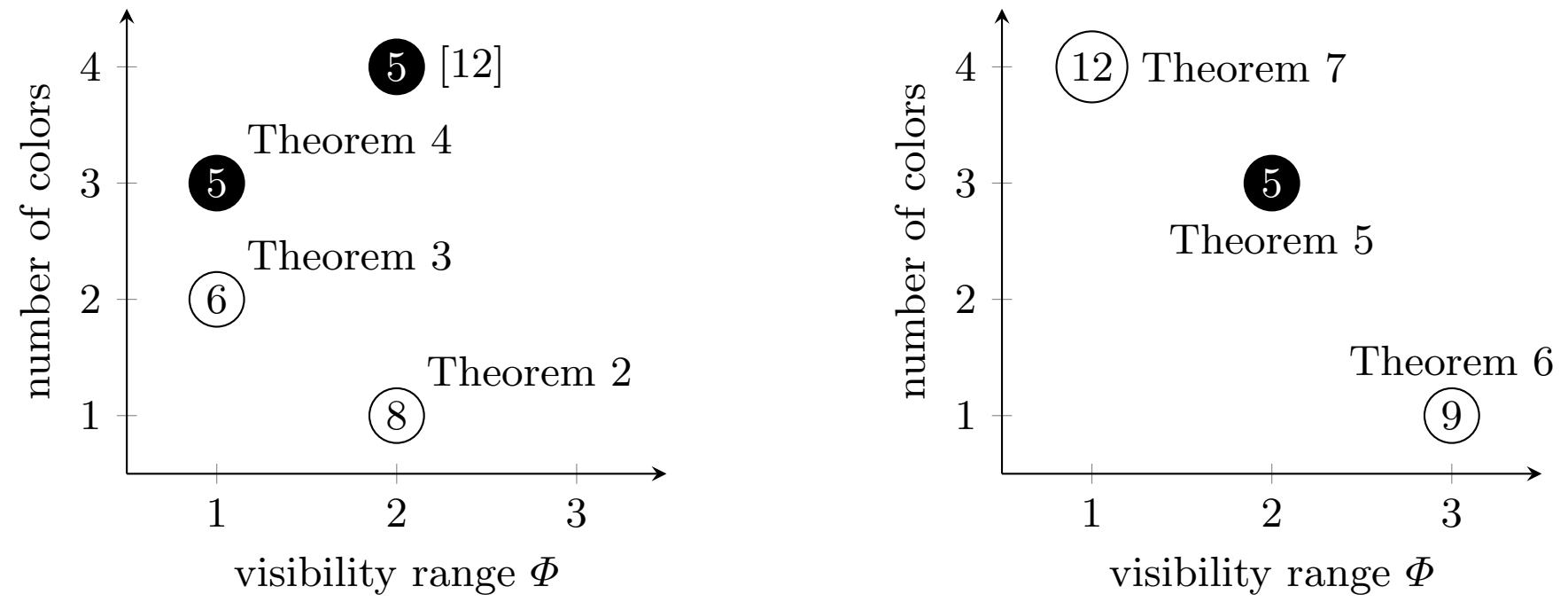
\begin{figure}[t]
    \centering
    \begin{subfigure}[c]{0.46\textwidth}
        \centering
        \begin{tikzpicture}
            \begin{axis}[
                width=\linewidth, height=0.9\linewidth,
                xlabel={visibility range $\Phi$},
                ylabel={number of colors},
                xmin=0.5, xmax=3.5,
                ymin=0.5, ymax=4.5,
                xtick={1,2,3},
                ytick={1,2,3,4},
                axis lines=left,
                clip=false,
                xlabel style={font=\small},
                ylabel style={font=\small},
                tick label style={font=\small},
            ]
                \node[circle, draw=black, fill=black, text=white, font=\small, inner sep=1.5pt] (tri5) at (axis cs:1,3) {5};
                \node[font=\footnotesize, fill=white, inner sep=0.5pt, above right=2pt of tri5] {Theorem~\ref{the:tri31}};
                \node[circle, draw=black, fill=black, text=white, font=\small, inner sep=1.5pt] (tridas) at (axis cs:2,4) {5};
                \node[font=\footnotesize, fill=white, inner sep=0.5pt, right=2pt of tridas] {\cite{DASRAJA}};
                \node[circle, draw=black, fill=white, font=\small, inner sep=1.5pt] (tri8) at (axis cs:2,1) {8};
                \node[font=\footnotesize, fill=white, inner sep=0.5pt, above right=2pt of tri8] {Theorem~\ref{the:tri12}};
                \node[circle, draw=black, fill=white, font=\small, inner sep=1.5pt] (tri6) at (axis cs:1,2) {6};
                \node[font=\footnotesize, fill=white, inner sep=0.5pt, above right=2pt of tri6] {Theorem~\ref{the:tri21}};
            \end{axis}
        \end{tikzpicture}
        \caption{The triangular grid}
        \label{fig:triplot}
    \end{subfigure}
    \hfill
    \begin{subfigure}[c]{0.46\textwidth}
        \centering
        \begin{tikzpicture}
            \begin{axis}[
                width=\linewidth, height=0.9\linewidth,
                xlabel={visibility range $\Phi$},
                ylabel={number of colors},
                xmin=0.5, xmax=3.5,
                ymin=0.5, ymax=4.5,
                xtick={1,2,3},
                ytick={1,2,3,4},
                axis lines=left,
                clip=false,
                xlabel style={font=\small},
                ylabel style={font=\small},
                tick label style={font=\small},
            ]
                \node[circle, draw=black, fill=black, text=white, font=\small, inner sep=1.5pt] (hex5) at (axis cs:2,3) {5};
                \node[font=\footnotesize, fill=white, inner sep=0.5pt, below=2pt of hex5] {Theorem~\ref{the:hex31}};
                \node[circle, draw=black, fill=white, font=\small, inner sep=1.5pt] (hex9) at (axis cs:3,1) {9};
                \node[font=\footnotesize, fill=white, inner sep=0.5pt, above=2pt of hex9] {Theorem~\ref{the:hex1color}};
                \node[circle, draw=black, fill=white, font=\small, inner sep=1.5pt] (hex12) at (axis cs:1,4) {12};
                \node[font=\footnotesize, fill=white, inner sep=0.5pt, right=2pt of hex12] {Theorem~\ref{the:hex1view}};
            \end{axis}
        \end{tikzpicture}
        \caption{The hexagonal grid}
        \label{fig:hexplot}
    \end{subfigure}
    \caption{The parameters of the algorithms discussed in this paper: each point corresponds to an algorithm, positioned by its visibility range and number of colors, with the number of robots inside the point and the corresponding theorem next to it. Filled points use the optimal number of robots.}
    \label{fig:algparams}
\end{figure}

\subsubsection{Our contribution}
We observe that the lower bound on the number of robots~\cite{BRAMAS2021} carries over to the triangular and hexagonal grids: four robots are insufficient, regardless of the number of colors and the visibility range used (Section~\ref{ch:impossibility}).

We then present three algorithms for the triangular grid, each optimal in a different one of three resources: the number of colors, the visibility range, and the number of robots.
The first uses eight robots of a single color under visibility range two (Theorem~\ref{the:tri12}), the second six robots of two colors under visibility range one (Theorem~\ref{the:tri21}), and the third five robots of three colors under visibility range one (Theorem~\ref{the:tri31}).
All three are deterministic and never require the robots to change their color.
Moreover, the five-robot algorithm strictly improves upon the only previously known algorithm of Das and Sau~\cite{DASRAJA} in both the visibility range and the number of colors, while still using the optimal number of five robots.

We also initiate the study of the IGE problem in the hexagonal grid, providing the first algorithms for it: one using five robots of three colors under visibility range two (Theorem~\ref{the:hex31}), one using nine robots of a single color under visibility range three (Theorem~\ref{the:hex1color}), and one using twelve robots of four colors under visibility range one (Theorem~\ref{the:hex1view}).
Again, they are optimal in the number of robots, the number of colors, and the visibility range, respectively.
The last algorithm is also the only one that is not well-defined, its execution depending on the adversary's choices.
We show this to be unavoidable: under visibility range one, any algorithm for the hexagonal grid must use an ambiguous rule (Lemma~\ref{lem:hexnowelldefined}).

The parameters of the algorithms discussed above are compared in Figure~\ref{fig:algparams}.
Since their executions are difficult to follow from static figures alone, we also provide an interactive visualization of all six algorithms.\footnote{\url{https://oplermic.pages.fit/demos/ige/}}

\section{Preliminaries}

We assume a non-empty finite set of robots evolving on the nodes of an infinite grid -- a straight-line drawing of a countably infinite planar graph $G = (V, E)$ with $V \subset \mathbb{R}^2$ where adjacent vertices will always have a Euclidean distance of one.
Throughout, we are given a \emph{visibility range} $\Phi \in \mathbb{N}^+$, which bounds the distance in the graph at which a robot can see other nodes, and a non-empty finite set of colors, denoted by $\mathit{X}$.

A \emph{configuration} $C$ is a set of pairs $(v, c)$, called \emph{states}, where $v \in V$ is the node occupied by a robot and $c \in \mathit{X}$ is the color of the robot. If there is only a single color in use or if the color is otherwise apparent from context, we may omit the color from the pair and only list a set of occupied nodes.

\paragraph{Views}
Let us consider a configuration $C$ and a robot~$r$ occupying a vertex~$v$ in this configuration.
The \emph{globally oriented view} of the robot~$r$ at~$v$, denoted by $V_C(v)$, is the subset of $C$ containing exactly the states of nodes whose distance from $v$ in the graph is at most $\Phi$, translated by $-v$ so that the robot~$r$ is placed at the origin.
Moreover, we consider opaque robots and thus, we also remove from the view the states of the nodes lying directly behind other robots.

It remains to model the disorientation of the robots.
To this end, we apply an \emph{indistinguishable transformation} to $V_C(v)$, chosen by an adversary.
The set of all indistinguishable transformations is denoted by $\mathcal{IT}$ and will be defined for each grid separately, consisting generally of all its symmetries.
A different transformation may be applied to different robots.
In addition, the robots are \emph{self-inconsistent}, meaning that a different transformation may be applied from round to round.
By applying a transformation to a global view, we obtain a \emph{local view}, based on which the robots compute their actions.
We denote by $\mathit{Views}$ the set of all obtainable local views.
We say that two views $W$ and $W'$ are \emph{indistinguishable} if $W = f(W')$ for some $f\in\mathcal{IT}$.

\paragraph{Algorithms}
An \emph{algorithm} is a tuple $(\mathit{X}, I, T)$, where $\mathit{X}$ is the set of all possible colors, $I$ is the initial configuration and \mbox{$T:\mathit{Views} \rightarrow N \times \mathit{X}$} is the \emph{transition function}, where $\mathit{Views}$ is the set of local views and $N \subset V$ is the set containing the vertex $(0, 0)$
and its neighbors.
We now consider a configuration $C$ and define all possible moves of a robot positioned on a vertex~$v$ where $(v,c) \in C$.
Formally, we say that a state $(w,c')$ is a \emph{successor of $(v,c)$ with respect to~$C$} if and only if there exists an indistinguishable transformation $f \in \mathcal{IT}$ such that the following holds:
\[f^{-1}( T( f( V_C(v) ) ) ) + v = (w, c').\]
We denote this by $(v,c) \mapsto_C (w,c')$.
In other words, we first obtain the globally oriented view of a robot located at $v$ and apply the transformation~$f$ to obtain a local view. The transition function then calculates the action taken, which is inversely transformed to make it globally oriented. This action is finally translated by $v$ to obtain the resulting position of the robot.

A configuration $C'$ is \emph{obtainable from $C$}, denoted as $C \mapsto C'$, if there is a bijection $\psi\colon C \to C'$ such that for every $(v,c) \in C$, we have $(v,c) \mapsto_C \psi((v,c))$.
Observe that we are not ruling out that two robots have swapped along an edge.
We will forbid such situations later, together with vertex collisions.
Finally, the \emph{execution} of an algorithm is an infinite sequence of configurations $(C_n)_{n \in \mathbb{N}}$ such that $C_0 = I$ and \mbox{$\forall i \ge 0,\ C_i \mapsto C_{i+1}$}.

We also need to avoid configurations where collisions are imminent.
We say that a configuration~$C$ is \emph{dangerous} if there are two states $(v_1, c_1)$ and $(v_2, c_2)$ in~$C$ and colors $c'_1, c'_2$ such that either both $(v_1, c_1) \mapsto_C (w, c'_1)$ and $(v_2, c_2) \mapsto_C (w, c'_2)$ for some vertex $w$ (vertex collision), or $(v_1, c_1) \mapsto_C (v_2, c'_1)$ and $(v_2, c_2) \mapsto_C (v_1, c'_2)$ (edge collision).

We say that an algorithm $\mathcal{A}$ \emph{solves} the \emph{infinite grid exploration} (IGE) problem if for every execution $(C_n)_{n \in \mathbb{N}}$ and every vertex $v \in V$, there exists a time $t \in \mathbb{N}$ such that $v$ is occupied in the configuration $C_t$, and such that no dangerous configuration is reachable from $I$; that is, there is no finite sequence of configurations $C_0, \dots, C_k$ such that $C_0 = I$, $C_k$ is dangerous and $C_{i-1} \mapsto C_{i}$ for each $i \in [k]$.

When describing algorithms in the upcoming sections, we will usually define the transition function $T$ as a set of rules where a \emph{rule} is a tuple $(W, v, c) \in \mathit{Views} \times N \times \mathit{X}$.
We say that the transition function includes the rule $(W, v, c)$ if $T(W) = (v, c)$.

The execution of an algorithm may vary if its transition function contains ambiguous rules (i.e., rules that have different outcomes depending on which indistinguishable transformation is used).
We say that an algorithm $(\mathit{X}, I, T)$ is \emph{well-defined} if it has a unique execution $(C_n)_{n \in \mathbb{N}}$.
Note that we do not require that all rules are unambiguous; we simply require that starting from the initial configuration~$I$, no ambiguous rule is used and the sequence of configurations is deterministic.
In our case, all well-defined algorithms will, in fact, not contain any ambiguous rules at all, which is an even stronger property.

\subsection{Grids}

We now formally define the coordinate systems and indistinguishable transformations for the triangular and hexagonal grids.

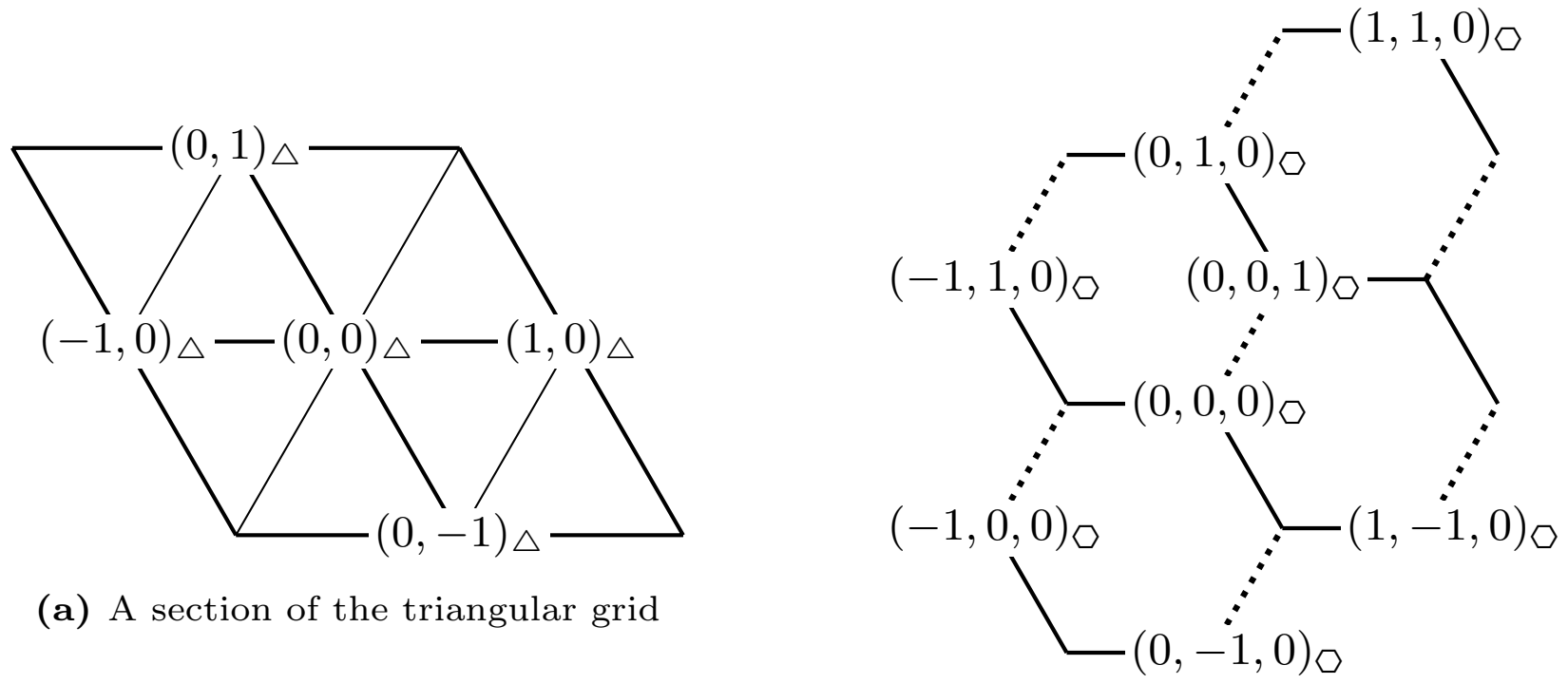
\begin{figure}[t]
    \hfill
    \centering
    \begin{subfigure}[c]{0.385\textwidth}
    \captionsetup{justification=centering}
    \centering
        \resizebox{\linewidth}{!}{\begin{tikzpicture}[triangular_slant, scale=1.6]
            \begin{scope}
                \draw[thick] (-1,-1) grid (1, 1);
                \foreach \i in {-1, ..., 0} {
                    \draw (-1, -\i) -- (\i, 1);
                    \draw (1, -\i) -- (\i, -1);
                }
            \end{scope}
            \begin{scope}[every node/.append style={fill=white, inner sep=1pt}]
                \node at (0,0) {$(0,0)_\triangle$};
                \node at (1,0) {$(1,0)_\triangle$};
                \node at (0,1) {$(0,1)_\triangle$};
                \node at (-1,0) {$(-1,0)_\triangle$};
                \node at (0,-1) {$(0,-1)_\triangle$};
            \end{scope}
        \end{tikzpicture}}\caption{A section of the triangular grid}
    \label{fig:triangle_grid}
    \end{subfigure}
    \hfill
    \begin{subfigure}[c]{0.39\textwidth}
    \captionsetup{justification=centering}
    \centering
        \resizebox{\linewidth}{!}{\begin{tikzpicture}[triangular_slant, scale=1.0]
            \begin{scope}
                \foreach \pos in {(0,2), (-1,0), (-2,-2),
                (2, 3), (1, 1), (0, -1)} {
                    \draw[thick] \pos -- +(1,0);
                }
                \foreach \pos in {(-2,-2), (-1,0), (0,-1),
                (1, 1), (2, 0), (3, 2)} {
                    \draw[thick] \pos -- +(0,1);
                }
                \foreach \pos in {(-2,-1), (-1,1),
                (-1, -2), (0, 0), (1, 2),
                (1, -1), (2, 1)
} {
                    \draw[dotted, very thick] \pos -- +(1,1);
                }

            \end{scope}
            \begin{scope}[every node/.append style={fill=white, inner sep=1pt}, l/.style={left=-.6cm}, r/.style={right=-.6cm}]
                \node[r] at (0,0) {$(0,0,0)_\hexagon$};
                \node[l] at (1,1) {$(0,0,1)_\hexagon$};

                \node[r] at (1,2) {$(0,1,0)_\hexagon$};

                \node at (-2,-1) {$(-1,0,0)_\hexagon$};

                \node[r] at (-1,-2) {$(0,-1,0)_\hexagon$};

                \node[r] at (3,3) {$(1,1,0)_\hexagon$};

                \node[r] at (1,-1) {$(1,-1,0)_\hexagon$};

                \node at (-1,1) {$(-1,1,0)_\hexagon$};

            \end{scope}
        \end{tikzpicture}}\caption{A section of the hexagonal grid}
    \label{fig:hexagon_grid}
    \end{subfigure}
    \hfill
    \caption{A visualization of the triangular and hexagonal coordinate systems}
    \label{fig:grids}
\end{figure}

\paragraph{Triangular grid}
To help define the coordinate system of the triangular grid, we introduce the following function $(\cdot)_\triangle: \mathbb{Z}^2 \rightarrow \mathbb{R}^2$, combining a horizontal shear transformation and a vertical scaling:
$$(x, y)_\triangle := \left(x - y\cdot\sin\left(\frac{\pi}{6}\right), \ y\cdot\cos\left(\frac{\pi}{6}\right)\right).$$
We then consider the graph $G$ with the vertex set $\{(x, y)_\triangle\ |\ (x, y) \in \mathbb{Z}\times\mathbb{Z}\}$ and an edge between two vertices if their Euclidean distance is equal to one. See Figure~\ref{fig:triangle_grid} for a better understanding of the coordinate system. Note that the robots cannot access these coordinates -- they are used purely in algorithm description and analysis.

The set of indistinguishable transformations $\mathcal{IT}$ contains all functions generated by all $\frac{k\pi}{3}$ rotations for $k\in\{0, \dots, 5\}$ (this includes the 0 radian rotation as the identity) together with the reflection of the plane along the vertical axis.

\paragraph{Hexagonal grid}
To help define the coordinate system of the hexagonal grid, we introduce the following function $(\cdot)_\hexagon: \mathbb{Z}\times\mathbb{Z}\times\{0,1\} \rightarrow \mathbb{R}^2$:
$$(x, y, z)_\hexagon := (x, y)_\triangle + (x + y + z) \cdot (1, 1)_\triangle
$$

We then consider the graph $G$ with the vertex set
$V = \{(x, y, z)_\hexagon\ |\ (x, y, z) \in \mathbb{Z}\times\mathbb{Z}\times\{0, 1\}\}$
and edges connecting vertices at Euclidean distance exactly one.

Alternatively, the construction of the grid can be thought of as follows. We begin with the set $\mathbb{Z}\times\mathbb{Z}$ and apply the $(\cdot)_\triangle$ function to all of its vertices, as shown in Figure~\ref{fig:triangle_grid}. Each vertex is then subdivided into two along the $\tri{1,1}$ direction, as is demonstrated in Figure~\ref{fig:hexagon_grid} by the dotted lines. The resulting vertices can then be thought of as a pair, only differing in the $z$ argument of the $\hex{x, y, z}$ function (with $z=0$ for the bottom left vertex and $z=1$ for the top right vertex). We refer to vertices of type $\hex{x, y, 0}$ as \emph{left-facing} and to vertices of type $\hex{x, y, 1}$ as \emph{right-facing}, as they have a neighboring vertex directly to the left and to the right, respectively.

The set of indistinguishable transformations $\mathcal{IT}$ contains all functions generated by  all $\frac{k\pi}{3}$ rotations for $k\in\{0, \dots, 5\}$ (this includes the 0 radian rotation as the identity) together with the reflection of the plane along the vertical axis.
This is the same set as in the triangular grid case. Note that the reflection and some rotations might change a left-facing vertex to a right-facing one and vice-versa.
This is desirable since we want rules to apply to both types.

\section{Impossibility result}
\label{ch:impossibility}

Bramas et al.~\cite{BRAMAS2021} showed that four robots cannot explore the infinite square grid, regardless of the number of colors and the visibility range.
We observe that their proof extends to the triangular and hexagonal grids, as it uses no property of the square grid beyond the following four: (i) there are finitely many vertices at a fixed distance from a common central vertex, (ii) the grid is periodic, i.e., there is a finite set of vertices such that every vertex can be moved onto one of them by a translation that maps the grid onto itself, (iii) the distance of two robots changes by at most two in a single round, and (iv) finitely many strips of finite width never cover all but finitely many vertices.
Together, (i) and (ii) imply that, up to translation, there are only finitely many configurations of robots lying within a fixed distance of each other.
All four properties hold in the triangular and hexagonal grids as well.
The only subtlety is (ii) in the hexagonal grid: there a single vertex does not suffice, as no translation maps a left-facing vertex onto a right-facing one, but two vertices, one of each kind, do.
We sketch the argument and the places where these properties are used in Appendix~\ref{ch:APimpossibility}.

\begin{theorem}\label{the:fiveoptimaltri}
There is no algorithm that solves the IGE problem in the infinite square, triangular or hexagonal grid using at most four robots.
\end{theorem}

\section{Triangular grid}
\label{ch:tri}

In this section, we propose three solutions for the IGE problem in the infinite triangular grid, denoted by \trialg{1}, \trialg{2} and \trialg{3}.
All of them follow the same scheme: a few robots form a \emph{moving group} that explores an ever larger part of the grid, while the remaining robots act as \emph{beacons} which delimit the explored area and reposition themselves only when approached by the moving group.
Due to space constraints, we describe in detail only the third algorithm and defer the other two, as well as all the correctness proofs, to Appendix~\ref{ch:APtri}.

The first algorithm~\trialg{1} is optimal with respect to the number of colors.
Five of its robots form the moving group, which explores the grid in an outward triangular spiral, and the remaining three act as beacons.

\begin{theorem}\label{the:tri12}Algorithm \trialg{1} solves the IGE problem using eight robots of a single color under visibility range $\Phi=2$. Moreover, it is well-defined.
\end{theorem}

The second algorithm~\trialg{2} solves the IGE problem using six robots of two colors under visibility range one, i.e., in the optimal visibility range.

\begin{theorem}\label{the:tri21}
Algorithm \trialg{2} solves the IGE problem using six robots of two fixed colors under visibility range $\Phi=1$. Moreover, it is well-defined.
\end{theorem}

\begin{figure}
\centering
    \begin{tikzpicture}[triangular_slant,scale=0.7]

        \triangulargrid{9}{5}
        \begin{scope}[shift={(1,0)}]
        \begin{scope}[every node/.style={robottiny,rred}]
        \node (front) at (5, 2) [rblue] {L};
        \node (back) at (4, 2) [rgreen] {F};
        \node (beacon1) at (4, 1) {B};
        \node (beacon2) at (6, 4) {B};
        \node (beacon3) at (2, 2) {B};
        \end{scope}
\begin{scope} [movestiny]
            \draw (front) -- +(0, 1);
            \draw (back) -- (front);
            \draw (beacon1) -- +(1,0);
        \end{scope}
        \end{scope}
    \end{tikzpicture}

    \caption[Initial configuration of algorithm \trialg{3}.]{Initial configuration of algorithm \trialg{3}, including the first movement}
    \label{fig:algotri3init}
\end{figure}
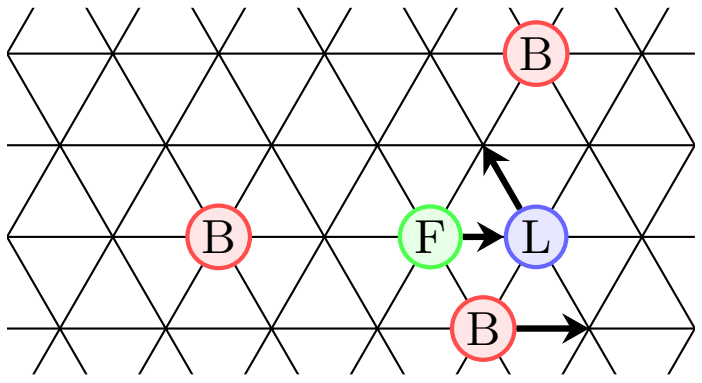

The final algorithm~\trialg{3} we propose for the triangular grid solves the IGE problem with five robots of three colors under visibility range one.
Recall that this is the minimum number of robots required by Theorem~\ref{the:fiveoptimaltri}.
The initial configuration is shown in Figure~\ref{fig:algotri3init}.

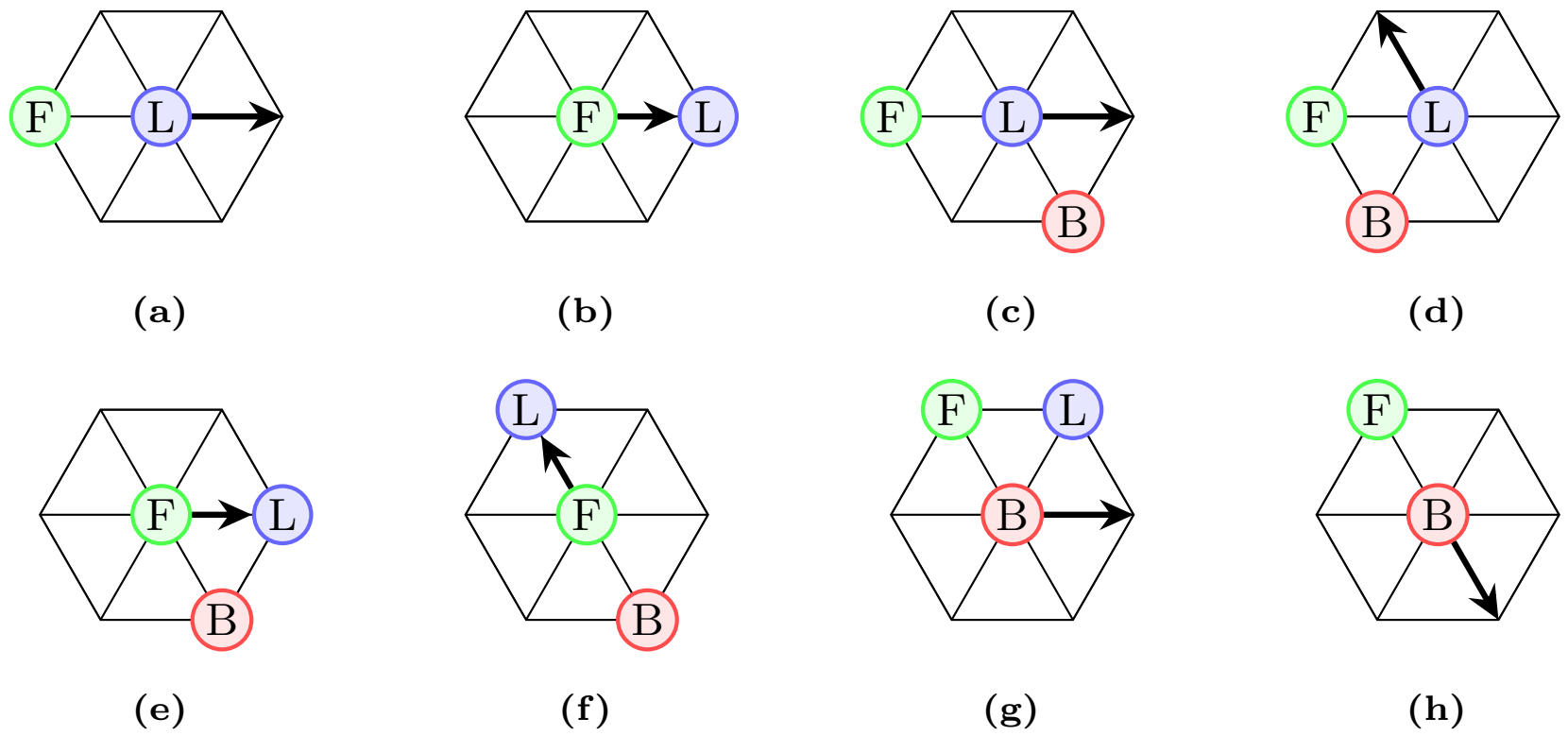
\begin{figure}[h!t]
\centering

\hspace*{\fill}\begin{subfigure}{2.7cm}
\centering
\captionsetup{justification=centering}
    \begin{tikzpicture}[triangular_slant,scale=\rulescale]
        \triangulargridview{2}

        \begin{scope}[shift={(1,1)}]
        \begin{scope}[every node/.style={robottiny,rblue}]
            \node (front) at (0, 0) {L};
            \node (back) at (-1, 0) [rgreen] {F};
        \end{scope}
        \begin{scope} [moves]
            \draw (front) -- +(1, 0);
        \end{scope}
        \end{scope}
    \end{tikzpicture}
    \caption{}
    \label{fig:tri3move1}
\end{subfigure}\hfill \begin{subfigure}{2.7cm}
\centering
\captionsetup{justification=centering}
    \begin{tikzpicture}[triangular_slant,scale=\rulescale]
        \triangulargridview{2}

        \begin{scope}[shift={(2,1)}]
        \begin{scope}[every node/.style={robottiny,rblue}]
            \node (front) at (0, 0) {L};
            \node (back) at (-1, 0) [rgreen] {F};
        \end{scope}
        \begin{scope} [moves]
            \draw (back) -- (front);
        \end{scope}
        \end{scope}
    \end{tikzpicture}
    \caption{}
    \label{fig:tri3move2}
\end{subfigure}\hfill \begin{subfigure}{2.7cm}
\centering
\captionsetup{justification=centering}
    \begin{tikzpicture}[triangular_slant,scale=\rulescale]
        \triangulargridview{2}

        \begin{scope}[shift={(1,1)}]
        \begin{scope}[every node/.style={robottiny,rblue}]
            \node (front) at (0, 0) {L};
            \node (back) at (-1, 0) [rgreen] {F};
            \node (beacon) at (0, -1) [rred] {B};
        \end{scope}
        \begin{scope} [moves]
            \draw (front) -- +(1, 0);
        \end{scope}
        \end{scope}
    \end{tikzpicture}
    \caption{}
    \label{fig:tri3turn1}
\end{subfigure}\hfill \begin{subfigure}{2.7cm}
\centering
\captionsetup{justification=centering}
    \begin{tikzpicture}[triangular_slant,scale=\rulescale]
        \triangulargridview{2}

        \begin{scope}[shift={(1,1)}]
        \begin{scope}[every node/.style={robottiny,rblue}]
            \node (front) at (0, 0) {L};
            \node (back) at (-1, 0) [rgreen] {F};
            \node (beacon) at (-1, -1) [rred] {B};
        \end{scope}
        \begin{scope} [moves]
            \draw (front) -- +(0, 1);
        \end{scope}
        \end{scope}
    \end{tikzpicture}
    \caption{}
    \label{fig:tri3turn2}
\end{subfigure}\hspace*{\fill}

\par\vspace{0.20cm}

\hspace*{\fill}\begin{subfigure}{2.7cm}
\centering
\captionsetup{justification=centering}
    \begin{tikzpicture}[triangular_slant,scale=\rulescale]
        \triangulargridview{2}

        \begin{scope}[shift={(2,1)}]
        \begin{scope}[every node/.style={robottiny,rblue}]
            \node (front) at (0, 0) {L};
            \node (back) at (-1, 0) [rgreen] {F};
            \node (beacon) at (-1, -1) [rred] {B};
        \end{scope}
        \begin{scope} [moves]
            \draw (back) -- (front);
        \end{scope}
        \end{scope}
    \end{tikzpicture}
    \caption{}
    \label{fig:tri3turn3}
\end{subfigure}\hfill \begin{subfigure}{2.7cm}
\centering
\captionsetup{justification=centering}
    \begin{tikzpicture}[triangular_slant,scale=\rulescale]
        \triangulargridview{2}

        \begin{scope}[shift={(1,1)}]
        \begin{scope}[every node/.style={robottiny,rblue}]
            \node (front) at (0, 1) {L};
            \node (back) at (0, 0) [rgreen] {F};
            \node (beacon) at (0, -1) [rred] {B};
        \end{scope}
        \begin{scope} [moves]
            \draw (back) -- (front);
        \end{scope}
        \end{scope}
    \end{tikzpicture}
    \caption{}
    \label{fig:tri3turn4}
\end{subfigure}\hfill \begin{subfigure}{2.7cm}
\centering
\captionsetup{justification=centering}
    \begin{tikzpicture}[triangular_slant,scale=\rulescale]
        \triangulargridview{2}

        \begin{scope}[shift={(2,2)}]
        \begin{scope}[every node/.style={robottiny,rblue}]
            \node (front) at (0, 0) {L};
            \node (back) at (-1, 0) [rgreen] {F};
            \node (beacon) at (-1, -1) [rred] {B};
        \end{scope}
        \begin{scope} [moves]
            \draw (beacon) -- +(1, 0);
        \end{scope}
        \end{scope}
    \end{tikzpicture}
    \caption{}
    \label{fig:tri3beacon1}
\end{subfigure}\hfill \begin{subfigure}{2.7cm}
\centering
\captionsetup{justification=centering}
    \begin{tikzpicture}[triangular_slant,scale=\rulescale]
        \triangulargridview{2}

        \begin{scope}[shift={(1,1)}]
        \begin{scope}[every node/.style={robottiny}]
            \node (beacon) at (0, 0) [rred] {B};
            \node (back) at (0, 1) [rgreen] {F};
        \end{scope}
        \begin{scope} [moves]
            \draw (beacon) -- +(0, -1);
        \end{scope}
        \end{scope}
    \end{tikzpicture}
    \caption{}
    \label{fig:tri3beacon2}
\end{subfigure}\hspace*{\fill}

\caption{Rules of algorithm \trialg{3}: the translation of the moving group (a)--(b), the adjustment of the moving group (c)--(f), and a repositioning of a beacon (g)--(h)}
\label{fig:trialg3rules}
\end{figure}

The first two rules in Figure~\ref{fig:trialg3rules} allow the moving group to move along a straight line.
The robots are colored based on their role, with the $L$-colored leader in the front (Figure~\ref{fig:tri3move1}) and the $F$-colored follower in the back (Figure~\ref{fig:tri3move2}).

When the $B$-colored beacon is first spotted, the moving group continues in the same direction (Figure~\ref{fig:tri3turn1}), until the beacon is seen by both robots.
At this point, the robots take on a triangle formation, in which all robots are of a different color, allowing for unambiguous action.
The leader makes a sharp turn away from the beacon (Figure~\ref{fig:tri3turn2}), with the follower trailing behind (Figure~\ref{fig:tri3turn3}). After this, the beacon is visible only to the follower robot, which moves towards the leader as if the beacon was not there (Figure~\ref{fig:tri3turn4}).

When the beacon is approached by the moving group, it idles until it can see both robots. Then, it moves onto the empty node adjacent to the leader robot (Figure~\ref{fig:tri3beacon1}). Finally, it completes its two-step readjustment by moving away from the follower, the only robot it sees (Figure~\ref{fig:tri3beacon2}).
See Figure~\ref{fig:algotri3adjustment}.

{
\begin{figure}[h!t]
\centering

\hspace*{\fill}\begin{subfigure}{0.3\textwidth}
\centering
\captionsetup{justification=centering}
    \begin{tikzpicture}[triangular_slant,scale=\adjscale]
        \triangulargrid{8}{6}
        \begin{scope}[shift={(1,1)}]
        \begin{scope}[every node/.style={robot,rred}]
        \node (front) at (3, 2) [rblue] {L};
        \node (back) at (2, 2) [rgreen] {F};
        \node (beacon) at (3, 1) {B};
        \end{scope}
        \begin{scope} [moves]
            \draw (front) -- +(1, 0);
            \draw (back) -- (front);
        \end{scope}
        \end{scope}
    \end{tikzpicture}
\caption{}
\end{subfigure}\hfill \begin{subfigure}{0.3\textwidth}
\centering
\captionsetup{justification=centering}
    \begin{tikzpicture}[triangular_slant,scale=\adjscale]
        \triangulargrid{8}{6}
        \begin{scope}[shift={(1,1)}]
        \begin{scope}[every node/.style={robot,rred}]
        \node (front) at (4, 2) [rblue] {L};
        \node (back) at (3, 2) [rgreen] {F};
        \node (beacon) at (3, 1) {B};
        \end{scope}
        \begin{scope} [moves]
            \draw (front) -- +(0, 1);
            \draw (back) -- (front);
            \draw (beacon) -- +(1,0);
        \end{scope}
        \end{scope}
    \end{tikzpicture}
\caption{}
\end{subfigure}\hfill \begin{subfigure}{0.3\textwidth}
\centering
\captionsetup{justification=centering}
    \begin{tikzpicture}[triangular_slant,scale=\adjscale]
        \triangulargrid{8}{6}
        \begin{scope}[shift={(1,1)}]
        \begin{scope}[every node/.style={robot,rred}]
        \node (front) at (4, 3) [rblue] {L};
        \node (back) at (4, 2) [rgreen] {F};
        \node (beacon) at (4, 1) {B};
        \end{scope}
        \begin{scope} [moves]
            \draw (front) -- +(0, 1);
            \draw (back) -- (front);
            \draw (beacon) -- +(0,-1);
        \end{scope}
        \end{scope}
    \end{tikzpicture}
\caption{}
\end{subfigure}\hspace*{\fill}

\caption{An adjustment being performed during the execution of \trialg{3}}
\label{fig:algotri3adjustment}
\end{figure}
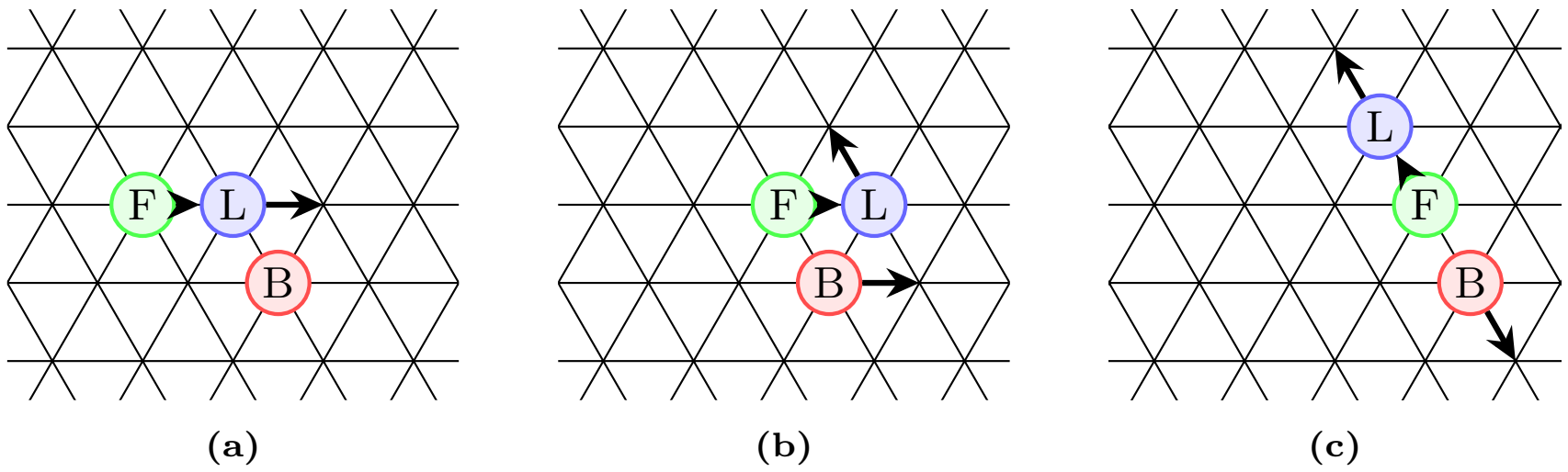
}

\begin{theorem}\label{the:tri31}
Algorithm \trialg{3} solves the IGE problem using five robots of three fixed colors under visibility range $\Phi=1$. Moreover, it is well-defined.
\end{theorem}

\proof[Sketch] The full proof can be found in Appendix~\ref{sec:APtri31}.
We divide the execution into phases, in each of which the moving group travels along the three sides of a triangle and repositions each of the three beacons further away from the center.
An induction on the number of phases shows that every phase is eventually reached and that the vertices visited during phase~$i$ are exactly those lying on the boundary of the $i$-th triangle.
These boundaries cover every vertex of the grid but one, lying in the middle of the spiral, which is occupied already in the initial configuration.
\qed

\renewcommand{\rulescale}{0.6}

\section{Hexagonal grid}

In this section, we study solutions to the IGE problem in the hexagonal grid, which is, to the best of our knowledge, the first time such a case has been considered.
Again, we describe in detail only the last of the three algorithms \hexalg{1}, \hexalg{2} and \hexalg{3} and defer the other two, together with all the correctness proofs, to Appendix~\ref{ch:APhex}.

The first algorithm~\hexalg{1} we propose for the hexagonal grid is optimal in the number of robots by Theorem~\ref{the:fiveoptimaltri}.
It is most comparable to \trialg{3}, as its moving group also consists of one $L$-colored leader and one $F$-colored follower, accompanied by three $B$-colored beacons.

\begin{theorem}\label{the:hex31}
Algorithm \hexalg{1} solves the IGE problem in the hexagonal grid using five robots of three colors under visibility range $\Phi=2$. It is well-defined.
\end{theorem}

The second algorithm~\hexalg{2} is optimal with respect to the number of colors.
Unlike previous algorithms, it uses six beacons instead of three, and so the spiraling path taken by its three-robot moving group is hexagonal rather than triangular.

\begin{theorem}\label{the:hex1color}
Algorithm \hexalg{2} solves the IGE problem in the hexagonal grid using nine robots of a single color under visibility range $\Phi=3$. It is well-defined.
\end{theorem}

\subsection{Exploration under visibility range one}
\label{sec:hex1view}

We begin this section by showing that any solution in the hexagonal grid using a visibility range of one will need to use at least one ambiguous rule.

\begin{lemma}\label{lem:hexnowelldefined}
    There is no well-defined algorithm that solves the IGE problem in the hexagonal grid under a visibility range of one using robots without chirality.
\end{lemma}

\begin{proof}
We proceed by contradiction. Suppose there exists a well-defined algorithm $\mathcal{A}$ that solves the IGE problem in the hexagonal grid under a visibility range of one. Let us assume, without loss of generality, that the position of every robot in the initial configuration $I$ of $\mathcal{A}$ can be written as $\hex{x, y, z}$, where $x\in\mathbb{Z}^-, y\in\mathbb{Z}$ and $z\in\{0,1\}$. Therefore, the $x$-coordinate of every robot is negative.
We define a set of right-facing vertices $W = \{\hex{0,y,1}\ |\ y\in\mathbb{Z}\}$, depicted in Figure~\ref{fig:hex1ambi}. Since the robots have to visit every vertex of the grid, they also have to eventually visit the vertices of $W$. We mark the first round in which a robot occupies a node in $W$ as round $d$.

\begin{figure}[t]
\centering
    \begin{subfigure}[c]{0.6\textwidth}
    \centering
    \begin{tikzpicture}[triangular_slant,scale=0.6]

        \hexgrid[(-0.5,-1)]{11.5}{6}
        \begin{scope}[shift={(6.5,0)}]
        \draw[dotted, thick] (0.25,0.5) -- (2.75, 5.5);
        \begin{scope}[every node/.style={robottiny,rviolet}]
            \node at (-3, 1) {};
            \node at (0, 4) {};
            \node at (-1, 5) {};
            \node at (-4, 2) {};
            \node at (-2, 5) {};
\end{scope}
        \foreach \i in {1, 2}
            \node at (\i,2*\i) [mark={8pt}, rectangle, thick] {};
        \end{scope}
    \end{tikzpicture}

    \caption{The set $W$ (squares) with all robots to the left}
    \label{fig:hex1ambi}
    \end{subfigure}
    \hfill
    \begin{subfigure}[c]{0.35\textwidth}
    \centering
    \begin{tikzpicture}[triangular_slant,scale=0.8]
        \hexviewoneL

        \begin{scope}[every node/.style={robot,rviolet}]
            \node (r) at (0, 0) [rblack] {$r$};
            \node at (-1, 0) {};
        \end{scope}
        \node (up) at (1,1) [mark={8pt}, rectangle, thick] {};
        \node (down) at (0,-1) [mark={8pt}, rectangle, thick] {};
        \begin{scope}[moves]
            \draw[dotted] (r) edge node[above left] {?} (up);
            \draw[dotted] (r) edge node[below left] {?} (down);
        \end{scope}
    \end{tikzpicture}

    \caption[The view of robot $r$ as it is about to visit a vertex of $W$.]{The view of a robot about to enter a vertex of $W$}
    \label{fig:hex1ambirobot}
    \end{subfigure}
\caption{Exploration with visibility range one requires ambiguous rules.}
\label{fig:hex1ambiguity}
\end{figure}
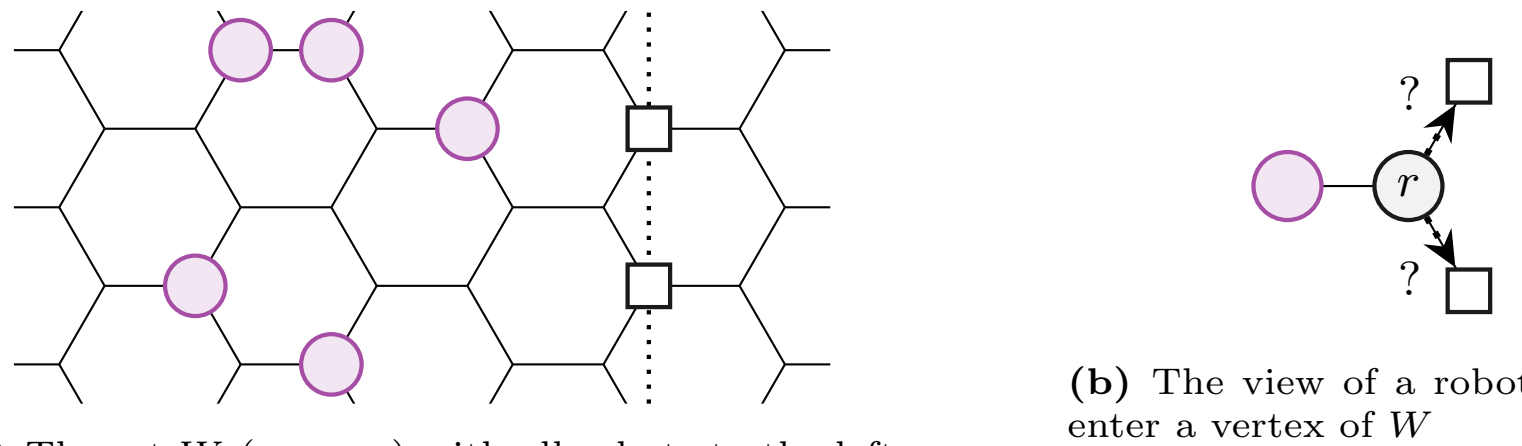

We now analyze the configuration at round $d-1$ and the moves of the robots in this configuration.
At round $d-1$, all robots still have to be to the left of~$W$, as if there were a robot to the right, it would have had to have crossed and occupied a node of $W$ before round $d-1$. For a robot to occupy a vertex of~$W$ at round $d$, there has to be a robot $r$ at position $\hex{0, y, 0}$ for some $y\in\mathbb{Z}$ in the previous round $d-1$. This robot cannot see any robots to the right, as they would be on the vertices of $W$.
Therefore, the only place where $r$ can see a different robot is to its left, at position $\hex{-1, y, 1}$.
If there were no robot at this position, the adversary could apply an indistinguishable transformation to make it move to the left.
Now, no matter what colors are used for the two robots, the view of robot $r$ is symmetric with respect to reflection along the $x$-axis (see Figure~\ref{fig:hex1ambirobot}), and therefore any action it takes to move onto a vertex of $W$ is ambiguous due to the lack of chirality.
\end{proof}

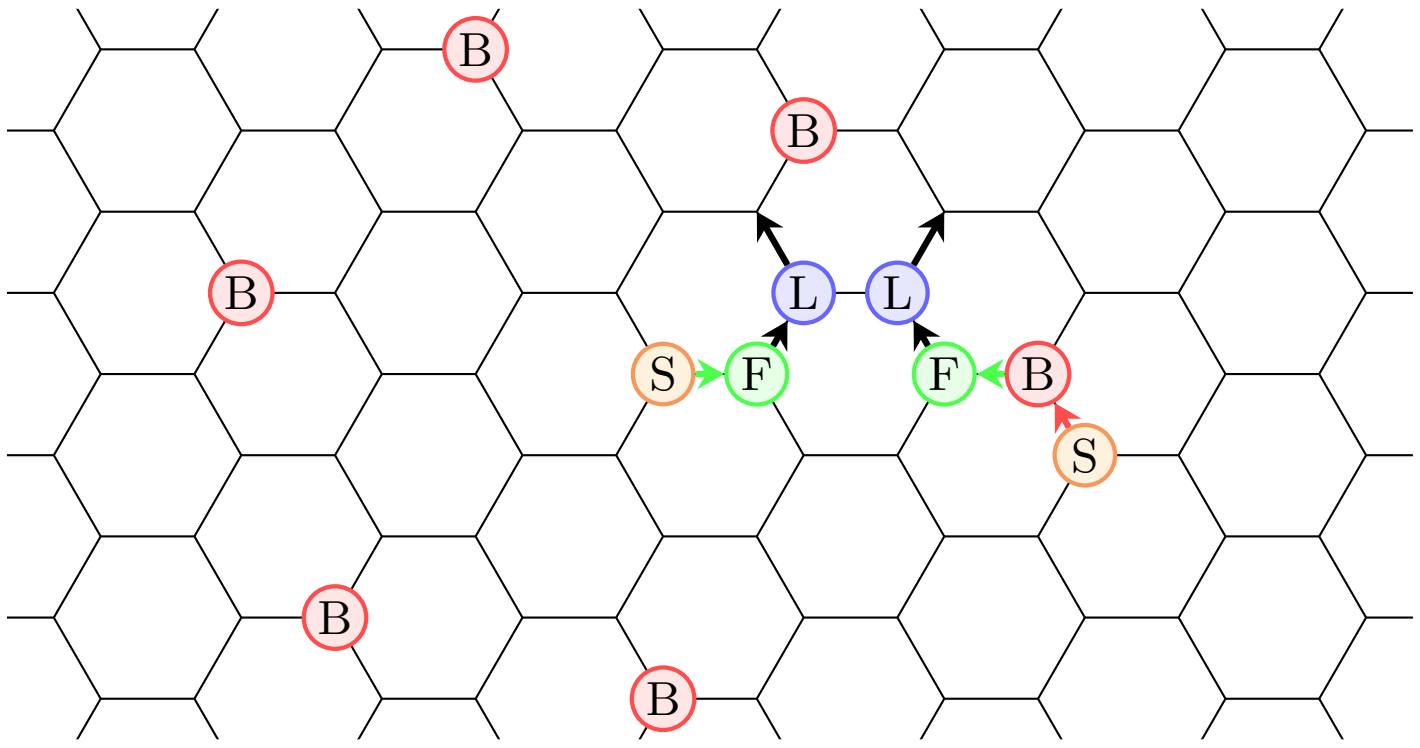
\begin{figure}
\centering
    \begin{tikzpicture}[triangular_slant,scale=0.62]

        \hexgrid[(-2,-1)]{19.5}{10}
        \begin{scope}[shift={(12,6)}]
        \begin{scope}[every node/.style={robottiny,rblack}]
            \node (1200) at (0, -1)[rgreen] {F};
            \node (-1553) at (-3, 3)[rred] {B};
            \node (-753) at (-5, -5)[rred] {B};
            \node (-691) at (-3, -1)[rorange] {S};
            \node (2130) at (1, -2)[rorange] {S};
            \node (276) at (-2, -1)[rgreen] {F};
            \node (1237) at (0, 0)[rblue] {L};
            \node (-3512) at (-7, 0)[rred] {B};
            \node (2167) at (1, -1)[rred] {B};
            \node (301) at (0, 2)[rred] {B};
            \node (270) at (-1, 0)[rblue] {L};
            \node (-2607) at (-8, -4)[rred] {B};
        \end{scope}
        \begin{scope} [movestiny]
            \draw (1200) -- (1237);
            \draw[rgreen] (-691) -- (276);
            \draw[rred] (2130) -- (2167);
            \draw (276) -- (270);
            \draw (1237) -- +(1, 1);
            \draw[rgreen] (2167) -- (1200);
            \draw (270) -- +(0, 1);
        \end{scope}
\end{scope}
    \end{tikzpicture}

    \caption[Initial configuration of algorithm \hexalg{3}.]{Initial configuration of~\hexalg{3}, including the first movement}\label{fig:algohex3init}
\end{figure}

Although a well-defined algorithm is not able to solve the IGE problem under these parameters, we propose an algorithm~\hexalg{3} that uses two ambiguous rules.
It uses four colors and twelve robots split into a moving group of six robots and six beacons.
The initial configuration is shown in Figure~\ref{fig:algohex3init}.

In the first round, the moving group overlaps with the position at which the right-most beacon is required.
To resolve this, the to-be beacon robot (the bottom right $S$-colored robot) is placed adjacent to the moving group and moves into position using an extra rule.
One robot of the moving group is recolored to allow this.

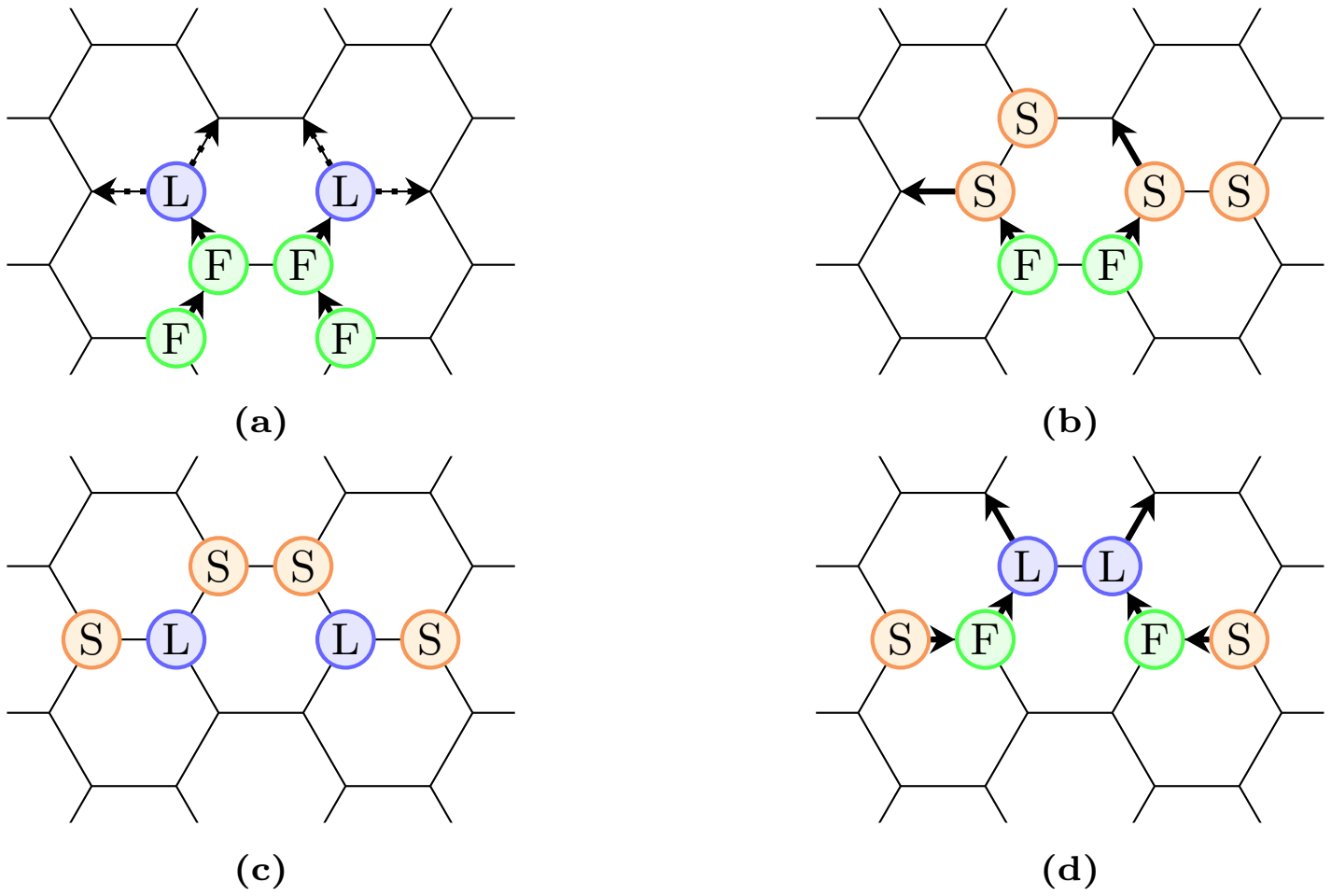
\begin{figure}[h!]
\centering
\hspace*{\fill}\begin{subfigure}{5cm}
\centering
\captionsetup{justification=centering}
    \begin{tikzpicture}[triangular_slant,scale=0.6]
    \hexgrid[(-1,-2)]{8.5}{6}
\begin{scope}[shift={(4,2)}]
    \begin{scope}[every node/.style={robottiny,rblack}]
        \node (1200) at (0, -1)[rgreen] {F};
        \node (307) at (-1, 1)[rblue] {L};
        \node (276) at (-2, -1)[rgreen] {F};
        \node (1237) at (0, 0)[rgreen] {F};
        \node (270) at (-1, 0)[rgreen] {F};
        \node (1231) at (1, 1)[rblue] {L};
    \end{scope}
    \begin{scope} [movestiny]
        \draw (1200) -- (1237);
        \draw[dotted] (307) -- +(1, 1);
        \draw[dotted] (307) -- +(-1, 0);
        \draw (276) -- (270);
        \draw (1237) -- (1231);
        \draw (270) -- (307);
        \draw[dotted] (1231) -- +(0, 1);
        \draw[dotted] (1231) -- +(1, 0);
    \end{scope}
    \end{scope}
\end{tikzpicture}
    \caption{}
    \label{fig:hex3move1}
\end{subfigure}\hfill \begin{subfigure}{5cm}
\centering
\captionsetup{justification=centering}
    \begin{tikzpicture}[triangular_slant,scale=0.6]
    \hexgrid[(-1,-2)]{8.5}{6}
\begin{scope}[shift={(4,2)}]
    \begin{scope}[every node/.style={robottiny,rblack}]
        \node (307) at (-1, 1)[rorange] {S};
        \node (1237) at (0, 0)[rgreen] {F};
        \node (2198) at (2, 1)[rorange] {S};
        \node (301) at (0, 2)[rorange] {S};
        \node (270) at (-1, 0)[rgreen] {F};
        \node (1231) at (1, 1)[rorange] {S};
    \end{scope}
    \begin{scope} [movestiny]
        \draw (307) -- +(-1, 0);
        \draw (1237) -- (1231);
        \draw (270) -- (307);
        \draw (1231) -- +(0, 1);
    \end{scope}
    \end{scope}
\end{tikzpicture}
    \caption{}
    \label{fig:hex3move2}
\end{subfigure}\hspace*{\fill}

\par\vspace{0.05cm}

\hspace*{\fill}\begin{subfigure}{5cm}
\centering
\captionsetup{justification=centering}
    \begin{tikzpicture}[triangular_slant,scale=0.6]
    \hexgrid[(-1,-2)]{8.5}{6}
\begin{scope}[shift={(4,2)}]
    \begin{scope}[every node/.style={robottiny,rblack}]
        \node (-660) at (-2, 1)[rorange] {S};
        \node (307) at (-1, 1)[rblue] {L};
        \node (1268) at (1, 2)[rorange] {S};
        \node (2198) at (2, 1)[rorange] {S};
        \node (301) at (0, 2)[rorange] {S};
        \node (1231) at (1, 1)[rblue] {L};
    \end{scope}
    \begin{scope} [movestiny]
\end{scope}
    \end{scope}
\end{tikzpicture}
    \caption{}
    \label{fig:hex3move3}
\end{subfigure}\hfill \begin{subfigure}{5cm}
\centering
\captionsetup{justification=centering}
    \begin{tikzpicture}[triangular_slant,scale=0.6]
    \hexgrid[(-1,-2)]{8.5}{6}
\begin{scope}[shift={(4,2)}]
    \begin{scope}[every node/.style={robottiny,rblack}]
        \node (-660) at (-2, 1)[rorange] {S};
        \node (307) at (-1, 1)[rgreen] {F};
        \node (1268) at (1, 2)[rblue] {L};
        \node (2198) at (2, 1)[rorange] {S};
        \node (301) at (0, 2)[rblue] {L};
        \node (1231) at (1, 1)[rgreen] {F};
    \end{scope}
    \begin{scope} [movestiny]
        \draw (-660) -- (307);
        \draw (307) -- (301);
        \draw (1268) -- +(1, 1);
        \draw (2198) -- (1231);
        \draw (301) -- +(0, 1);
        \draw (1231) -- (1268);
    \end{scope}
    \end{scope}
\end{tikzpicture}
    \caption{}
    \label{fig:hex3move4}
\end{subfigure}\hspace*{\fill}

\caption{The moving group of algorithm \hexalg{3}}
\label{fig:hex3movinggroup}
\end{figure}

The moving group then travels by periodic translations with a period equal to four, as shown in Figure~\ref{fig:hex3movinggroup}. It is essentially made up of two symmetrical groups of three robots, which periodically diverge and meet again, thus overcoming the limitation of the small visibility range.
Here, one of the ambiguous rules is used to advance the leader of the moving group, shown in Figure~\ref{fig:hex3move1} by dotted lines. The configuration in Figure~\ref{fig:hex3move2} is one of four possible configurations. Nonetheless, they all lead to the configuration in Figure~\ref{fig:hex3move3}. From there, the front $S$-colored robots can signal to the rest of the moving group about its orientation (Figure~\ref{fig:hex3move4}), after which the cycle repeats.

\begin{figure}
\centering
\hspace*{\fill}\begin{subfigure}{4cm}
\centering
\captionsetup{justification=centering}
    \begin{tikzpicture}[triangular_slant,scale=0.6]
    \hexgrid[(-1,-2)]{9}{7}
    \begin{scope}[shift={(4,2)}]
    \begin{scope}[every node/.style={robottiny,rblack}]
        \node (1200) at (0, -1)[rgreen] {F};
        \node (307) at (-1, 1)[rblue] {L};
        \node (276) at (-2, -1)[rgreen] {F};
        \node (1237) at (0, 0)[rgreen] {F};
        \node (301) at (0, 2)[rred] {B};
        \node (270) at (-1, 0)[rgreen] {F};
        \node (1231) at (1, 1)[rblue] {L};
    \end{scope}
    \begin{scope} [movestiny]
        \draw (1200) -- (1237);
        \draw (307) -- +(-1, 0);
        \draw (276) -- (270);
        \draw (1237) -- (1231);
        \draw (270) -- (307);
        \draw[dotted] (1231) -- +(0, 1);
        \draw[dotted] (1231) -- +(1, 0);
        \end{scope}
    \end{scope}
\end{tikzpicture}
    \caption{}
    \label{fig:hex3adj1}
\end{subfigure}\hfill \begin{subfigure}{4cm}
\centering
\captionsetup{justification=centering}
    \begin{tikzpicture}[triangular_slant,scale=0.6]
    \hexgrid[(-1,-2)]{9}{7}
    \begin{scope}[shift={(4,2)}]
    \begin{scope}[every node/.style={robottiny,rblack}]
        \node (-660) at (-2, 1)[rblue] {L};
        \node (307) at (-1, 1)[rorange] {S};
        \node (1237) at (0, 0)[rgreen] {F};
        \node (2198) at (2, 1)[rorange] {S};
        \node (301) at (0, 2)[rred] {B};
        \node (270) at (-1, 0)[rgreen] {F};
        \node (1231) at (1, 1)[rorange] {S};
    \end{scope}
    \begin{scope} [movestiny]
        \draw[dotted] (-660) -- +(0, 1);
        \draw[dotted] (-660) -- +(-1, -1);
        \draw (307) -- (-660);
        \draw (1237) -- (1231);
        \draw (270) -- (307);
        \draw (1231) -- +(0, 1);
    \end{scope}
    \end{scope}
\end{tikzpicture}
    \caption{}
    \label{fig:hex3adj2}
\end{subfigure}\hfill \begin{subfigure}{4cm}
\centering
\captionsetup{justification=centering}
    \begin{tikzpicture}[triangular_slant,scale=0.6]
    \hexgrid[(-1,-2)]{9}{7}
    \begin{scope}[shift={(4,2)}]
    \begin{scope}[every node/.style={robottiny,rblack}]
        \node (-660) at (-2, 1)[rgreen] {F};
        \node (307) at (-1, 1)[rblue] {L};
        \node (1268) at (1, 2)[rorange] {S};
        \node (2198) at (2, 1)[rorange] {S};
        \node (-654) at (-3, 0)[rgreen] {F};
        \node (301) at (0, 2)[rred] {B};
        \node (1231) at (1, 1)[rblue] {L};
    \end{scope}
    \begin{scope} [movestiny]
        \draw (-660) -- (307);
        \draw (307) -- +(0, -1);
\draw (-654) -- (-660);
\end{scope}
    \end{scope}
\end{tikzpicture}
    \caption{}
    \label{fig:hex3adj3}
\end{subfigure}\hspace*{\fill}

\par\vspace{0.05cm}

\hspace*{\fill}\begin{subfigure}{4cm}
\centering
\captionsetup{justification=centering}
    \begin{tikzpicture}[triangular_slant,scale=0.6]
    \hexgrid[(-1,-2)]{9}{7}
    \begin{scope}[shift={(4,2)}]
    \begin{scope}[every node/.style={robottiny,rblack}]
\node (307) at (-1, 1)[rred] {B};
        \node (-660) at (-2, 1)[rgreen] {F};
        \node (1268) at (1, 2)[rred] {B};
        \node (2198) at (2, 1)[rorange] {S};
\node (301) at (0, 2)[rred] {B};
        \node (270) at (-1, 0)[rblue] {L};
\node (1231) at (1, 1)[rgreen] {F};
    \end{scope}
    \begin{scope} [movestiny]
\draw (1268) -- (301);
        \draw (2198) -- (1231);
        \draw (301) -- +(0, 1);
\draw (1231) -- (1268);
    \end{scope}
    \end{scope}
\end{tikzpicture}
    \caption{}
    \label{fig:hex3adj4}
\end{subfigure}\hfill \begin{subfigure}{4cm}
\centering
\captionsetup{justification=centering}
    \begin{tikzpicture}[triangular_slant,scale=0.6]
    \hexgrid[(-1,-2)]{9}{7}
    \begin{scope}[shift={(4,2)}]
    \begin{scope}[every node/.style={robottiny,rblack}]
\node (338) at (0, 3)[rred] {B};
        \node (307) at (-1, 1)[rblue] {L};
        \node (-660) at (-2, 1)[rblue] {L};
        \node (1268) at (1, 2)[rorange] {S};
\node (301) at (0, 2)[rgreen] {F};
        \node (270) at (-1, 0)[rgreen] {F};
\node (1231) at (1, 1)[rgreen] {F};
    \end{scope}
    \begin{scope} [movestiny]
\draw (1268) -- +(1, 1);
        \draw (1231) -- (1268);
    \end{scope}
    \end{scope}
\end{tikzpicture}
    \caption{}
    \label{fig:hex3adj5}
\end{subfigure}\hfill \begin{subfigure}{4cm}
\centering
\captionsetup{justification=centering}
    \begin{tikzpicture}[triangular_slant,scale=0.6]
    \hexgrid[(-1,-2)]{9}{7}
    \begin{scope}[shift={(4,2)}]
    \begin{scope}[every node/.style={robottiny,rblack}]
\node (338) at (0, 3)[rblue] {L};
        \node (307) at (-1, 1)[rgreen] {F};
        \node (-660) at (-2, 1)[rblue] {L};
        \node (1268) at (1, 2)[rgreen] {F};
\node (301) at (0, 2)[rgreen] {F};
        \node (1262) at (2, 3)[rred] {B};
        \node (270) at (-1, 0)[rgreen] {F};
\end{scope}
    \begin{scope} [movestiny]
        \draw[dotted] (338) -- +(1, 1);
        \draw[dotted] (338) -- +(-1, 0);
        \draw (307) -- (-660);
        \draw[dotted] (-660) -- +(0, 1);
        \draw[dotted] (-660) -- +(-1, -1);
        \draw (1268) -- (301);
        \draw (301) -- (338);
\draw (270) -- (307);
    \end{scope}
    \end{scope}
\end{tikzpicture}
    \caption{}
    \label{fig:hex3adj6}
\end{subfigure}\hspace*{\fill}

\caption{An adjustment during the execution of \hexalg{3}}
\label{fig:hex3adjustment}
\end{figure}
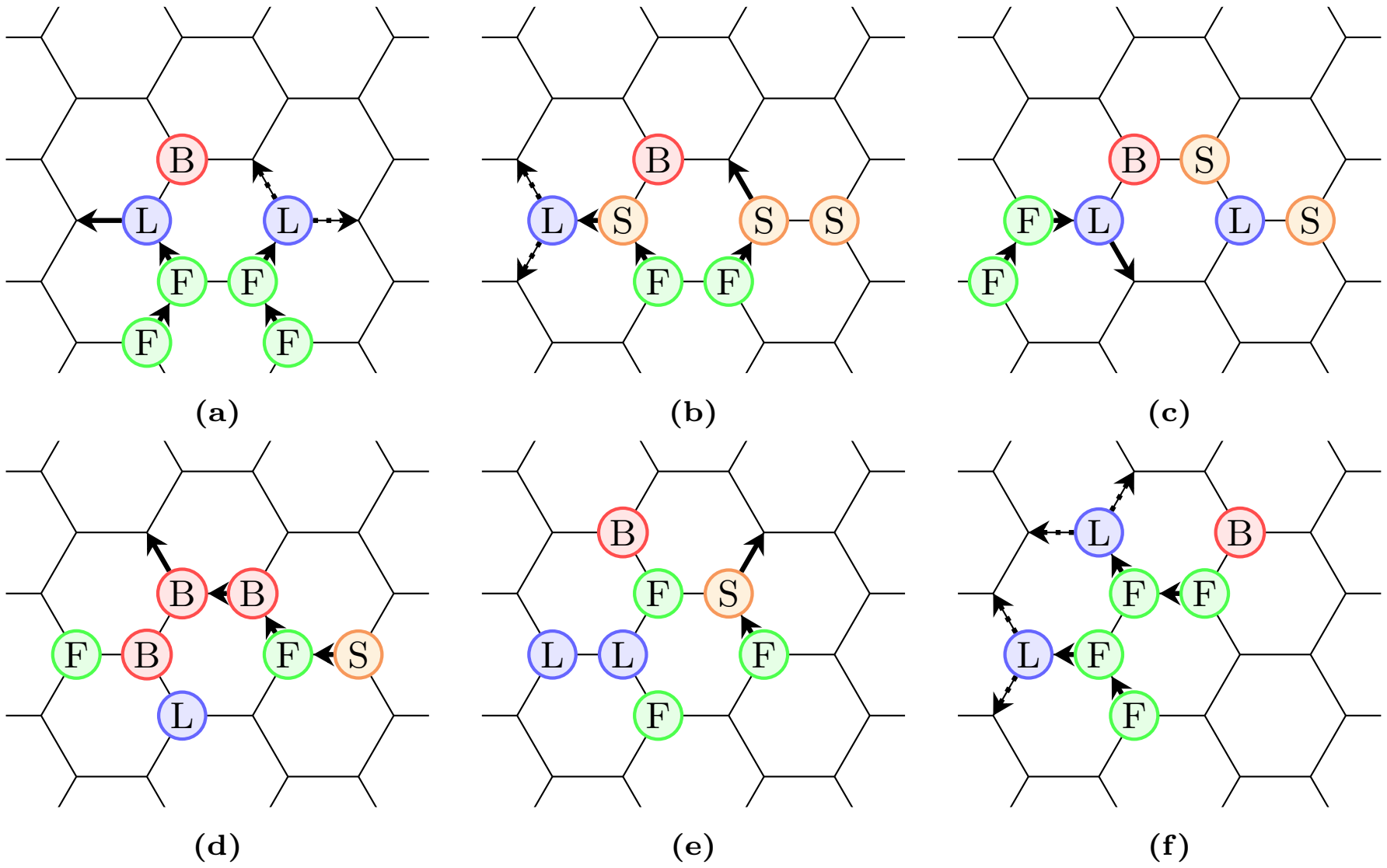

The algorithm uses six beacons, and so the moving group explores in an outward hexagonal spiral.
The moving group always encounters a beacon when in the H-like shape (Figure~\ref{fig:hex3adj1}). The outer side continues as if the beacon was not present, as it might not notice it until two rounds later. The inner side continues past the beacon robot, until all three robots have seen the beacon (Figures~\ref{fig:hex3adj1}, \ref{fig:hex3adj2} and \ref{fig:hex3adj3}). Here, the second ambiguous rule is used, as the front inner-side robot is required to move away to prevent a collision. No matter the indistinguishable transformation used, the configuration in Figure~\ref{fig:hex3adj4} is reached and no collision occurs. Also, the exploration does not depend on the outcome of this ambiguous rule, as both vertices are eventually visited in subsequent rounds.

In Figure~\ref{fig:hex3adj4}, the inner side has moved into position and only changes its colors from now on. The outer side also moves into position, leaving behind a beacon robot in the correct position for the next adjustment (Figures~\ref{fig:hex3adj4}, \ref{fig:hex3adj5} and \ref{fig:hex3adj6}). In the round following the configuration in Figure~\ref{fig:hex3adj6}, the beacon robot changes its color to $L$ due to a rule required in the previous round. An additional rule is then necessary to change its color to $B$ again.

\begin{theorem}\label{the:hex1view}
Algorithm \hexalg{3} solves the IGE problem in the hexagonal grid using twelve robots of four colors under visibility range $\Phi=1$.
\end{theorem}

\section{Open questions}

While we have extended the previous impossibility result regarding the number of robots, we have not considered the other parameters. Further research may  explore the optimality of the number of colors in relation to the visibility range.
One example of such a question is whether the triangular grid admits exploration using a single color under visibility range one.

Additionally, the solutions proposed for the hexagonal grid may have room for improvement. In particular, \hexalg{2} and \hexalg{3} use six beacon robots. Could they be modified to use only three beacons like the other algorithms? Finally, Bramas et al.~\cite{BRAMAS2025193} propose an algorithm in the square grid that solves the problem of \emph{perpetual infinite grid exploration}, a stronger form of IGE in which every vertex is to be visited infinitely often. What would such an algorithm look like in the triangular or hexagonal grid?

\begin{credits}
\subsubsection{\ackname}
This work was supported by the Student
Summer Research Program 2026 of FIT CTU in Prague.
Michal Opler was co-funded by the European Union under the project Robotics and advanced industrial production (reg.\ no.\ CZ.02.01.01/00/22\_008/0004590).
\end{credits}

\appendix

\newcommand{\translate}[2]{
\overrightarrow{t}_{#1}(#2)
}

\section{Justification of Theorem~\ref{the:fiveoptimaltri}}
\label{ch:APimpossibility}

We sketch the argument of Bramas et al.~\cite[Theorem 3]{BRAMAS2021} and indicate where it uses the four properties (i)--(iv) listed in Section~\ref{ch:impossibility}.

Since an algorithm has to solve the IGE problem against every adversary, it suffices to exhibit a single adversary against which it fails.
The adversary is fixed so that a robot seeing no other robot alternates between two adjacent vertices: the first time such a robot is instructed to move, the adversary applies the identity and anchors the robot to the edge it has just used, and afterwards it always makes the robot move to the other endpoint of this anchor, discarding the anchor as soon as the robot sees another robot again.
In particular, a robot that never sees another robot again occupies at most two vertices, and the execution of an algorithm against this adversary is uniquely determined.

Suppose that the maximum distance of two robots stays below a bound~$B$ in infinitely many rounds.
All the robots then lie within a ball of radius~$B$ and by property~(i), there are only finitely many configurations of this form up to a translation mapping the grid onto itself.
Property~(ii) is what makes this quotient finite in the right way: two configurations may be identified only if the translation taking one to the other is an automorphism of the grid, which in the hexagonal grid additionally requires the two classes of vertices to be matched.
Hence, two of the rounds carry the same configuration up to such a translation~$\mathbf{u}$, and as the execution is deterministic, it is periodic up to~$\mathbf{u}$ from that moment on.
The robots therefore never leave a strip of finite width and by property~(iv), infinitely many vertices are never visited.

Consequently, the maximum distance of two robots tends to infinity in every execution that visits all the vertices, and the rest of the argument is a case analysis of the ways in which four robots can be spread out.
By property~(iii), two robots far enough from each other can never come back within sight while they alternate between two vertices.
A group of at most two robots that stays isolated from the rest is again subject to the periodicity argument above and thus never leaves a strip, while a single isolated robot occupies at most two vertices.
The only remaining possibility is that two of the robots move back and forth between the other two, which are stationed far apart.
Ruling out this scenario is the technical heart of the argument and we refer the reader to~\cite[Theorem 3]{BRAMAS2021}.

\section{Triangular grid}
\label{ch:APtri}

We denote the translation of a configuration $C$ by a vector $\mathbf{u} \in \mathbb{R}^2$ as $\translate{\mathbf{u}}{C}$, i.e., every pair $(v, c) \in C$ becomes $(v+\mathbf{u}, c)$; recall that the vertices of a grid are points in the plane.

In this section, we propose three solutions for the IGE problem in the infinite triangular grid.
The first one uses the minimal number of colors at the cost of using a visibility range of two.
The second solution does the opposite, using the smallest possible visibility range with two colors.
The last solution is optimal in the number of robots.
All three algorithms proposed in this section are well-defined.

\subsection{Eight robots, single color, visibility range two}
\label{sec:APtri12}

\begin{figure}[h!t]
\centering
    \begin{tikzpicture}[triangular_slant,scale=0.85]

        \triangulargrid{12}{8}
        \begin{scope}[shift={(2,0)}]
        \begin{scope}[every node/.style={robot,draw=black!90,fill=black!2}]
            \node (front) at (5, 3) {R};
            \node (leftside) at (5, 4) {R};
            \node (rightside) at (4, 2) {R};
            \node (leftback) at (4, 4) {R};
            \node (rightback) at (3, 2) {R};
            \node (beacon1) at (5, 1) {R};
            \node (beacon2) at (7, 7) {R};
            \node (beacon3) at (0, 2) {R};
        \end{scope}
        \node (origin) at (0, 1) [mark] {};
        \begin{scope} [moves]
            \draw (front) -- +(1, 0);
            \draw (leftside) -- +(1, 0);
            \draw (rightside) -- +(0, 1);
            \draw (leftback) -- (leftside);
            \draw (rightback) -- (rightside);
            \draw (beacon1) -- +(1,0);
\end{scope}
        \end{scope}
    \end{tikzpicture}

    \caption[Initial configuration of algorithm \trialg{1}.]{Initial configuration of algorithm \trialg{1}, including the actions taken in the first round and the origin $\tri{0,0}$ marked by a circle}
    \label{fig:APalgotri1init}
\end{figure}
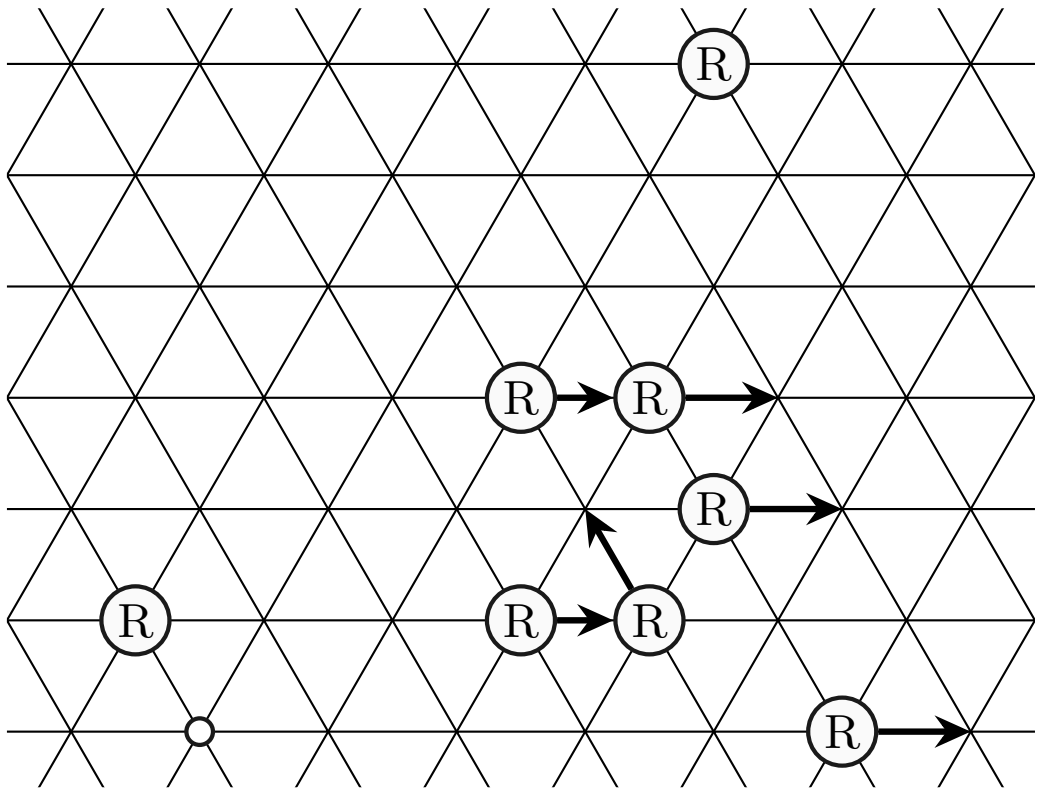

The first algorithm, which we will denote by \trialg{1}, solves the IGE problem in the triangular grid using eight robots of a single color and a visibility range of~two.
The existence of the algorithm proves that it is possible to solve the IGE problem in the triangular grid using robots of a single color.
It does this while also keeping the visibility range almost optimal. The initial configuration $I$ is shown in Figure~\ref{fig:APalgotri1init}. Five robots form the \emph{moving group}, which explores the grid in an outward triangular spiral. The other three act as separate \emph{beacons} that delimit the explored area, only repositioning when approached by the moving group. These two robot categories remain fixed throughout the entire execution. With the visibility range being two, robots may obstruct other robots that would otherwise be visible. This, however, does not happen at any point during the execution.

The algorithm \trialg{1} can be compared to the algorithm of Bramas et al.~\cite{BRAMAS2025193}, which solves the problem in the square grid under the exact same parameters. One key difference is in the sizes of each group, as their algorithm uses one less robot for the moving group, and instead requires an additional beacon.

The three rules shown in Figure~\ref{fig:APtrialg1movealong} allow the moving group to move in a straight line. Each robot in the group can be described as being either at the front (Figure~\ref{fig:APtri1move1}), on the side (Figure~\ref{fig:APtri1move2}), or at the back (Figure~\ref{fig:APtri1move3}), based on their unique views, as apparent from the diagrams. The single robot at the front is directly adjacent to the two side robots. Each of the two back robots is adjacent only to a single side robot. Due to the moving group being symmetric along its axis and the lack of chirality, both side robots move according to the same rule, as do the back robots.

\renewcommand{\rulescale}{0.63}

\begin{figure}[h!t]
\centering

\hspace*{\fill}\begin{subfigure}{3.5cm}
\centering
\captionsetup{justification=centering}
    \begin{tikzpicture}[triangular_slant,scale=\rulescale]
        \triangulargridview{4}

        \begin{scope}[shift={(2,2)}]
        \begin{scope}[every node/.style={robottiny,rblack}]
            \node (front) at (0, 0) {R};
            \node (leftside) at (0, 1) {R};
            \node (rightside) at (-1, -1) {R};
            \node (leftback) at (-1, 1) {R};
            \node (rightback) at (-2, -1) {R};
        \end{scope}
        \begin{scope} [moves]
            \draw (front) -- +(1, 0);
        \end{scope}
        \end{scope}
    \end{tikzpicture}
    \caption{}
    \label{fig:APtri1move1}
\end{subfigure}\hfill \begin{subfigure}{3.5cm}
\centering
\captionsetup{justification=centering}
    \begin{tikzpicture}[triangular_slant,scale=\rulescale]
        \triangulargridview{4}

        \begin{scope}[shift={(2,1)}]
        \begin{scope}[every node/.style={robottiny,rblack}]
            \node (front) at (0, 0) {R};
            \node (leftside) at (0, 1) {R};
            \node (rightside) at (-1, -1) {R};
            \node (leftback) at (-1, 1) {R};
            \node (rightback) at (-2, -1) {R};
        \end{scope}
        \begin{scope} [moves]
            \draw (leftside) -- +(1, 0);
        \end{scope}
        \end{scope}
    \end{tikzpicture}
    \caption{}
    \label{fig:APtri1move2}
\end{subfigure}\hfill \begin{subfigure}{3.5cm}
\centering
\captionsetup{justification=centering}
    \begin{tikzpicture}[triangular_slant,scale=\rulescale]
        \triangulargridview{4}

        \begin{scope}[shift={(3,1)}]
        \begin{scope}[every node/.style={robottiny,rblack}]
            \node (front) at (0, 0) {R};
            \node (leftside) at (0, 1) {R};
            \node (rightside) at (-1, -1) {R};
            \node (leftback) at (-1, 1) {R};
            \node (rightback) at (-2, -1) {R};
        \end{scope}
        \begin{scope} [moves]
            \draw (leftback) -- (leftside);
        \end{scope}
        \end{scope}
    \end{tikzpicture}
    \caption{}
    \label{fig:APtri1move3}
\end{subfigure}\hspace*{\fill}

\caption{Rules for the moving group in \trialg{1} to move along a straight line}
\label{fig:APtrialg1movealong}
\end{figure}
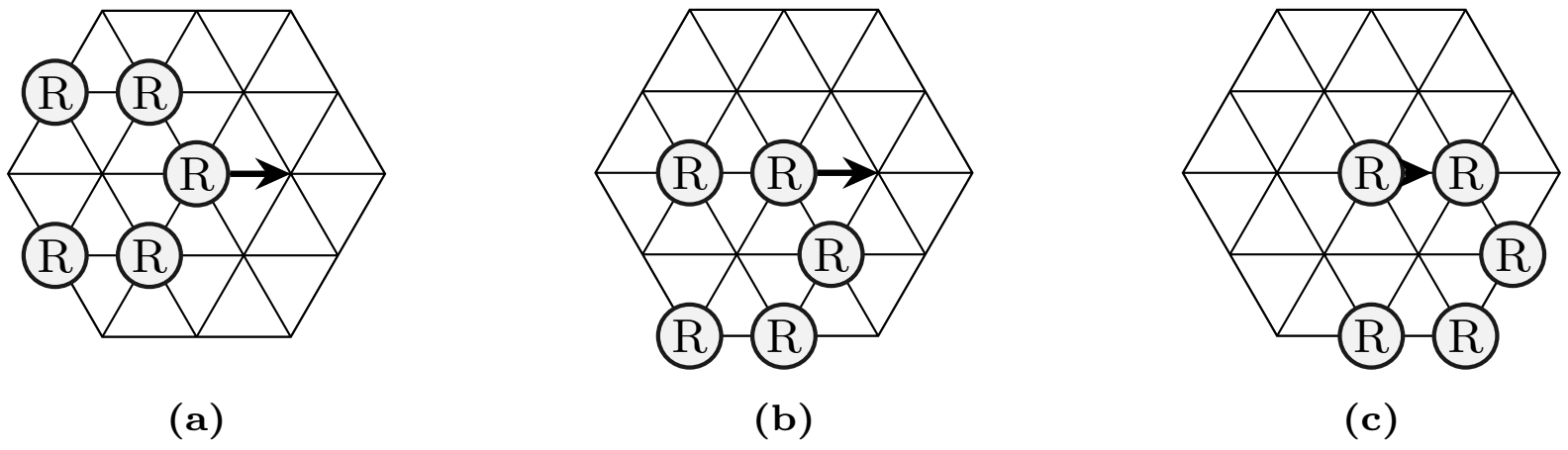

\begin{figure}[h!t]
\centering

\hspace*{\fill}\begin{subfigure}{3.5cm}
\centering
\captionsetup{justification=centering}
    \begin{tikzpicture}[triangular_slant,scale=\rulescale]
        \triangulargridview{4}

        \begin{scope}[shift={(2,2)}]
        \begin{scope}[every node/.style={robottiny,rblack}]
            \node (front) at (0, 0) {R};
            \node (leftside) at (0, 1) {R};
            \node (rightside) at (-1, -1) {R};
            \node (leftback) at (-1, 1) {R};
            \node (rightback) at (-2, -1) {R};
            \node (beacon) at (0, -2) {R};
        \end{scope}
        \begin{scope} [moves]
            \draw (front) -- +(1, 0);
        \end{scope}
        \end{scope}
    \end{tikzpicture}
    \caption{}
    \label{fig:APtri1turn1}
\end{subfigure}\hfill \begin{subfigure}{3.5cm}
\centering
\captionsetup{justification=centering}
    \begin{tikzpicture}[triangular_slant,scale=\rulescale]
        \triangulargridview{4}

        \begin{scope}[shift={(3,3)}]
        \begin{scope}[every node/.style={robottiny,rblack}]
            \node (front) at (0, 0) {R};
            \node (leftside) at (0, 1) {R};
            \node (rightside) at (-1, -1) {R};
            \node (leftback) at (-1, 1) {R};
            \node (rightback) at (-2, -1) {R};
            \node (beacon) at (0, -2) {R};
        \end{scope}
        \begin{scope} [moves]
            \draw (rightside) -- +(0, 1);
        \end{scope}
        \end{scope}
    \end{tikzpicture}
    \caption{}
    \label{fig:APtri1turn2}
\end{subfigure}\hfill \begin{subfigure}{3.5cm}
\centering
\captionsetup{justification=centering}
    \begin{tikzpicture}[triangular_slant,scale=\rulescale]
        \triangulargridview{4}

        \begin{scope}[shift={(1,3)}]
        \begin{scope}[every node/.style={robottiny,rblack}]
            \node (front) at (0, 0) {R};
            \node (leftside) at (-1, -1) {R};
            \node (rightside) at (1, 0) {R};
            \node (leftback) at (-1, -2) {R};
            \node (rightback) at (1, -1) {R};
            \node (beacon) at (1, -3) {R};
        \end{scope}
        \begin{scope} [moves]
            \draw (rightback) -- (rightside);
        \end{scope}
        \end{scope}
    \end{tikzpicture}
    \caption{}
    \label{fig:APtri1turn3}
\end{subfigure}\hspace*{\fill}

\caption{Rules for the moving group in \trialg{1} during an adjustment}
\label{fig:APtrialg1moveturn}
\end{figure}
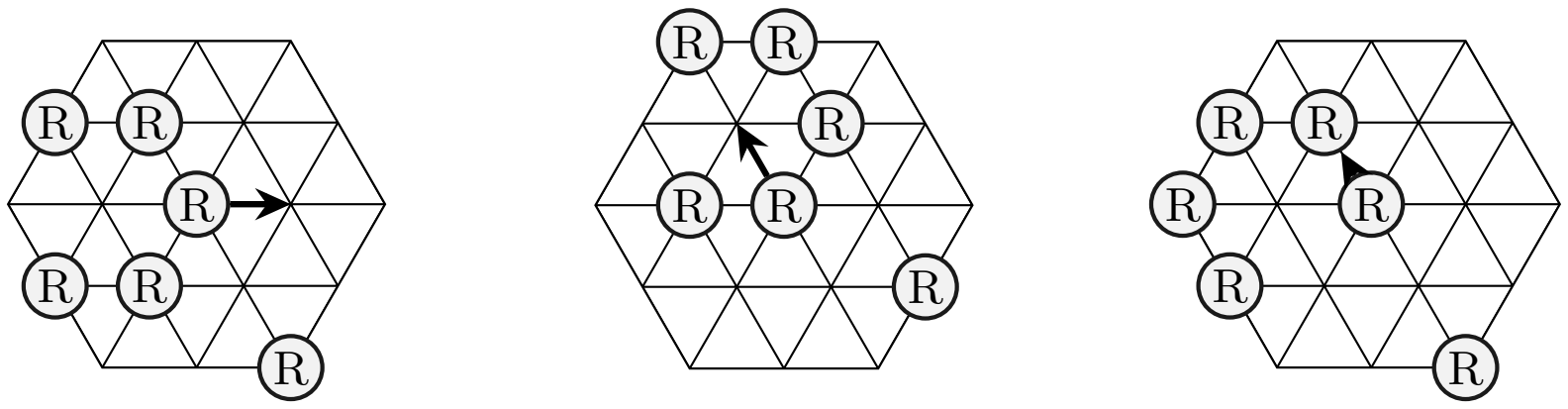

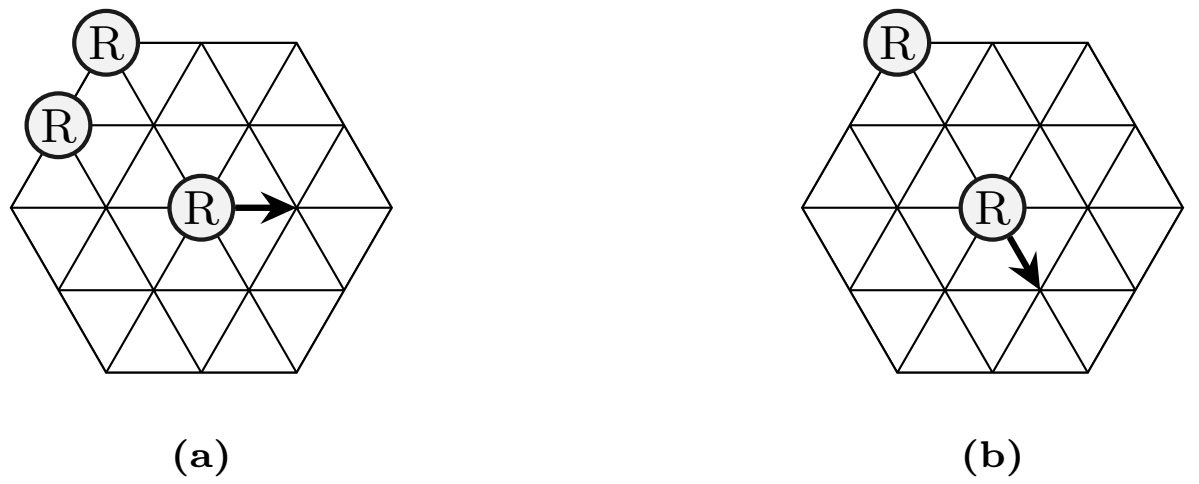
\begin{figure}[h!t]
\centering

\hspace*{\fill}\begin{subfigure}{3.5cm}
\centering
\captionsetup{justification=centering}
    \begin{tikzpicture}[triangular_slant,scale=\rulescale]
        \triangulargridview{4}

        \begin{scope}[shift={(2,4)}]
        \begin{scope}[every node/.style={robottiny,rblack}]
            \node (front) at (0, 0) {R};
            \node (rightside) at (-1, -1) {R};
            \node (beacon) at (0, -2) {R};
        \end{scope}
        \begin{scope} [moves]
            \draw (beacon) -- +(1, 0);
        \end{scope}
        \end{scope}
    \end{tikzpicture}
    \caption{}
    \label{fig:APtri1beacon1}
\end{subfigure}\hfill \begin{subfigure}{3.5cm}
\centering
\captionsetup{justification=centering}
    \begin{tikzpicture}[triangular_slant,scale=\rulescale]
        \triangulargridview{4}

        \begin{scope}[shift={(1,5)}]
        \begin{scope}[every node/.style={robottiny,rblack}]
            \node (rightback) at (1, -1) {R};
            \node (beacon) at (1, -3) {R};
        \end{scope}
        \begin{scope} [moves]
            \draw (beacon) -- +(0,-1);
        \end{scope}
        \end{scope}
    \end{tikzpicture}
    \caption{}
    \label{fig:APtri1beacon2}
\end{subfigure}\hspace*{\fill}

\caption{Rules for the beacons in \trialg{1} when spotting the moving group}
\label{fig:APtrialg1beaconmove}
\end{figure}

The next three rules depicted in Figure~\ref{fig:APtrialg1moveturn} apply when the moving group approaches a beacon, allowing an \emph{adjustment}. When a beacon comes into view of the moving group, only the front and one side robot are able to see it. The front robot continues moving as if the beacon was not there (Figure~\ref{fig:APtri1turn1}), as do the robots that cannot see the beacon. The side robot moves towards the empty space in the center of the moving group (Figure~\ref{fig:APtri1turn2}). This rearranges the moving group, preserving its shape, yet making it turn away sharply from the beacon. During this turn, some robots switch roles -- the front robot becomes a back robot and one of the back robots becomes a front robot. The beacon also moves along with the moving group during this, as shown later (Figure~\ref{fig:APtri1beacon1}). After the turn, the beacon is still in view of the back robot that was previously at the front. This requires an additional rule to allow the back robot to move with the moving group, away from the beacon (Figure~\ref{fig:APtri1turn3}).

The beacons themselves stay idle when the moving group is not in view. When approached by the moving group, rules in Figure~\ref{fig:APtrialg1beaconmove} apply. First, the beacon sees the front and side robot of the moving group, and moves along with the group (Figure~\ref{fig:APtri1beacon1}). During this, the side robot disappears from view, while the front robot stays, becoming a back robot in the process. Then, the beacon moves away from the only robot it sees, completing its repositioning (Figure~\ref{fig:APtri1beacon2}). After the adjustment, the explored perimeter is larger and the beacon is in position for the next adjustment. The entire sequence of moves made during an adjustment can be seen in Figure~\ref{fig:APalgotri1adjustment}.

\renewcommand{\adjscale}{0.75}

{
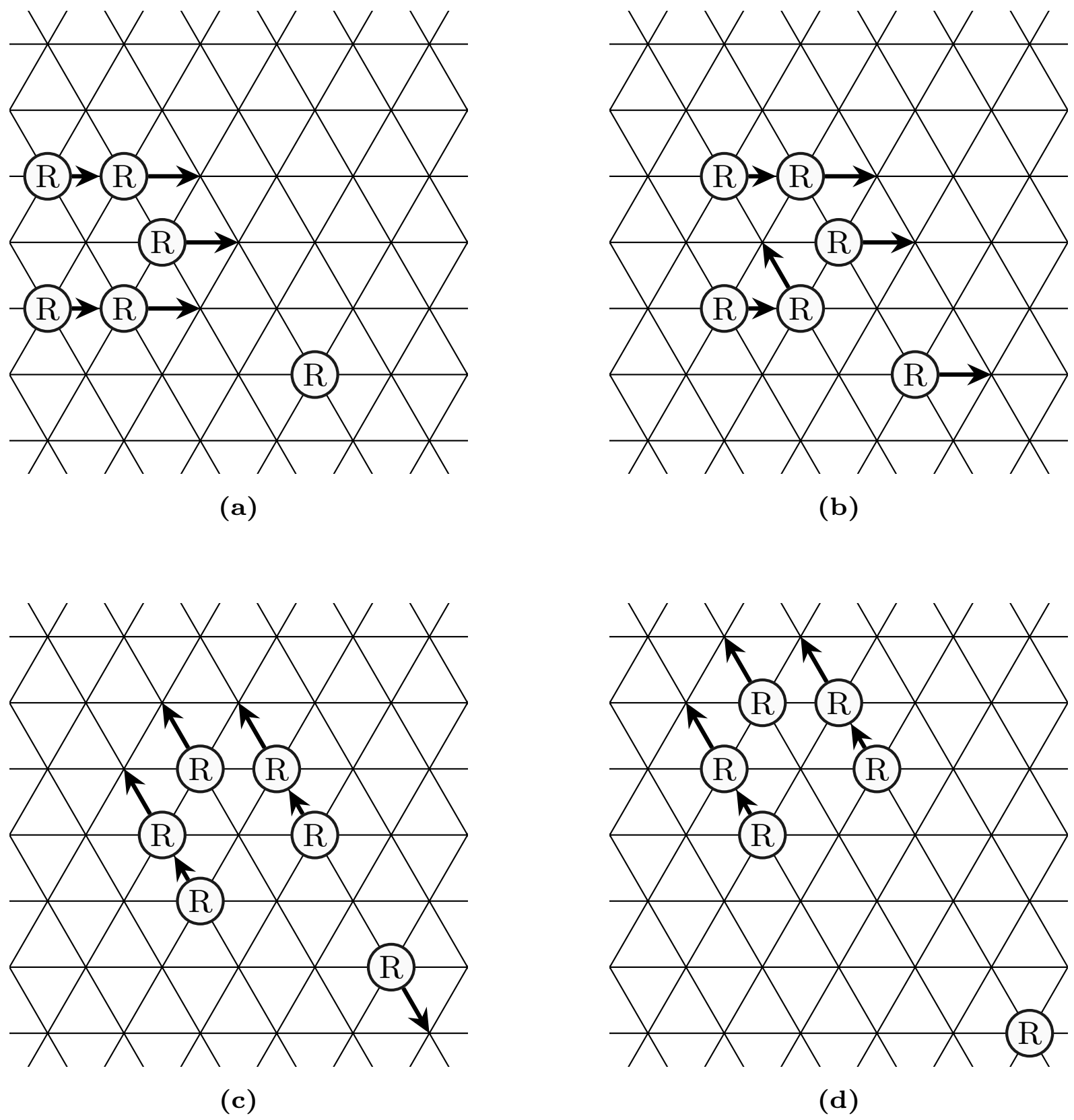
\begin{figure}[h!t]
\centering

\hspace*{\fill}\begin{subfigure}{0.45\textwidth}
\centering
\captionsetup{justification=centering}
    \begin{tikzpicture}[triangular_slant,scale=\adjscale]
        \triangulargrid{10}{8}
        \begin{scope}[shift={(0,1)}]
        \begin{scope}[every node/.style={robot,draw=black!90,fill=black!2}]
            \node (front) at (4, 3) {R};
            \node (leftside) at (4, 4) {R};
            \node (rightside) at (3, 2) {R};
            \node (leftback) at (3, 4) {R};
            \node (rightback) at (2, 2) {R};
            \node (beacon1) at (5, 1) {R};
        \end{scope}
        \begin{scope} [moves]
            \draw (front) -- +(1, 0);
            \draw (leftside) -- +(1, 0);
            \draw (rightside) -- +(1, 0);
            \draw (leftback) -- (leftside);
            \draw (rightback) -- (rightside);
        \end{scope}
        \end{scope}
    \end{tikzpicture}
    \caption{}
\end{subfigure}\hfill \begin{subfigure}{0.45\textwidth}
\centering
\captionsetup{justification=centering}
    \begin{tikzpicture}[triangular_slant,scale=\adjscale]
        \triangulargrid{10}{8}
        \begin{scope}[shift={(0,1)}]
        \begin{scope}[every node/.style={robot,draw=black!90,fill=black!2}]
            \node (front) at (5, 3) {R};
            \node (leftside) at (5, 4) {R};
            \node (rightside) at (4, 2) {R};
            \node (leftback) at (4, 4) {R};
            \node (rightback) at (3, 2) {R};
            \node (beacon1) at (5, 1) {R};
        \end{scope}
        \begin{scope} [moves]
            \draw (front) -- +(1, 0);
            \draw (leftside) -- +(1, 0);
            \draw (rightside) -- +(0, 1);
            \draw (leftback) -- (leftside);
            \draw (rightback) -- (rightside);
            \draw (beacon1) -- +(1,0);
        \end{scope}
        \end{scope}
    \end{tikzpicture}
    \caption{}
\end{subfigure}\hspace*{\fill}

\par\vspace{0.75cm}

\hspace*{\fill}\begin{subfigure}{0.45\textwidth}
\centering
\captionsetup{justification=centering}
    \begin{tikzpicture}[triangular_slant,scale=\adjscale]
        \triangulargrid{10}{8}
        \begin{scope}[shift={(0,1)}]
        \begin{scope}[every node/.style={robot,draw=black!90,fill=black!2}]
            \node (front) at (6, 3) {R};
            \node (leftside) at (6, 4) {R};
            \node (rightside) at (4, 3) {R};
            \node (leftback) at (5, 4) {R};
            \node (rightback) at (4, 2) {R};
            \node (beacon1) at (6, 1) {R};
        \end{scope}
        \begin{scope} [moves]
            \draw (front) -- (leftside);
            \draw (leftside) -- +(0, 1);
            \draw (rightside) -- +(0, 1);
            \draw (leftback) -- +(0, 1);
            \draw (rightback) -- (rightside);
            \draw (beacon1) -- +(0,-1);
        \end{scope}
        \end{scope}
    \end{tikzpicture}
    \caption{}
\end{subfigure}\hfill \begin{subfigure}{0.45\textwidth}
\centering
\captionsetup{justification=centering}
    \begin{tikzpicture}[triangular_slant,scale=\adjscale]
        \triangulargrid{10}{8}
        \begin{scope}[shift={(0,1)}]
        \begin{scope}[every node/.style={robot,draw=black!90,fill=black!2}]
            \node (front) at (6, 4) {R};
            \node (leftside) at (6, 5) {R};
            \node (rightside) at (4, 4) {R};
            \node (leftback) at (5, 5) {R};
            \node (rightback) at (4, 3) {R};
            \node (beacon1) at (6, 0) {R};
        \end{scope}
        \begin{scope} [moves]
            \draw (front) -- (leftside);
            \draw (leftside) -- +(0, 1);
            \draw (rightside) -- +(0, 1);
            \draw (leftback) -- +(0, 1);
            \draw (rightback) -- (rightside);
        \end{scope}
        \end{scope}
    \end{tikzpicture}
\caption{}
\end{subfigure}\hspace*{\fill}

\caption{An adjustment being performed during the execution of \trialg{1}}
\label{fig:APalgotri1adjustment}
\end{figure}
}

\paragraph{Algorithm \trialg{1}.} To summarize, we formally define algorithm \trialg{1} as the triplet $(\{R\}, I, T)$, where $I$ is the initial configuration as shown in Figure~\ref{fig:APalgotri1init} and $T$ is the transition function defined as the set of rules depicted in Figures~\ref{fig:APtrialg1movealong},~\ref{fig:APtrialg1moveturn}~and~\ref{fig:APtrialg1beaconmove}.

\begin{theorem}\label{the:APtri12}Algorithm \trialg{1} solves the IGE problem using eight robots of a single color under visibility range $\Phi=2$. Moreover, it is well-defined.
\end{theorem}

\proof The proof is structured as follows. First, we divide the execution into \emph{phases}. We then show by induction that every phase is reached. Finally, we demonstrate that during each phase, a particular set of vertices is visited, which, when combined across all phases, yields the entirety of $V$.

To establish a concrete coordinate system, we assume the right vertex below the leftmost robot in the initial configuration to be $(0,0)_\triangle$ (marked in Figure~\ref{fig:APalgotri1init}). This assumption is arbitrary and only helps us describe the positions of the robots; the robots cannot access this coordinate system. The initial configuration is then declared as the beginning of phase 0 and partitioned as
$$C^0 = M^0 \cup C^0_0 \cup C^0_1 \cup C^0_2,$$
where
\begin{align*}
    M^0 &= \{\tri{3, 1}, \tri{4, 1}, \tri{5, 2}, \tri{5, 3}, \tri{4, 3}\},\\
    C^0_0 &= \{(5, 0)_\triangle\},\\
    C^0_1 &= \{(7, 6)_\triangle\},\\
    C^0_2 &= \{(0, 1)_\triangle\}.
\end{align*}
Configuration $M^0$ contains the robots that make up the moving group. Configurations $C^0_0$, $C^0_1$ and $C^0_2$ contain the beacons, beginning with the one in the bottom right and continuing counterclockwise. This is the order in which they will be visited during each phase.

We then define the beginning of phase $i \in \mathbb{N}$ as the configuration
$$C^i = M^i \cup C^i_0 \cup C^i_1 \cup C^i_2,$$
where
\begin{align*}
    M^i &= \translate{\tri{i,-i}}{M^0},\\
    C^i_0 &= \translate{\tri{i,-i}}{C^0_0},\\
    C^i_1 &= \translate{\tri{i,2i}}{C^0_1},\\
    C^i_2 &= \translate{\tri{-2i,-i}}{C^0_2}.
\end{align*}

We now assume that we are at the beginning of phase $i$ and the configuration $C^i$ has been reached at round $d$. Let us analyze the configurations in the following rounds and prove that $C^{i+1}$ is eventually reached. We begin with the configuration
$$
C^i = \translate{\tri{i, -i}}{M^0}
\cup C^i_0 \cup C^i_1 \cup C^i_2.
$$
After two rounds, at time $d+2$, an adjustment has been completed using the beacon in $C^i_0$.  The moving group is now facing in the $\tri{0,1}$ direction and the configuration is now
$$
\translate{\tri{i, -i}}
{M^\nwarrow}
\cup C^{i+1}_0 \cup C^i_1 \cup C^i_2
,
$$
where $M^\nwarrow = \{\tri{4, 2},\allowbreak \tri{4, 3},\allowbreak \tri{6, 3},\allowbreak \tri{6, 4},\allowbreak \tri{5, 4}\}$.
Notice that, as a result, the beacon-side robot of the moving group has been translated by $\tri{0,2}$ and the rest of the moving group has been translated by $\tri{1,1}$.

The moving group now begins to travel in the $\tri{0,1}$ direction towards the second beacon. It comes into its view in $3i$ rounds at time $d + 3i + 2$, reaching the configuration
$$
\translate{\tri{i, 2i}}
{M^\nwarrow}
\cup C^{i+1}_0 \cup C^i_1 \cup C^i_2
.
$$
Let us point out the position of the second beacon robot, located at
$\tri{7, 6}+\tri{i, 2i}$, in relation to the front and beacon-side robots of the moving group, located at $\tri{5, 4}+\tri{i, 2i}$ and $\tri{6, 4}+\tri{i, 2i}$, respectively. Notice that they are all visible to each other and aligned correctly for the adjustment to take place.

After two rounds, at time $d +3i +4$, the second adjustment of the phase is completed with the configuration
$$
\translate{\tri{i, 2i}}
{M^\swarrow}
\cup C^{i+1}_0 \cup C^{i+1}_1 \cup C^i_2
,
$$
where $M^\swarrow=\{\tri{3, 2},\allowbreak \tri{3, 3},\allowbreak \tri{5, 3},\allowbreak \tri{4, 2},\allowbreak \tri{4, 4}\}$. The moving group then travels in the $\tri{-1,-1}$ direction towards the last beacon.

After $3i+1$ rounds, the front robot is at position $\tri{2, 1} + \tri{-2i, -i}$, in view of the last beacon, which is positioned at $\tri{0, 1} + \tri{-2i, -i}$. At time $d +6i +5$, the configuration is
$$
\translate{\tri{-2i-1, -i-1}}
{M^\swarrow}
\cup C^{i+1}_0 \cup C^{i+1}_1 \cup C^i_2
.
$$
After two rounds, the final adjustment is complete and the moving group is facing the same direction as at the beginning of the phase. The configuration at time $d +6i +7$ is
$$
\translate{\tri{-2i-1, -i-1}}
{M^0}
\cup C^{i+1}_0 \cup C^{i+1}_1 \cup C^{i+1}_2
.
$$

The moving group finally starts traveling in the $\tri{1, 0}$ direction towards the first beacon. It reaches it in $3i+2$ rounds at time $d +9i +9$, at which point phase $i+1$ begins and the configuration becomes
\begin{align*}
\translate{\tri{i+1, -i-1}}{M^0} \cup C^{i+1}_0 \cup C^{i+1}_1 \cup C^{i+1}_2 \\
= M^{i+1} \cup C^{i+1}_0 \cup C^{i+1}_1 \cup C^{i+1}_2
= C^{i+1}.
\end{align*}

\begin{figure}[h!t]
\centering
    \begin{tikzpicture}[triangular_slant,scale=0.9]

        \triangulargrid{12}{8}
        \begin{scope}[shift={(2,0)}]
        \begin{scope} [moves,shift={(0, 1)}]
            \draw (5, 2) -- (5,4);
            \draw (5,4) -- (2, 1);
            \draw (2, 1) -- (5, 1);
        \end{scope}
        \begin{scope}[every node/.style={robot,draw=black!90,fill=black!2}]
            \node (front) at (5, 3) {R};
            \node (leftside) at (5, 4) {R};
            \node (rightside) at (4, 2) {R};
            \node (leftback) at (4, 4) {R};
            \node (rightback) at (3, 2) {R};
            \node (beacon1) at (5, 1) {R};
            \node (beacon2) at (7, 7) {R};
            \node (beacon3) at (0, 2) {R};
        \end{scope}
        \node (origin) at (0, 1) [mark] {};
        \end{scope}
    \end{tikzpicture}

    \caption{Path taken by the front robot in the execution of \trialg{1} during phase 0}
    \label{fig:APalgotri1triangle}
\end{figure}
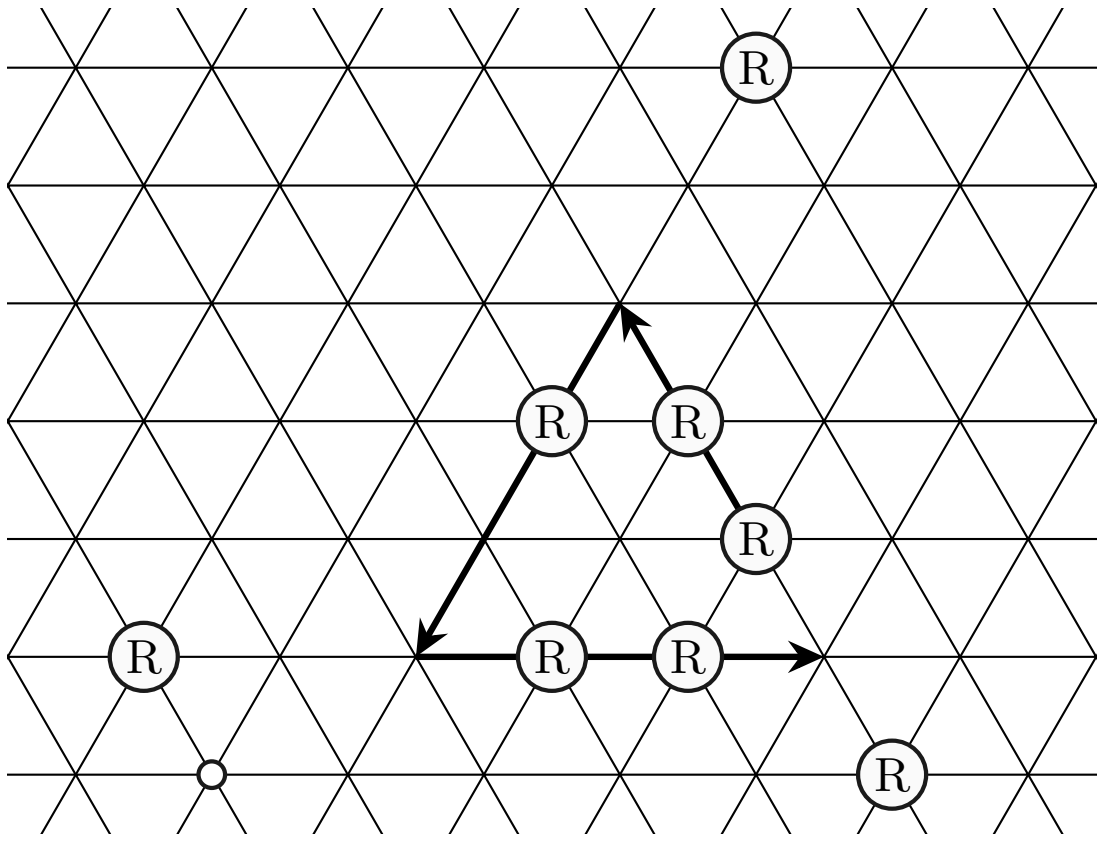

By induction, every phase eventually takes place, since the robots begin in configuration $C^0$ and reach configuration $C^i$ within a finite amount of time. When we take note of the vertices occupied by the front of the moving group during phase $i$, we can observe that the visited nodes form a triangle defined by the vertices $\tri{5,1}+\tri{i,-i}$, $\tri{5,4}+\tri{i,2i}$ and $\tri{2,1}+\tri{-2i,-i}$. The set of vertices on the edges of this triangle in phase $i$ will be denoted by~$S_i$. See Figure~\ref{fig:APalgotri1triangle}, which depicts the path taken in phase 0. The vertex $\tri{4,2}$ is not part of the triangle in any phase, but it is visited in phase 0 nonetheless. Since $\bigcup_{i\ge0}S_i \cup \{\tri{4,2}\} = V$, every vertex is visited and the algorithm \trialg{1} solves the IGE problem.
\qed

\subsection{Six robots, two colors, visibility range one}
\label{sec:APtri21}

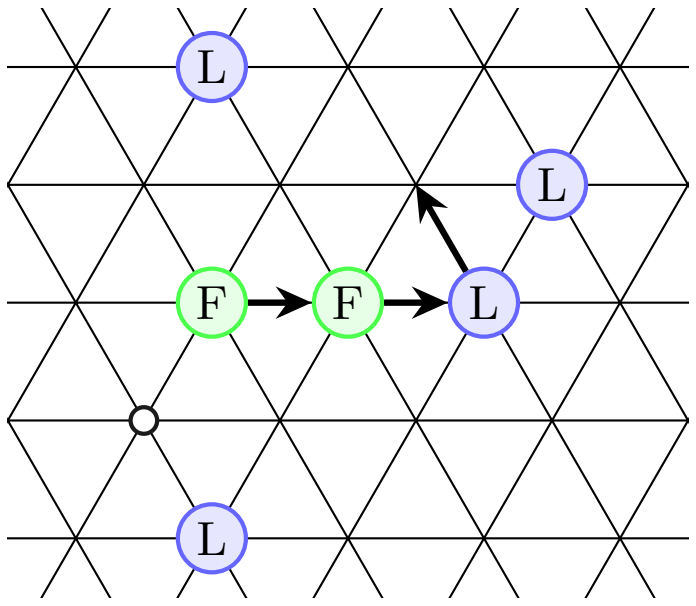
\begin{figure}[h!t]
\centering
\begin{tikzpicture}[triangular_slant,scale=0.9]

    \triangulargrid{8}{6}
    \begin{scope}[shift={(5,3)}]
    \begin{scope}[every node/.style={robot,rblue}]
        \node (front) at (0, 0) {L};
        \node (middle) at (-1, 0) [rgreen] {F};
        \node (back) at (-2, 0) [rgreen] {F};
        \node (beacon1) at (1, 1) {L};
        \node (beacon2) at (-1, 2) {L};
        \node (beacon3) at (-3, -2) {L};
    \end{scope}
    \node (origin) at (-3, -1) [mark] {};
    \begin{scope} [moves]
        \draw (front) -- +(0, 1);
        \draw (middle) -- (front);
        \draw (back) -- (middle);
    \end{scope}
    \end{scope}
\end{tikzpicture}

\caption[Initial configuration of algorithm \trialg{2}.]{Initial configuration of algorithm \trialg{2}, including the actions taken in the first round and the origin $\tri{0,0}$ marked by a circle}
\label{fig:APalgotri2init}
\end{figure}

The second algorithm, denoted as \trialg{2}, behaves similarly to algorithm \trialg{1}. It solves the IGE problem using six robots of two colors under visibility range one. The existence of the algorithm proves that it is possible to solve the IGE problem in the triangular grid using the smallest possible visibility range $\Phi = 1$. It achieves this while also keeping the number of colors nearly optimal. The initial configuration $I$ is shown in Figure~\ref{fig:APalgotri2init}. The moving group is smaller, being made up of three robots. The other three robots act as beacons and serve the same purpose as in \trialg{1}.

The algorithm \trialg{2} is once again comparable to an algorithm previously proposed by Bramas et al.~\cite{BRAMAS2025193}, which solves the IGE problem in the square grid using the same number of robots and colors and under an equal visibility range. There, the moving group consists of only two robots while using an additional beacon, for a total of four.
Additionally, their algorithm requires its robots to change colors during some of their moves, while in \trialg{2}, the color of all robots remains fixed throughout the execution.
In this sense, algorithm \trialg{2} solves the IGE problem under a stricter set of constraints.

\begin{figure}[h!t]
\centering

\hspace*{\fill}\begin{subfigure}{2.7cm}
\centering
\captionsetup{justification=centering}
    \begin{tikzpicture}[triangular_slant,scale=\rulescale]
        \triangulargridview{2}

        \begin{scope}[shift={(1,1)}]
        \begin{scope}[every node/.style={robottiny,rblue}]
            \node (front) at (0, 0) {L};
            \node (middle) at (-1, 0) [rgreen] {F};
        \end{scope}
        \begin{scope} [moves]
            \draw (front) -- +(1, 0);
        \end{scope}
        \end{scope}
    \end{tikzpicture}
    \caption{}
    \label{fig:APtri2move1}
\end{subfigure}\hfill \begin{subfigure}{2.7cm}
\centering
\captionsetup{justification=centering}
    \begin{tikzpicture}[triangular_slant,scale=\rulescale]
        \triangulargridview{2}

        \begin{scope}[shift={(2,1)}]
        \begin{scope}[every node/.style={robottiny,rblue}]
            \node (front) at (0, 0) {L};
            \node (middle) at (-1, 0) [rgreen] {F};
            \node (back) at (-2, 0) [rgreen] {F};
        \end{scope}
        \begin{scope} [moves]
            \draw (middle) -- (front);
        \end{scope}
        \end{scope}
    \end{tikzpicture}
    \caption{}
    \label{fig:APtri2move2}
\end{subfigure}\hfill \begin{subfigure}{2.7cm}
\centering
\captionsetup{justification=centering}
    \begin{tikzpicture}[triangular_slant,scale=\rulescale]
        \triangulargridview{2}

        \begin{scope}[shift={(3,1)}]
        \begin{scope}[every node/.style={robottiny,rgreen}]
            \node (middle) at (-1, 0) {F};
            \node (back) at (-2, 0) {F};
        \end{scope}
        \begin{scope} [moves]
            \draw (back) -- (middle);
        \end{scope}
        \end{scope}
    \end{tikzpicture}
    \caption{}
    \label{fig:APtri2move3}
\end{subfigure}\hspace*{\fill}

\caption{Rules for the moving group in \trialg{2} to move along a straight line}
\label{fig:APtrialg2movealong}
\end{figure}
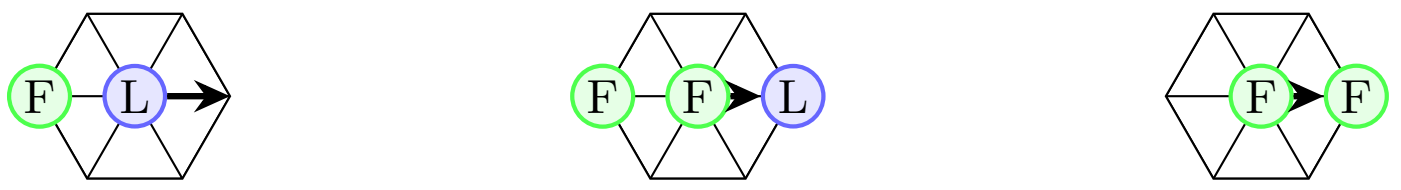

The three rules depicted in Figure~\ref{fig:APtrialg2movealong} allow the moving group to move in a straight line when not adjacent to a beacon. The two colors used are marked as $L$ and $F$ (as in \emph{leader} and \emph{follower}, respectively). These colors describe the roles of the robots in the moving group, which is made up of one leader robot in the front (Figure~\ref{fig:APtri2move1}) and two trailing follower robots, one in the center (Figure~\ref{fig:APtri2move2}) and the other at the back of the group (Figure~\ref{fig:APtri2move3}). The beacons use the color $L$.

\begin{figure}[h!t]
\centering

\hspace*{\fill}\begin{subfigure}{2.7cm}
\centering
\captionsetup{justification=centering}
    \begin{tikzpicture}[triangular_slant,scale=\rulescale]
        \triangulargridview{2}

        \begin{scope}[shift={(1,1)}]
        \begin{scope}[every node/.style={robottiny,rblue}]
            \node (front) at (0, 0) {L};
            \node (middle) at (-1, 0) [rgreen] {F};
            \node (beacon) at (1, 1) {L};
        \end{scope}
        \begin{scope} [moves]
            \draw (front) -- +(0, 1);
        \end{scope}
        \end{scope}
    \end{tikzpicture}
    \caption{}
    \label{fig:APtri2turn1}
\end{subfigure}\hfill \begin{subfigure}{2.7cm}
\centering
\captionsetup{justification=centering}
    \begin{tikzpicture}[triangular_slant,scale=\rulescale]
        \triangulargridview{2}

        \begin{scope}[shift={(1,1)}]
        \begin{scope}[every node/.style={robottiny,rblue}]
            \node (front) at (0, 0) {L};
            \node (middle) at (0, -1) [rgreen] {F};
            \node (back) at (-1, -1) [rgreen] {F};
            \node (beacon) at (1, 0) {L};
        \end{scope}
        \begin{scope} [moves]
            \draw (front) -- +(0, 1);
        \end{scope}
        \end{scope}
    \end{tikzpicture}
    \caption{}
    \label{fig:APtri2turn2}
\end{subfigure}\hfill \begin{subfigure}{2.7cm}
\centering
\captionsetup{justification=centering}
    \begin{tikzpicture}[triangular_slant,scale=\rulescale]
        \triangulargridview{2}

        \begin{scope}[shift={(1,2)}]
        \begin{scope}[every node/.style={robottiny,rblue}]
            \node (front) at (0, 0) {L};
            \node (middle) at (0, -1) [rgreen] {F};
            \node (back) at (-1, -1) [rgreen] {F};
            \node (beacon) at (1, 0) {L};
        \end{scope}
        \begin{scope} [moves]
            \draw (middle) -- (front);
        \end{scope}
        \end{scope}
    \end{tikzpicture}
    \caption{}
    \label{fig:APtri2turn3}
\end{subfigure}\hspace*{\fill}

\par\medskip

\hspace*{\fill}\begin{subfigure}{2.7cm}
\centering
\captionsetup{justification=centering}
    \begin{tikzpicture}[triangular_slant,scale=\rulescale]
        \triangulargridview{2}

        \begin{scope}[shift={(2,2)}]
        \begin{scope}[every node/.style={robottiny,rblue}]
            \node (front) at (0, 0) {L};
            \node (middle) at (0, -1) [rgreen] {F};
            \node (back) at (-1, -1) [rgreen] {F};
        \end{scope}
        \begin{scope} [moves]
            \draw (back) -- (middle);
        \end{scope}
        \end{scope}
    \end{tikzpicture}
    \caption{}
    \label{fig:APtri2turn4}
\end{subfigure}\hfill \begin{subfigure}{2.7cm}
\centering
\captionsetup{justification=centering}
    \begin{tikzpicture}[triangular_slant,scale=\rulescale]
        \triangulargridview{2}

        \begin{scope}[shift={(1,1)}]
        \begin{scope}[every node/.style={robottiny,rblue}]
            \node (middle) at (0, 1) [rgreen] {F};
            \node (back) at (0, 0) [rgreen] {F};
            \node (beacon) at (1, 0) {L};
        \end{scope}
        \begin{scope} [moves]
            \draw (back) -- (middle);
        \end{scope}
        \end{scope}
    \end{tikzpicture}
    \caption{}
    \label{fig:APtri2turn5}
\end{subfigure}\hspace*{\fill}

\caption{Rules for the moving group in \trialg{2} during an adjustment}
\label{fig:APtrialg2moveturn}
\end{figure}
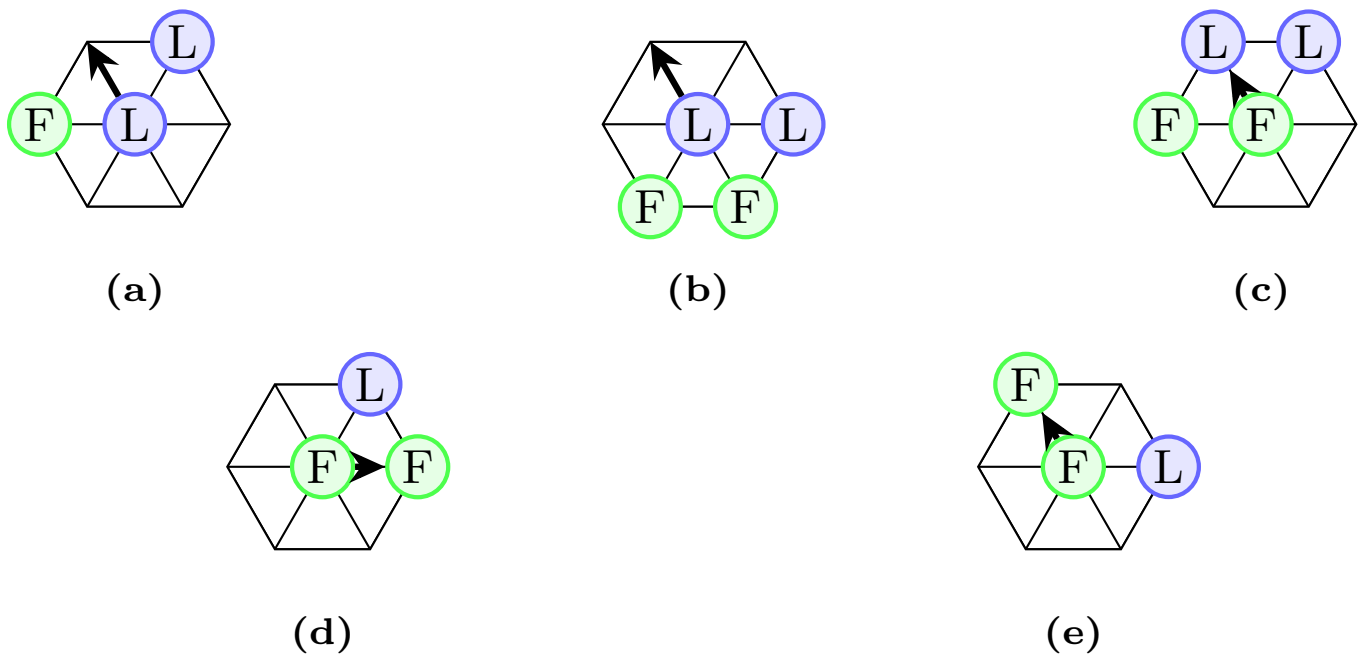

Rules in Figure~\ref{fig:APtrialg2moveturn} apply to the robots of the moving group when adjacent to a beacon, making it perform an adjustment. The beacon robot is first spotted by the leader robot from the side. The leader then moves into the free space directly between the beacon and adjacent follower robot (Figure~\ref{fig:APtri2turn1}). The follower robots continue in a straight line, as they cannot see the beacon yet. This forms a diamond shape, which is made asymmetric by the colors of the robots, thus allowing each one to perform a unique action. The moving group leader continues in the same direction as in the previous round (Figure~\ref{fig:APtri2turn2}), with the rest of the group following behind (Figures~\ref{fig:APtri2turn3} and \ref{fig:APtri2turn4}). The moving group then forms a straight line again, with the beacon being visible only to the back robot. Since the back robot sees the $L$-colored beacon at an angle, it can choose to move towards the other robot instead, staying with the group (Figure~\ref{fig:APtri2turn5}).

Note that it is possible to conserve one robot action during the adjustment by using slightly different rules for the follower robots while in the diamond formation. By making the center follower idle and making the other follow the leader instead, the same exact configuration is reached as with the rules we have chosen.

\begin{figure}[h!t]
\centering

\hspace*{\fill}\begin{subfigure}{2.7cm}
\centering
\captionsetup{justification=centering}
    \begin{tikzpicture}[triangular_slant,scale=\rulescale]
        \triangulargridview{2}

        \begin{scope}[shift={(0,0)}]
        \begin{scope}[every node/.style={robot,rblue}]
            \node (front) at (0, 1) {L};
            \node (middle) at (0, 0) [rgreen] {F};
            \node (beacon) at (1, 1) {L};
        \end{scope}
        \begin{scope} [moves]
            \draw (beacon) -- +(0, -1);
        \end{scope}
        \end{scope}
    \end{tikzpicture}
    \caption{}
    \label{fig:APtri2beacon1}
\end{subfigure}\hfill \begin{subfigure}{2.7cm}
\centering
\captionsetup{justification=centering}
    \begin{tikzpicture}[triangular_slant,scale=\rulescale]
        \triangulargridview{2}

        \begin{scope}[shift={(1,1)}]
        \begin{scope}[every node/.style={robot,rblue}]
            \node (back) at (-1, 0) [rgreen] {F};
            \node (beacon) at (0, 0) {L};
        \end{scope}
        \begin{scope} [moves]
            \draw (beacon) -- +(1, 0);
        \end{scope}
        \end{scope}
    \end{tikzpicture}
    \caption{}
    \label{fig:APtri2beacon2}
\end{subfigure}\hspace*{\fill}

\caption{Rules for the beacons in \trialg{2} during an adjustment}
\label{fig:APtrialg2beaconmove}
\end{figure}
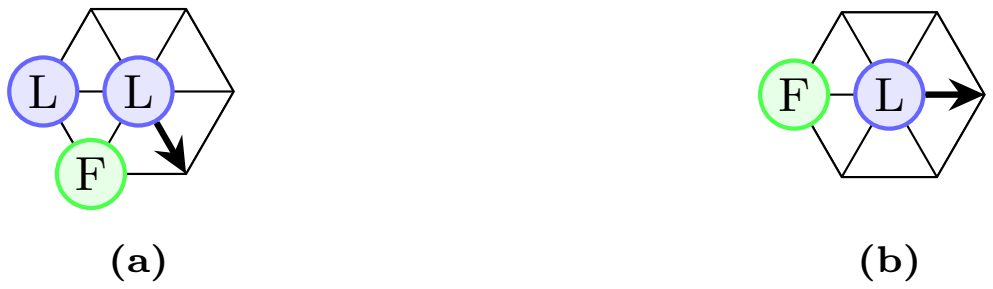

The beacons stay idle when alone, only moving according to the rules in Figure~\ref{fig:APtrialg2beaconmove} during an adjustment. The beacon waits until two robots, a leader and a follower, are in view, moving towards the empty node next to the follower (Figure~\ref{fig:APtri2beacon1}). In the next round, it only sees a follower robot, moving away from it (Figure~\ref{fig:APtri2beacon2}). Note that it uses the same rule as the leader of the moving group, shown in Figure~\ref{fig:APtri2move1}. The entire adjustment is shown in Figure~\ref{fig:APalgotri2adjustment}.

\renewcommand{\adjscale}{0.8}

{
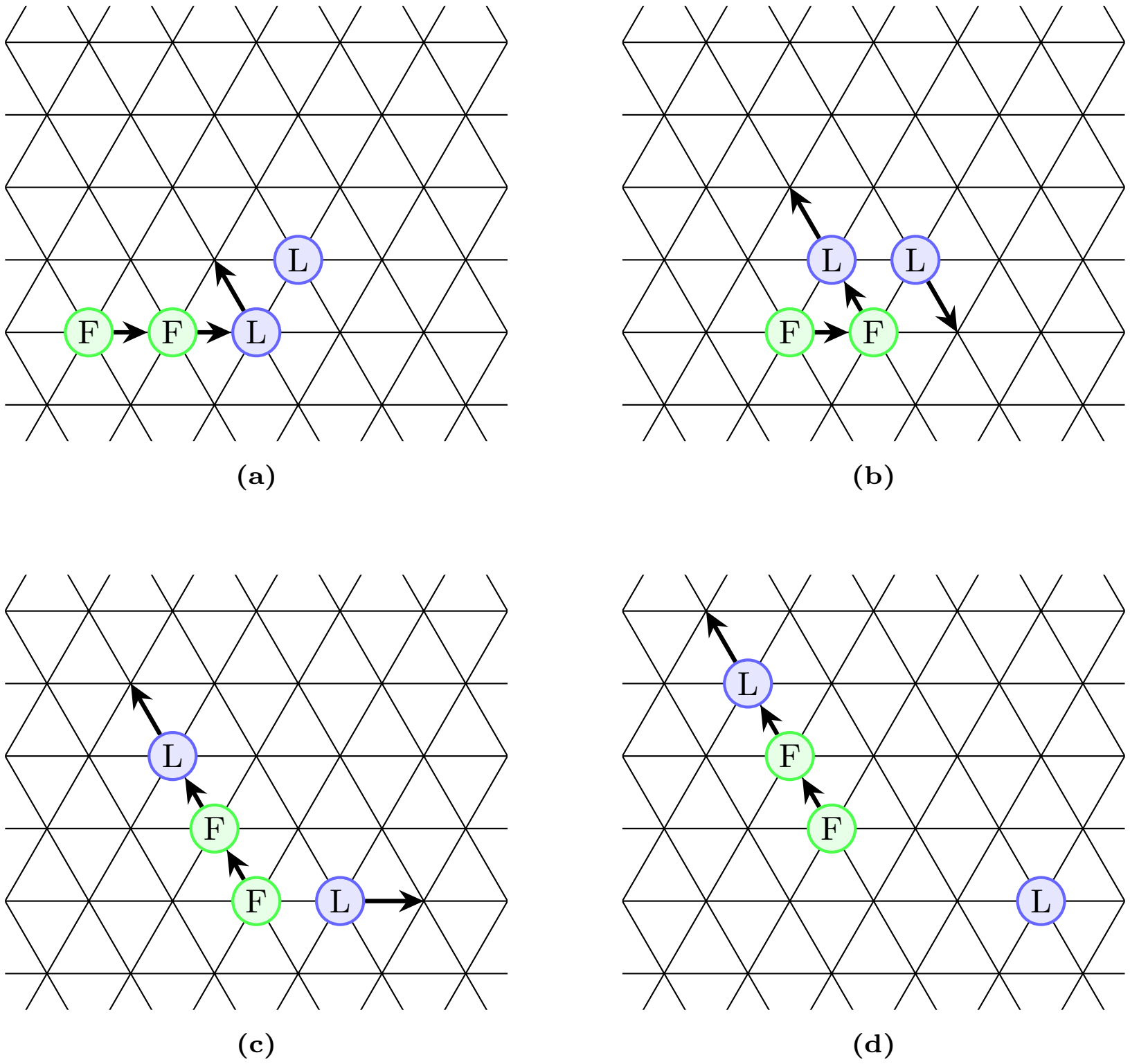
\begin{figure}[h!t]
\centering

\hspace*{\fill}\begin{subfigure}{0.45\textwidth}
\centering
\captionsetup{justification=centering}
    \begin{tikzpicture}[triangular_slant,scale=\adjscale]
        \triangulargrid{9.5}{7}
        \begin{scope}[shift={(4,2)}]
        \begin{scope}[every node/.style={robot,rblue}]
            \node (front) at (0, 0) {L};
            \node (middle) at (-1, 0) [rgreen] {F};
            \node (back) at (-2, 0) [rgreen] {F};
            \node (beacon) at (1, 1) {L};
        \end{scope}
        \begin{scope} [moves]
            \draw (front) -- +(0, 1);
            \draw (middle) -- (front);
            \draw (back) -- (middle);
        \end{scope}
        \end{scope}
    \end{tikzpicture}
\caption{}
\end{subfigure}\hfill \begin{subfigure}{0.45\textwidth}
\centering
\captionsetup{justification=centering}
    \begin{tikzpicture}[triangular_slant,scale=\adjscale]
        \triangulargrid{9.5}{7}
        \begin{scope}[shift={(4,2)}]
        \begin{scope}[every node/.style={robot,rblue}]
            \node (front) at (0, 1) {L};
            \node (middle) at (0, 0) [rgreen] {F};
            \node (back) at (-1, 0) [rgreen] {F};
            \node (beacon) at (1, 1) {L};
        \end{scope}
        \begin{scope} [moves]
            \draw (front) -- +(0, 1);
            \draw (middle) -- (front);
            \draw (back) -- (middle);
            \draw (beacon) -- +(0, -1);
        \end{scope}
        \end{scope}
    \end{tikzpicture}
\caption{}
\end{subfigure}\hspace*{\fill}

\par\vspace{0.75cm}

\hspace*{\fill}\begin{subfigure}{0.45\textwidth}
\centering
\captionsetup{justification=centering}
    \begin{tikzpicture}[triangular_slant,scale=\adjscale]
        \triangulargrid{9.5}{7}
        \begin{scope}[shift={(4,2)}]
        \begin{scope}[every node/.style={robot,rblue}]
            \node (front) at (0, 2) {L};
            \node (middle) at (0, 1) [rgreen] {F};
            \node (back) at (0, 0) [rgreen] {F};
            \node (beacon) at (1, 0) {L};
        \end{scope}
        \begin{scope} [moves]
            \draw (front) -- +(0, 1);
            \draw (middle) -- (front);
            \draw (back) -- (middle);
            \draw (beacon) -- +(1, 0);
        \end{scope}
        \end{scope}
    \end{tikzpicture}
\caption{}
\end{subfigure}\hfill \begin{subfigure}{0.45\textwidth}
\centering
\captionsetup{justification=centering}
    \begin{tikzpicture}[triangular_slant,scale=\adjscale]
        \triangulargrid{9.5}{7}
        \begin{scope}[shift={(4,2)}]
        \begin{scope}[every node/.style={robot,rblue}]
            \node (front) at (0, 3) {L};
            \node (middle) at (0, 2) [rgreen] {F};
            \node (back) at (0, 1) [rgreen] {F};
            \node (beacon) at (2, 0) {L};
        \end{scope}
        \begin{scope} [moves]
            \draw (front) -- +(0, 1);
            \draw (middle) -- (front);
            \draw (back) -- (middle);
        \end{scope}
        \end{scope}
    \end{tikzpicture}
\caption{}
\end{subfigure}\hspace*{\fill}

\caption{An adjustment being performed during the execution of \trialg{2}}
\label{fig:APalgotri2adjustment}
\end{figure}
}

\paragraph{Algorithm \trialg{2}.} To summarize, we formally define algorithm \trialg{2} as the triplet $(\{L,F\}, I, T)$, where $I$ is the initial configuration as shown in Figure~\ref{fig:APalgotri2init} and $T$ is the transition function defined as the set of rules depicted in Figures~\ref{fig:APtrialg2movealong},~\ref{fig:APtrialg2moveturn}~and~\ref{fig:APtrialg2beaconmove}.

\begin{theorem}\label{the:APtri21}
Algorithm \trialg{2} solves the IGE problem using six robots of two colors under visibility range $\Phi=1$. Moreover, it is well-defined and uses robots of fixed colors.
\end{theorem}

\proof The proof follows the same structure as for the previous algorithm. We divide the execution into phases, inductively prove that every phase is eventually reached and then show that every vertex is visited in some phase. We fix a concrete coordinate system with the vertex $\tri{0, 0}$ above and to the left of the bottommost beacon, as marked in Figure~\ref{fig:APalgotri2init}. We declare the configuration reached from $I$ after one round of execution as the beginning of phase 0, mark it as $C^0$ and partition it as
$$
C^0 = M^0\cup C^0_0 \cup C^0_1 \cup C^0_2,
$$
where
\begin{align*}
    M^0 &= \{(\tri{2, 1},F), (\tri{3, 1},F), (\tri{3, 2},L)\},\\
    C^0_0 &= \{(\tri{4, 2}, L)\},\\
    C^0_1 &= \{(\tri{2, 3}, L)\},\\
    C^0_2 &= \{(\tri{0, -1}, L)\}.
\end{align*}
Configuration $M^0$ contains the moving group, forming the diamond shape alongside the $C^0_0$ beacon. Configurations $C^0_0$, $C^0_1$ and $C^0_2$ contain the beacon robots, in the order in which they are visited during each phase.

The beginning of phase $i \in \mathbb{N}$ is then defined as the configuration
$$C^i = M^i \cup C^i_0 \cup C^i_1 \cup C^i_2,$$
where
\begin{align*}
    M^i &= \translate{\tri{i,-i}}{M^0},\\
    C^i_0 &= \translate{\tri{i,-i}}{C^0_0},\\
    C^i_1 &= \translate{\tri{i,2i}}{C^0_1},\\
    C^i_2 &= \translate{\tri{-2i,-i}}{C^0_2}.
\end{align*}

We now prove the inductive step by assuming that phase $i$ beginning with configuration $C^i$ has been reached at round $d$ and demonstrating that $C^{i+1}$ is reached at a later time. We begin with the configuration
$$
C^i = \translate{\tri{i, -i}}{M^0}
\cup C^i_0 \cup C^i_1 \cup C^i_2.
$$

After two rounds, at time $d+2$, the first adjustment of the phase has been completed with the help of the first beacon, and the configuration becomes
$$
\translate{\tri{i, -i}}
{\{(\tri{3, 2},F), (\tri{3, 3},F), (\tri{2, 2},L)\}}
\cup C^{i+1}_0 \cup C^i_1 \cup C^i_2
,
$$
if the current phase is 0, or
$$
\translate{\tri{i, -i}}
{\{(\tri{3, 2},F), (\tri{3, 3},F), (\tri{3, 4},L)\}}
\cup C^{i+1}_0 \cup C^i_1 \cup C^i_2
,
$$
otherwise. Notice the difference in the placement of the group leader, as in phase 0, the moving group has already managed to form the diamond shape.

In either case, after another $3i$ rounds, the moving group has reached the second beacon and formed the diamond shape. At round $d + 3i + 2$, the configuration is
$$
\translate{\tri{i, 2i}}
{\{(\tri{3, 2},F), (\tri{3, 3},F), (\tri{2, 2},L)\}}
\cup C^{i+1}_0 \cup C^i_1 \cup C^i_2
.
$$
The moving group and the second beacon are aligned properly, with the beacon in $C^i_1$ being positioned at $\tri{2,3}+\tri{i, 2i}$.

After two rounds, at time $d +3i +4$, the second adjustment of the phase is completed. The moving group is facing down and to the left, towards the third beacon. The current configuration is
$$
\translate{\tri{i, 2i}}
{\{(\tri{2, 2},F), (\tri{1, 1},F), (\tri{0, 0},L)\}}
\cup C^{i+1}_0 \cup C^{i+1}_1 \cup C^i_2
.
$$

After $3i+1$ rounds, the final beacon is reached and the diamond shape is formed again. At time $d + 6i + 5$, the final adjustment of the phase is about to take place and the configuration is
$$
\translate{\tri{-2i, -i}}
{\{(\tri{1, 1},F), (\tri{0, 0},F), (\tri{1, 0},L)\}}
\cup C^{i+1}_0 \cup C^{i+1}_1 \cup C^i_2
.
$$

After two rounds, the adjustment is finished and at time $d + 6i + 7$ the configuration is
$$
\translate{\tri{-2i, -i}}
{\{(\tri{1, 0},F), (\tri{2, 0},F), (\tri{3, 0},L)\}}
\cup C^{i+1}_0 \cup C^{i+1}_1 \cup C^{i+1}_2
.
$$

The moving group finally starts traveling to the right towards the first beacon. It reaches it in $3i+2$ rounds at time $d +9i +9$, at which point phase $i+1$ begins. The configuration of the moving group can be written as
\begin{align*}
&\translate{\tri{i, -i}}{\{(\tri{3, 0},F), (\tri{4, 0},F), (\tri{4, 1},L)\}} \\
& = \translate{\tri{i+1, -i-1}}{\{(\tri{2, 1},F), (\tri{3, 1},F), (\tri{3, 2},L)\}} \\
& = \translate{\tri{i+1, -i-1}}{M^0} \\
& = M^{i+1}.
\end{align*}
Therefore, the entire configuration is
\begin{multline*}
\translate{\tri{i, -i}}{\{(\tri{3, 0},F), (\tri{4, 0},F), (\tri{4, 1},L)\}} \cup C^{i+1}_0 \cup C^{i+1}_1 \cup C^{i+1}_2 \\
= M^{i+1} \cup C^{i+1}_0 \cup C^{i+1}_1 \cup C^{i+1}_2 = C^{i+1}.
\end{multline*}

Inductively, every phase eventually takes place, since configuration $C^0$ takes place after one round of execution from $I$ and configuration $C^i$ is reached within a finite amount of time. When we take note of the vertices visited during phase $i$, we can observe that they form a triangle defined by the vertices $\tri{3,0}+\tri{i,-i}$, $\tri{3,3}+\tri{i,2i}$ and $\tri{0,0}+\tri{-2i,-i}$. We denote the set of vertices on the edges of this triangle in phase $i$ as $S_i$. The vertex $\tri{2,1}$, while not on the edge of the triangle in phase 0, is occupied in the initial configuration. Since $\bigcup_{i\ge0}S_i \cup \{\tri{2,1}\} = V$, every vertex is visited and the algorithm \trialg{2} solves the IGE problem.
\qed

\subsection{Five robots, three colors, visibility range one}
\label{sec:APtri31}

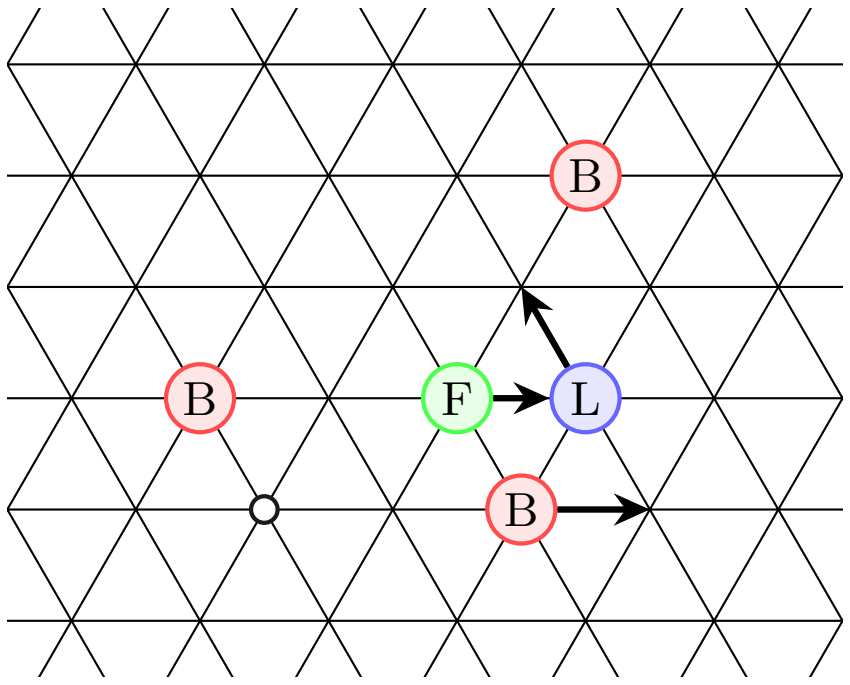
\begin{figure}[h!t]
\centering
    \begin{tikzpicture}[triangular_slant,scale=0.85]

        \triangulargrid{10}{7}
        \begin{scope}[shift={(1,1)}]
        \begin{scope}[every node/.style={robot,rred}]
        \node (front) at (5, 2) [rblue] {L};
        \node (back) at (4, 2) [rgreen] {F};
        \node (beacon1) at (4, 1) {B};
        \node (beacon2) at (6, 4) {B};
        \node (beacon3) at (2, 2) {B};
        \end{scope}
        \node (origin) at (2, 1) [mark] {};
        \begin{scope} [moves]
            \draw (front) -- +(0, 1);
            \draw (back) -- (front);
            \draw (beacon1) -- +(1,0);
        \end{scope}
        \end{scope}
    \end{tikzpicture}

    \caption[Initial configuration of algorithm \trialg{3}.]{Initial configuration of algorithm \trialg{3}, including the actions taken in the first round}
    \label{fig:APalgotri3init}
\end{figure}

The final algorithm we propose for the triangular grid, denoted as \trialg{3}, solves the IGE problem using only five robots of three colors under visibility range one. This is the optimal number of robots as per Theorem~\ref{the:fiveoptimaltri}.
The initial configuration is shown in Figure~\ref{fig:APalgotri3init}. In comparison to \trialg{2}, algorithm \trialg{3} is able to use one less robot in the moving group by making the beacons a different color. As such, it uses two robots -- colored $L$ and $F$ -- for the moving group, and three $B$-colored beacons.

Let us once again draw a comparison between \trialg{3} and other algorithms proposed in the past. Bramas et al.~\cite{BRAMAS2025193} propose an algorithm for the square grid which uses the optimal number of five robots under visibility range one, however, with twelve colors, compared to the three used in \trialg{3}. Its robots also require their colors to change between rounds, while in \trialg{3}, all colors remain fixed. Additionally, it uses one ambiguous rule. Their algorithm does however solve a stronger specification of IGE, being the first to solve the \emph{perpetual} infinite exploration problem, in which every node is to be visited infinitely often.

Algorithm \trialg{3} can also be seen as an improvement to the algorithm of Das and Sau~\cite{DASRAJA}, which uses five robots, four colors and a visibility range of two -- an extra color and a visibility range one larger when compared to \trialg{3}.

\begin{figure}[h!t]
\centering

\hspace*{\fill}\begin{subfigure}{2.7cm}
\centering
\captionsetup{justification=centering}
    \begin{tikzpicture}[triangular_slant,scale=\rulescale]
        \triangulargridview{2}

        \begin{scope}[shift={(1,1)}]
        \begin{scope}[every node/.style={robottiny,rblue}]
            \node (front) at (0, 0) {L};
            \node (back) at (-1, 0) [rgreen] {F};
        \end{scope}
        \begin{scope} [moves]
            \draw (front) -- +(1, 0);
        \end{scope}
        \end{scope}
    \end{tikzpicture}
    \caption{}
    \label{fig:APtri3move1}
\end{subfigure}\hfill \begin{subfigure}{2.7cm}
\centering
\captionsetup{justification=centering}
    \begin{tikzpicture}[triangular_slant,scale=\rulescale]
        \triangulargridview{2}

        \begin{scope}[shift={(2,1)}]
        \begin{scope}[every node/.style={robottiny,rblue}]
            \node (front) at (0, 0) {L};
            \node (back) at (-1, 0) [rgreen] {F};
        \end{scope}
        \begin{scope} [moves]
            \draw (back) -- (front);
        \end{scope}
        \end{scope}
    \end{tikzpicture}
    \caption{}
    \label{fig:APtri3move2}
\end{subfigure}\hspace*{\fill}

\caption{Rules for the moving group in \trialg{3} to move along a straight line}
\label{fig:APtrialg3movealong}
\end{figure}
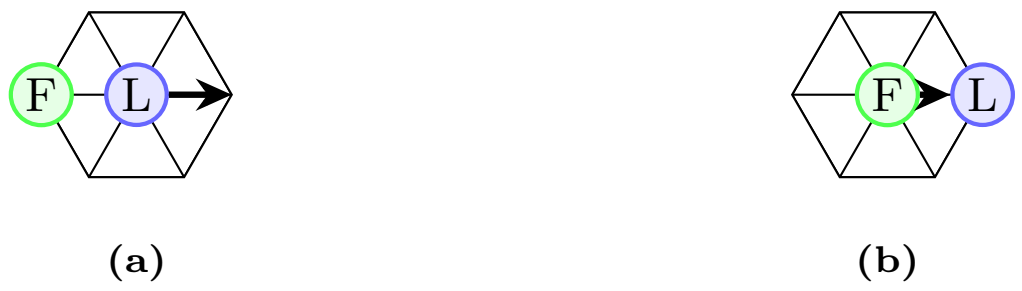

Rules in Figure~\ref{fig:APtrialg3movealong} allow the moving group to move along a straight line. As in algorithm \trialg{2}, the robots are colored based on their role, with the $L$-colored leader in the front (Figure~\ref{fig:APtri3move1}) and $F$-colored follower in the back (Figure~\ref{fig:APtri3move2}). The moving group closely resembles the one used by Bramas et al.~\cite{BRAMAS2025193} in the square grid.

\begin{figure}[h!t]
\centering

\hspace*{\fill}\begin{subfigure}{2.7cm}
\centering
\captionsetup{justification=centering}
    \begin{tikzpicture}[triangular_slant,scale=\rulescale]
        \triangulargridview{2}

        \begin{scope}[shift={(1,1)}]
        \begin{scope}[every node/.style={robottiny,rblue}]
            \node (front) at (0, 0) {L};
            \node (back) at (-1, 0) [rgreen] {F};
            \node (beacon) at (0, -1) [rred] {B};
        \end{scope}
        \begin{scope} [moves]
            \draw (front) -- +(1, 0);
        \end{scope}
        \end{scope}
    \end{tikzpicture}
    \caption{}
    \label{fig:APtri3turn1}
\end{subfigure}\hfill \begin{subfigure}{2.7cm}
\centering
\captionsetup{justification=centering}
    \begin{tikzpicture}[triangular_slant,scale=\rulescale]
        \triangulargridview{2}

        \begin{scope}[shift={(1,1)}]
        \begin{scope}[every node/.style={robottiny,rblue}]
            \node (front) at (0, 0) {L};
            \node (back) at (-1, 0) [rgreen] {F};
            \node (beacon) at (-1, -1) [rred] {B};
        \end{scope}
        \begin{scope} [moves]
            \draw (front) -- +(0, 1);
        \end{scope}
        \end{scope}
    \end{tikzpicture}
    \caption{}
    \label{fig:APtri3turn2}
\end{subfigure}\hfill \begin{subfigure}{2.7cm}
\centering
\captionsetup{justification=centering}
    \begin{tikzpicture}[triangular_slant,scale=\rulescale]
        \triangulargridview{2}

        \begin{scope}[shift={(2,1)}]
        \begin{scope}[every node/.style={robottiny,rblue}]
            \node (front) at (0, 0) {L};
            \node (back) at (-1, 0) [rgreen] {F};
            \node (beacon) at (-1, -1) [rred] {B};
        \end{scope}
        \begin{scope} [moves]
            \draw (back) -- (front);
        \end{scope}
        \end{scope}
    \end{tikzpicture}
    \caption{}
    \label{fig:APtri3turn3}
\end{subfigure}\hfill \begin{subfigure}{2.7cm}
\centering
\captionsetup{justification=centering}
    \begin{tikzpicture}[triangular_slant,scale=\rulescale]
        \triangulargridview{2}

        \begin{scope}[shift={(1,1)}]
        \begin{scope}[every node/.style={robottiny,rblue}]
            \node (front) at (0, 1) {L};
            \node (back) at (0, 0) [rgreen] {F};
            \node (beacon) at (0, -1) [rred] {B};
        \end{scope}
        \begin{scope} [moves]
            \draw (back) -- (front);
        \end{scope}
        \end{scope}
    \end{tikzpicture}
    \caption{}
    \label{fig:APtri3turn4}
\end{subfigure}\hspace*{\fill}

\caption{Rules for the moving group in \trialg{3} during an adjustment}
\label{fig:APtrialg3moveturn}
\end{figure}
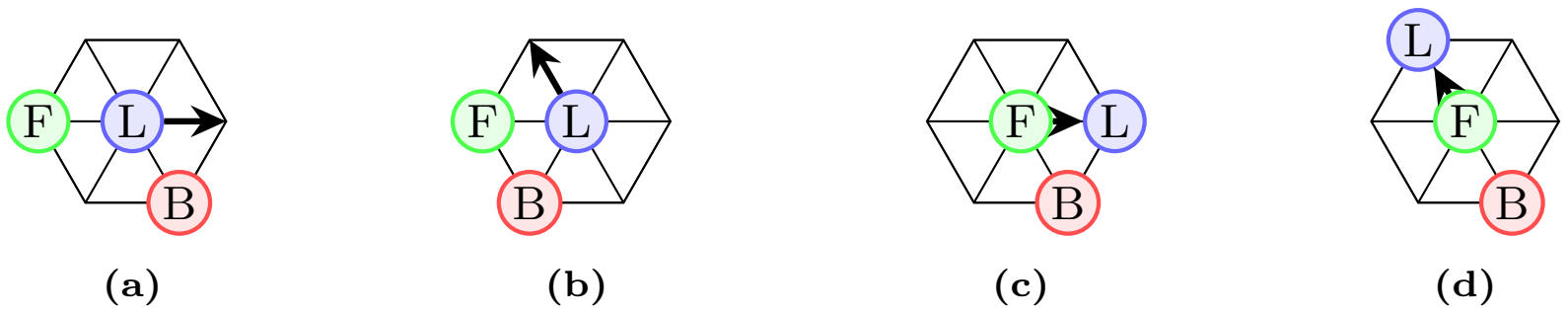

Rules in Figure~\ref{fig:APtrialg3moveturn} depict the actions taken by the moving group during an adjustment. When the beacon is first spotted, the moving group continues in the same direction (Figure~\ref{fig:APtri3turn1}), until the beacon is seen by both robots. At this point, the robots take on a triangle formation, in which all robots are of a different color. This allows each one to take a unique and unambiguous action. The leader makes a sharp turn away from the beacon (Figure~\ref{fig:APtri3turn2}), with the follower trailing behind (Figure~\ref{fig:APtri3turn3}). After this, the beacon is visible only to the follower robot, which moves towards the leader as if the beacon was not there (Figure~\ref{fig:APtri3turn4}).

\begin{figure}[h!t]
\centering

\hspace*{\fill}\begin{subfigure}{2.7cm}
\centering
\captionsetup{justification=centering}
    \begin{tikzpicture}[triangular_slant,scale=\rulescale]
        \triangulargridview{2}

        \begin{scope}[shift={(2,2)}]
        \begin{scope}[every node/.style={robottiny,rblue}]
            \node (front) at (0, 0) {L};
            \node (back) at (-1, 0) [rgreen] {F};
            \node (beacon) at (-1, -1) [rred] {B};
        \end{scope}
        \begin{scope} [moves]
            \draw (beacon) -- +(1, 0);
        \end{scope}
        \end{scope}
    \end{tikzpicture}
    \caption{}
    \label{fig:APtri3beacon1}
\end{subfigure}\hfill \begin{subfigure}{2.7cm}
\centering
\captionsetup{justification=centering}
    \begin{tikzpicture}[triangular_slant,scale=\rulescale]
        \triangulargridview{2}

        \begin{scope}[shift={(1,1)}]
        \begin{scope}[every node/.style={robottiny}]
            \node (beacon) at (0, 0) [rred] {B};
            \node (back) at (0, 1) [rgreen] {F};
        \end{scope}
        \begin{scope} [moves]
            \draw (beacon) -- +(0, -1);
        \end{scope}
        \end{scope}
    \end{tikzpicture}
    \caption{}
    \label{fig:APtri3beacon2}
\end{subfigure}\hspace*{\fill}

\caption{Rules for the beacons in \trialg{3} during an adjustment}
\label{fig:APtrialg3beaconmove}
\end{figure}
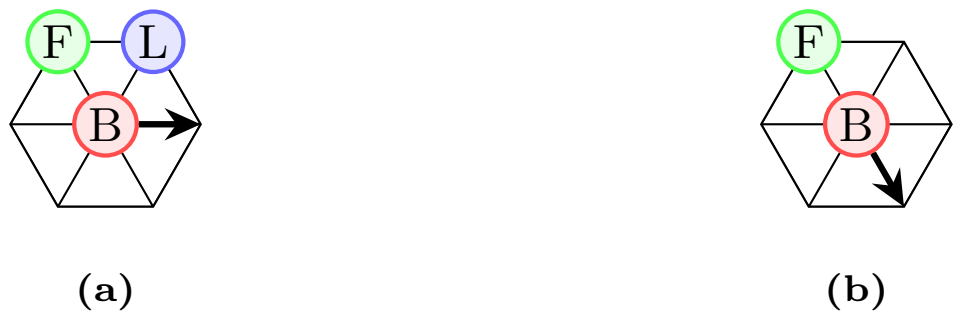

When the beacon is approached by the moving group, it idles until it can see both robots. Then, it moves onto the empty node adjacent to the leader robot (Figure~\ref{fig:APtri3beacon1}). Finally, it completes its two-step readjustment by moving away from the follower, the only robot it sees (Figure~\ref{fig:APtri3beacon2}). The rules describing these actions are shown in Figure~\ref{fig:APtrialg3beaconmove}. The entire adjustment can be seen in Figure~\ref{fig:APalgotri3adjustment}.

{
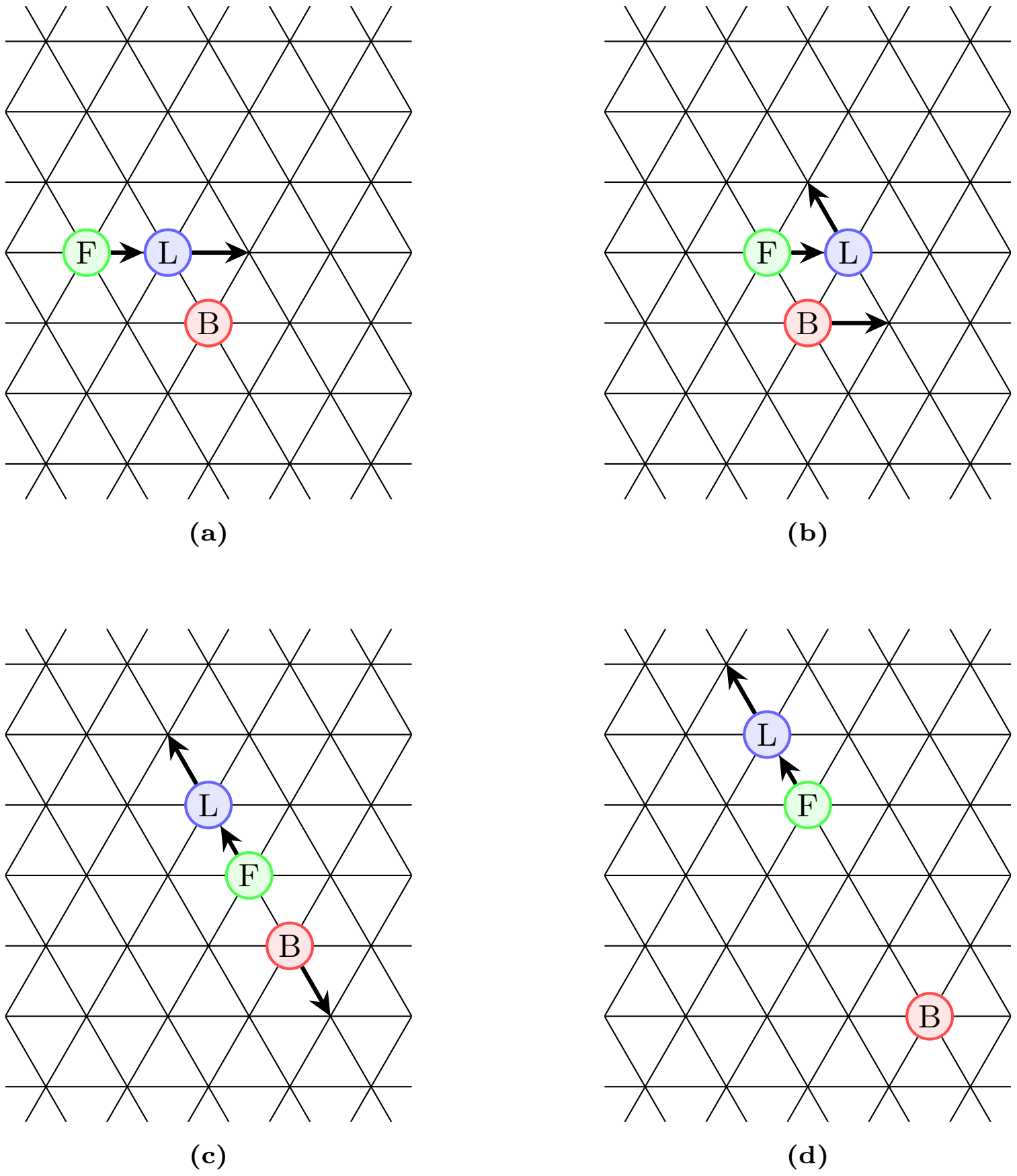
\begin{figure}[h!t]
\centering

\hspace*{\fill}\begin{subfigure}{0.45\textwidth}
\centering
\captionsetup{justification=centering}
    \begin{tikzpicture}[triangular_slant,scale=\adjscale]
        \triangulargrid{9}{8}
        \begin{scope}[shift={(1,2)}]
        \begin{scope}[every node/.style={robot,rred}]
        \node (front) at (3, 2) [rblue] {L};
        \node (back) at (2, 2) [rgreen] {F};
        \node (beacon) at (3, 1) {B};
        \end{scope}
        \begin{scope} [moves]
            \draw (front) -- +(1, 0);
            \draw (back) -- (front);
        \end{scope}
        \end{scope}
    \end{tikzpicture}
\caption{}
\end{subfigure}\hfill \begin{subfigure}{0.45\textwidth}
\centering
\captionsetup{justification=centering}
    \begin{tikzpicture}[triangular_slant,scale=\adjscale]
        \triangulargrid{9}{8}
        \begin{scope}[shift={(1,2)}]
        \begin{scope}[every node/.style={robot,rred}]
        \node (front) at (4, 2) [rblue] {L};
        \node (back) at (3, 2) [rgreen] {F};
        \node (beacon) at (3, 1) {B};
        \end{scope}
        \begin{scope} [moves]
            \draw (front) -- +(0, 1);
            \draw (back) -- (front);
            \draw (beacon) -- +(1,0);
        \end{scope}
        \end{scope}
    \end{tikzpicture}
\caption{}
\end{subfigure}\hspace*{\fill}

\par\vspace{0.75cm}

\hspace*{\fill}\begin{subfigure}{0.45\textwidth}
\centering
\captionsetup{justification=centering}
    \begin{tikzpicture}[triangular_slant,scale=\adjscale]
        \triangulargrid{9}{8}
        \begin{scope}[shift={(1,2)}]
        \begin{scope}[every node/.style={robot,rred}]
        \node (front) at (4, 3) [rblue] {L};
        \node (back) at (4, 2) [rgreen] {F};
        \node (beacon) at (4, 1) {B};
        \end{scope}
        \begin{scope} [moves]
            \draw (front) -- +(0, 1);
            \draw (back) -- (front);
            \draw (beacon) -- +(0,-1);
        \end{scope}
        \end{scope}
    \end{tikzpicture}
\caption{}
\end{subfigure}\hfill \begin{subfigure}{0.45\textwidth}
\centering
\captionsetup{justification=centering}
    \begin{tikzpicture}[triangular_slant,scale=\adjscale]
        \triangulargrid{9}{8}
        \begin{scope}[shift={(1,2)}]
        \begin{scope}[every node/.style={robot,rred}]
        \node (front) at (4, 4) [rblue] {L};
        \node (back) at (4, 3) [rgreen] {F};
        \node (beacon) at (4, 0) {B};
        \end{scope}
        \begin{scope} [moves]
            \draw (front) -- +(0, 1);
            \draw (back) -- (front);
        \end{scope}
        \end{scope}
    \end{tikzpicture}
\caption{}
\end{subfigure}\hspace*{\fill}

\caption{An adjustment being performed during the execution of \trialg{3}}
\label{fig:APalgotri3adjustment}
\end{figure}
}

\paragraph{Algorithm \trialg{3}.} To summarize, we formally define algorithm \trialg{3} as the triplet $(\{L,F,B\}, I, T)$, where $I$ is the initial configuration as shown in Figure~\ref{fig:APalgotri3init} and $T$ is the transition function defined as the set of rules depicted in Figures~\ref{fig:APtrialg3movealong},~\ref{fig:APtrialg3moveturn}~and~\ref{fig:APtrialg3beaconmove}.

\begin{theorem}\label{the:APtri31}
Algorithm \trialg{3} solves the IGE problem using five robots of three colors under visibility range $\Phi=1$. Moreover, it is well-defined and uses robots with fixed colors.
\end{theorem}

\proof The proof once again works by dividing the execution into phases, demonstrating that each phase is eventually reached and then showing that each node is visited during some phase.

We assume the vertex below and to the right of the leftmost beacon to be the origin $\tri{0,0}$, as is marked in Figure~\ref{fig:APalgotri3init}. We declare configuration $I$ as the beginning of phase 0, denoted by $C^0$. Configuration $C^0$ is then partitioned as
$$
C^0 = M^0\cup C^0_0 \cup C^0_1 \cup C^0_2,
$$
where
\begin{align*}
    M^0 &= \{(\tri{2, 1},F), (\tri{3, 1},L)\},\\
    C^0_0 &= \{(\tri{2, 0}, B)\},\\
    C^0_1 &= \{(\tri{4, 3}, B)\},\\
    C^0_2 &= \{(\tri{0, 1}, B)\}.
\end{align*}

The beginning of phase $i \in \mathbb{N}$ is then defined as the configuration
$$C^i = M^i \cup C^i_0 \cup C^i_1 \cup C^i_2,$$
where
\begin{align*}
    M^i &= \translate{\tri{i,-i}}{M^0},\\
    C^i_0 &= \translate{\tri{i,-i}}{C^0_0},\\
    C^i_1 &= \translate{\tri{i,2i}}{C^0_1},\\
    C^i_2 &= \translate{\tri{-2i,-i}}{C^0_2}.
\end{align*}

Let us now prove that, starting at configuration $C^i$ in phase $i$, configuration $C^{i+1}$ is eventually reached. We begin at round $d$ with the configuration
$$
C^i = \translate{\tri{i, -i}}{M^0}
\cup C^i_0 \cup C^i_1 \cup C^i_2.
$$

Two rounds later, the first adjustment is complete and the configuration at time $d+2$ is
$$
\translate{\tri{i, -i}}
{\{(\tri{3, 2},F), (\tri{3, 3},L)\}}
\cup C^{i+1}_0 \cup C^i_1 \cup C^i_2
.
$$

The moving group then travels in the $\tri{0,1}$ direction towards the second beacon, and after $3i$ rounds, it is in view of both the follower and leader robot. Thus, at time $d+3i+2$, the configuration is
$$
\translate{\tri{i, 2i}}
{\{(\tri{3, 2},F), (\tri{3, 3},L)\}}
\cup C^{i+1}_0 \cup C^i_1 \cup C^i_2
.
$$
The second adjustment is over in two rounds, at time $d+3i+4$, with the configuration
$$
\translate{\tri{i, 2i}}
{\{(\tri{2, 2},F), (\tri{1, 1},L)\}}
\cup C^{i+1}_0 \cup C^{i+1}_1 \cup C^i_2
.
$$

The moving group travels towards the third beacon, in the $\tri{-1, -1}$ direction. After $3i+1$ rounds, the moving group and the beacon form the triangle shape. At time $d+6i+5$, the configuration is
$$
\translate{\tri{-2i, -i}}
{\{(\tri{1, 1},F), (\tri{0, 0},L)\}}
\cup C^{i+1}_0 \cup C^{i+1}_1 \cup C^i_2
.
$$
After two rounds, at time $d+6i+7$, the third adjustment of the phase is over with the configuration
$$
\translate{\tri{-2i, -i}}
{\{(\tri{1, 0},F), (\tri{2, 0},L)\}}
\cup C^{i+1}_0 \cup C^{i+1}_1 \cup C^{i+1}_2
.
$$

Finally, the group travels in the $\tri{1, 0}$ direction towards the first beacon and after $3i+2$ rounds, at time $d+9i+9$, the configuration $C^{i+1}$ is achieved:
\begin{multline*}
\translate{\tri{i, -i}}
{\{(\tri{3, 0},F), (\tri{4, 0},L)\}}
\cup C^{i+1}_0 \cup C^{i+1}_1 \cup C^{i+1}_2\\
= \translate{\tri{i+1, -i-1}}{M^0}
\cup C^{i+1}_0 \cup C^{i+1}_1 \cup C^{i+1}_2
= C^{i+1}.
\end{multline*}

Thus, every phase eventually takes place within a finite amount of time. During phase $i$, a set of vertices $S_i$ is visited, which can be defined as the vertices that lie on the edges of the triangle $\tri{3,0}+\tri{i,-i}$, $\tri{3,3}+\tri{i,2i}$, $\tri{0,0}+\tri{-2i,-i}$. The vertex $\tri{2,1}$ is also visited in the initial configuration. Since $\bigcup_{i\ge0}S_i \cup \{\tri{2,1}\} = V$, every vertex is visited and the algorithm \trialg{3} solves the IGE problem.
\qed

\renewcommand{\rulescale}{0.6}

\section{Hexagonal grid}
\label{ch:APhex}

In this section, we study solutions to the IGE problem in the hexagonal grid. This is, to the best of our knowledge, the first time such a case has been considered. We propose three different solutions.

\subsection{Five robots, three colors, visibility range two}
\label{sec:APhex31}

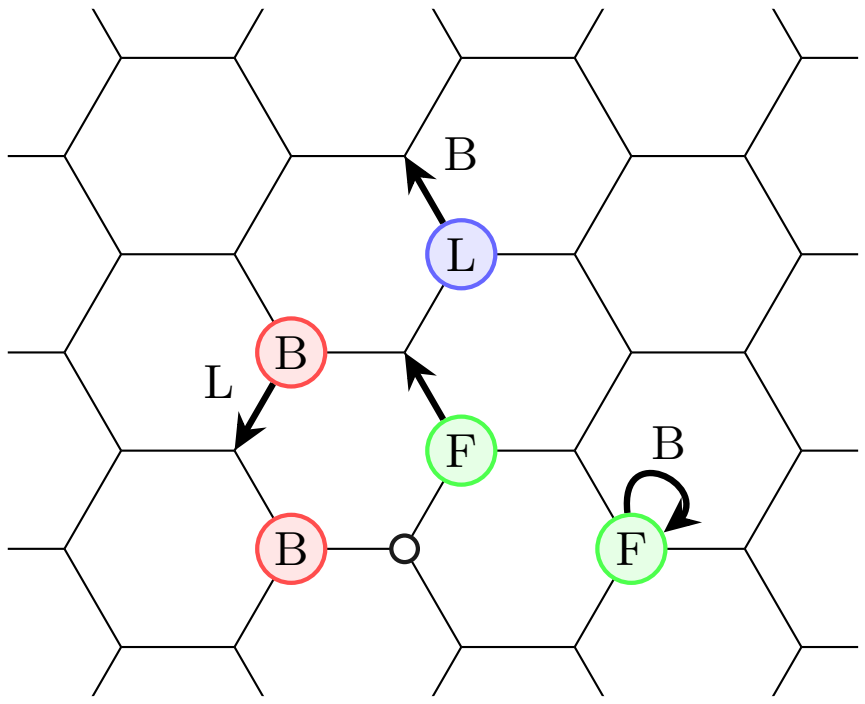
\begin{figure}[h!t]
\centering
    \begin{tikzpicture}[triangular_slant,scale=0.75]

        \hexgrid{11}{8}
        \begin{scope}[shift={(5,4)}]
        \begin{scope}[every node/.style={robot,rred}]
            \node (follower) at (0, -1) [rgreen] {F};
            \node (beacon3) at (1, -2) [rgreen] {F};
            \node (beacon1) at (-1, 0) {B};
            \node (leader) at (1, 1) [rblue] {L};
            \node at (-2, -2) {B};
        \end{scope}
        \node (origin) at (-1, -2) [mark] {};
        \begin{scope} [moves]
            \draw (follower) -- +(0, 1);
            \draw (beacon3) edge[out=75,in=25,looseness=10] node[above] {B} (beacon3);
            \draw (beacon1) edge node[above left] {L} +(-1, -1);
            \draw (leader) edge node[above right] {B} +(0, 1);

        \end{scope}
        \end{scope}
    \end{tikzpicture}

    \caption[Initial configuration of algorithm \hexalg{1}.]{Initial configuration of algorithm \hexalg{1}, including the actions taken in the first round and the origin $\hex{0,0,0}$ marked by a circle. Note the color changes of some robots.}
    \label{fig:APalgohex1init}
\end{figure}

The first algorithm we propose for the hexagonal grid, denoted as \hexalg{1}, solves the IGE problem using only five robots of three colors under visibility range two. While being the first algorithm to solve the problem in the hexagonal grid, it is also optimal in the number of robots needed, as per Theorem~\ref{the:fiveoptimaltri}. The initial configuration is shown in Figure~\ref{fig:APalgohex1init}. It is most comparable to algorithm \trialg{3}, as it also uses a moving group made up of one $L$-colored leader and one $F$-colored follower, and three $B$-colored beacons.

\begin{figure}[h!t]
\centering

\hspace*{\fill}\begin{subfigure}{3.6cm}
\centering
\captionsetup{justification=centering}
    \begin{tikzpicture}[triangular_slant,scale=\rulescale]
        \hexviewtwoR
        \begin{scope}[shift={(1,0)}]
        \begin{scope}[every node/.style={robottiny,rred}]
            \node (follower) at (0, -1) [rgreen] {F};
            \node (leader) at (1, 1) [rblue] {L};
            \node (beacon1) at (-1, 0) {B};
            \node (beacon2) at (-2, -2) {B};
        \end{scope}
        \begin{scope} [moves]
            \draw (beacon1) edge node[above left] {L} +(-1, -1);
        \end{scope}
        \end{scope}
    \end{tikzpicture}
    \caption{}
    \label{fig:APhex1uniq1}
\end{subfigure}\hfill \begin{subfigure}{3.6cm}
\centering
\captionsetup{justification=centering}
    \begin{tikzpicture}[triangular_slant,scale=\rulescale]
        \hexviewtwoR
        \begin{scope}[shift={(0,1)}]
        \begin{scope}[every node/.style={robottiny,rred}]
            \node (follower) at (0, -1) [rgreen] {F};
            \node (leader) at (1, 1) [rblue] {L};
            \node (beacon1) at (-1, 0) {B};
            \node (beacon2) at (-2, -2) {B};
            \node (beacon3) at (1, -2) [rgreen] {F};
        \end{scope}
        \begin{scope} [moves]
            \draw (follower) -- +(0, 1);
        \end{scope}
        \end{scope}
    \end{tikzpicture}
    \caption{}
    \label{fig:APhex1uniq2}
\end{subfigure}\hfill \begin{subfigure}{3.6cm}
\centering
\captionsetup{justification=centering}
    \begin{tikzpicture}[triangular_slant,scale=\rulescale]
        \hexviewtwoR
        \begin{scope}[shift={(-1,2)}]
        \begin{scope}[every node/.style={robottiny,rred}]
            \node (follower) at (0, -1) [rgreen] {F};
            \node (beacon3) at (1, -2) [rgreen] {F};
        \end{scope}
        \begin{scope} [moves]
            \draw (beacon3) edge[out=75,in=25,looseness=10] node[above] {B} (beacon3);
        \end{scope}
        \end{scope}
    \end{tikzpicture}
    \caption{}
    \label{fig:APhex1uniq3}
\end{subfigure}\hspace*{\fill}

\caption{Rules unique to the first round of execution in \hexalg{1}}
\label{fig:APhexalg1initunique}
\end{figure}
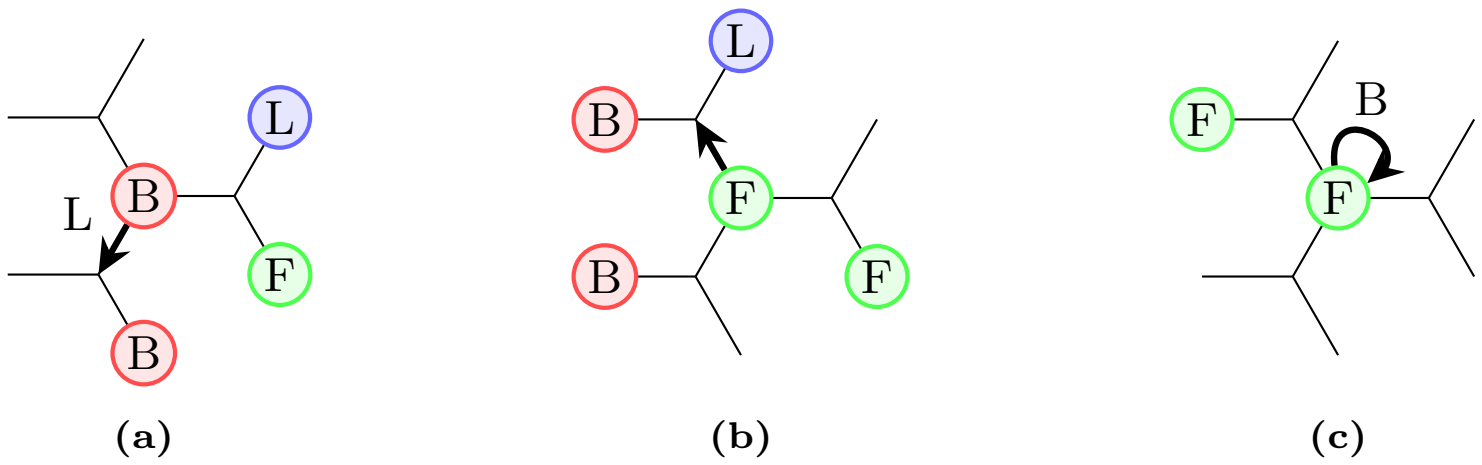

The initial configuration is constructed in such a way that no vertices stay unvisited after the explored area expands outwards.
This requires three additional rules, shown in Figure~\ref{fig:APhexalg1initunique}.
The first rule (Figure~\ref{fig:APhex1uniq1}) makes the top left beacon act the same as during an adjustment, discussed later. Its view is different because of the beacon below it. The second rule (Figure~\ref{fig:APhex1uniq2}) allows the center $F$-colored robot, which belongs to the moving group, to move towards the leader as if the bottom two robots were absent. The final rule (Figure~\ref{fig:APhex1uniq3}) changes the color of the bottom right robot from $F$ to $B$ while staying put.
We require a beacon at that exact position, however, it cannot be $B$-colored from the beginning as it would move due to a rule required later.
This way, it is allowed to stay in place while also becoming the desired color.

\begin{figure}[h!t]
\centering

\hspace*{\fill}\begin{subfigure}{3.6cm}
\centering
\captionsetup{justification=centering}
    \begin{tikzpicture}[triangular_slant,scale=\rulescale]
        \hexviewtwoR
        \begin{scope}[shift={(-1,-1)}]
        \begin{scope}[every node/.style={robottiny}]
            \node (follower) at (0, -1) [rgreen] {F};
            \node (leader) at (1, 1) [rblue] {L};
        \end{scope}
        \begin{scope} [moves]
            \draw (leader) -- +(0, 1);
        \end{scope}
        \end{scope}
    \end{tikzpicture}
    \caption{}
    \label{fig:APhex1move1}
\end{subfigure}\hfill \begin{subfigure}{3.6cm}
\centering
\captionsetup{justification=centering}
    \begin{tikzpicture}[triangular_slant,scale=\rulescale]
        \hexviewtwoR
        \begin{scope}[shift={(0,1)}]
        \begin{scope}[every node/.style={robottiny}]
            \node (follower) at (0, -1) [rgreen] {F};
            \node (leader) at (1, 1) [rblue] {L};
        \end{scope}
        \begin{scope} [moves]
            \draw (follower) -- +(0, 1);
        \end{scope}
        \end{scope}
    \end{tikzpicture}
    \caption{}
    \label{fig:APhex1move2}
\end{subfigure}\hspace*{\fill}

\caption{Rules for the moving group in \hexalg{1}}
\label{fig:APhexalg1movealong}
\end{figure}
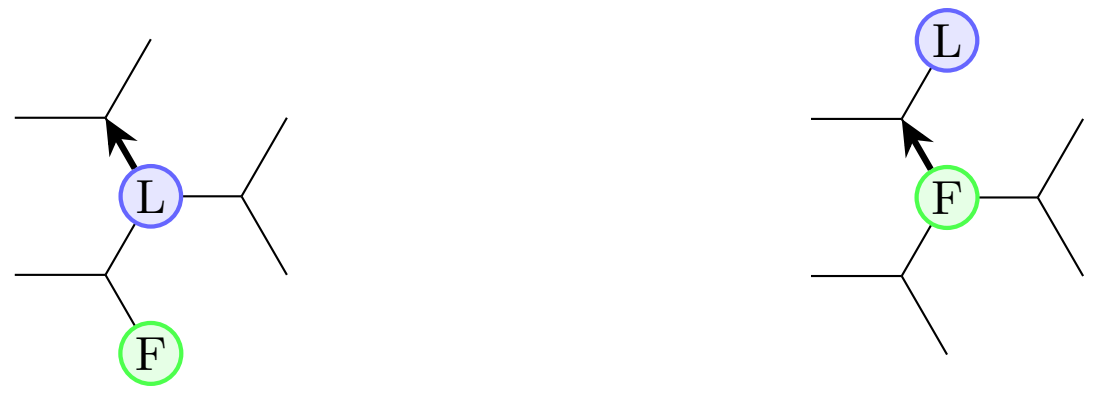

The exploration is then done by the moving group, which periodically visits each delimiting beacon. As a result, the path taken by the moving group somewhat resembles a triangular outward spiral. However, due to the hexagonal grid having no straight lines to travel along, the moving group moves in a zigzag manner, switching between left-facing and right-facing vertices each round. The rules for the moving group are shown in Figure~\ref{fig:APhexalg1movealong}. It moves based on the same principles as in algorithm \trialg{3}, with the leader moving away from the follower (Figure~\ref{fig:APhex1move1}) and the follower moving towards the leader (Figure~\ref{fig:APhex1move2}), only this time, they are separated by one vertex, allowed by the visibility range of two. Thanks to this, the rules are unambiguous.

\begin{figure}[h!t]
\centering

\hspace*{\fill}\begin{subfigure}{3.6cm}
\centering
\captionsetup{justification=centering}
    \begin{tikzpicture}[triangular_slant,scale=\rulescale]
        \hexviewtwoR
        \begin{scope}[shift={(1,3)}]
        \begin{scope}[every node/.style={robottiny}]
            \node (leader) at (-1, -3) [rblue] {L};
            \node (beacon) at (1, -2) [rred] {B};
            \node (follower) at (-2, -2) [rgreen] {F};
        \end{scope}
        \begin{scope} [moves]
            \draw (leader) -- +(1, 0);
        \end{scope}
        \end{scope}
    \end{tikzpicture}
    \caption{}
    \label{fig:APhex1adj1}
\end{subfigure}\hfill \begin{subfigure}{3.6cm}
\captionsetup{justification=centering}
\centering
    \begin{tikzpicture}[triangular_slant,scale=\rulescale]
        \hexviewtwoL
        \begin{scope}[shift={(0,3)}]
        \begin{scope}[every node/.style={robottiny}]
            \node (leader) at (0, -3) [rblue] {L};
            \node (beacon) at (1, -2) [rred] {B};
            \node (follower) at (-1, -2) [rgreen] {F};
        \end{scope}
        \begin{scope} [moves]
            \draw (leader) -- +(0, -1);
        \end{scope}
        \end{scope}
    \end{tikzpicture}
    \caption{}
    \label{fig:APhex1adj2}
\end{subfigure}\hspace*{\fill}

\par\medskip

\hspace*{\fill}\begin{subfigure}{3.6cm}
\centering
\captionsetup{justification=centering}
    \begin{tikzpicture}[triangular_slant,scale=\rulescale]
        \hexviewtwoR
        \begin{scope}[shift={(0,1)}]
        \begin{scope}[every node/.style={robottiny}]
            \node (leader) at (0, -1) [rblue] {L};
            \node (beacon) at (1, 1) [rred] {B};
            \node (follower) at (-1, 0) [rgreen] {F};
        \end{scope}
        \begin{scope} [moves]
            \draw (leader) edge node[above] {B} +(1, 0);
        \end{scope}
        \end{scope}
    \end{tikzpicture}
    \caption{}
    \label{fig:APhex1adj3}
\end{subfigure}\hfill \begin{subfigure}{3.6cm}
\centering
\captionsetup{justification=centering}
    \begin{tikzpicture}[triangular_slant,scale=\rulescale]
        \hexviewtwoR
        \begin{scope}[shift={(1,0)}]
        \begin{scope}[every node/.style={robottiny}]
            \node (leader) at (0, -1) [rblue] {L};
            \node (beacon) at (1, 1) [rred] {B};
            \node (follower) at (-1, 0) [rgreen] {F};
        \end{scope}
        \begin{scope} [moves]
            \draw (follower) -- +(1, 0);
        \end{scope}
        \end{scope}
    \end{tikzpicture}
    \caption{}
    \label{fig:APhex1adj4}
\end{subfigure}\hfill \begin{subfigure}{3.6cm}
\centering
\captionsetup{justification=centering}
    \begin{tikzpicture}[triangular_slant,scale=\rulescale]
        \hexviewtwoR
        \begin{scope}[shift={(-1,-1)}]
        \begin{scope}[every node/.style={robottiny}]
            \node (leader) at (0, -1) [rblue] {L};
            \node (beacon) at (1, 1) [rred] {B};
            \node (follower) at (-1, 0) [rgreen] {F};
        \end{scope}
        \begin{scope} [moves]
            \draw (beacon) edge node[above right] {L} +(0, 1);
        \end{scope}
        \end{scope}
    \end{tikzpicture}
    \caption{}
    \label{fig:APhex1adj5}
\end{subfigure}\hspace*{\fill}

\par\medskip

\hspace*{\fill}\begin{subfigure}{3.6cm}
\centering
\captionsetup{justification=centering}
    \begin{tikzpicture}[triangular_slant,scale=\rulescale]
        \hexviewtwoL
        \begin{scope}[shift={(-1,1)}]
        \begin{scope}[every node/.style={robottiny}]
            \node (leader) at (1, -1) [rred] {B};
            \node (follower) at (0, 0) [rgreen] {F};
        \end{scope}
        \begin{scope} [moves]
            \draw (leader) -- +(0, -1);
        \end{scope}
        \end{scope}
    \end{tikzpicture}
    \caption{}
    \label{fig:APhex1adj6}
\end{subfigure}\hfill \begin{subfigure}{3.6cm}
\centering
\captionsetup{justification=centering}
    \begin{tikzpicture}[triangular_slant,scale=\rulescale]
        \hexviewtwoL
        \begin{scope}[shift={(0,0)}]
        \begin{scope}[every node/.style={robottiny}]
            \node (leader) at (1, -1) [rred] {B};
            \node (beacon) at (1, 2) [rblue] {L};
            \node (follower) at (0, 0) [rgreen] {F};
        \end{scope}
        \begin{scope} [moves]
            \draw (follower) -- +(1, 1);
        \end{scope}
        \end{scope}
    \end{tikzpicture}
    \caption{}
    \label{fig:APhex1adj7}
\end{subfigure}\hspace*{\fill}

\caption{Rules used during an adjustment in \hexalg{1}}
\label{fig:APhexalg1adjustment}
\end{figure}
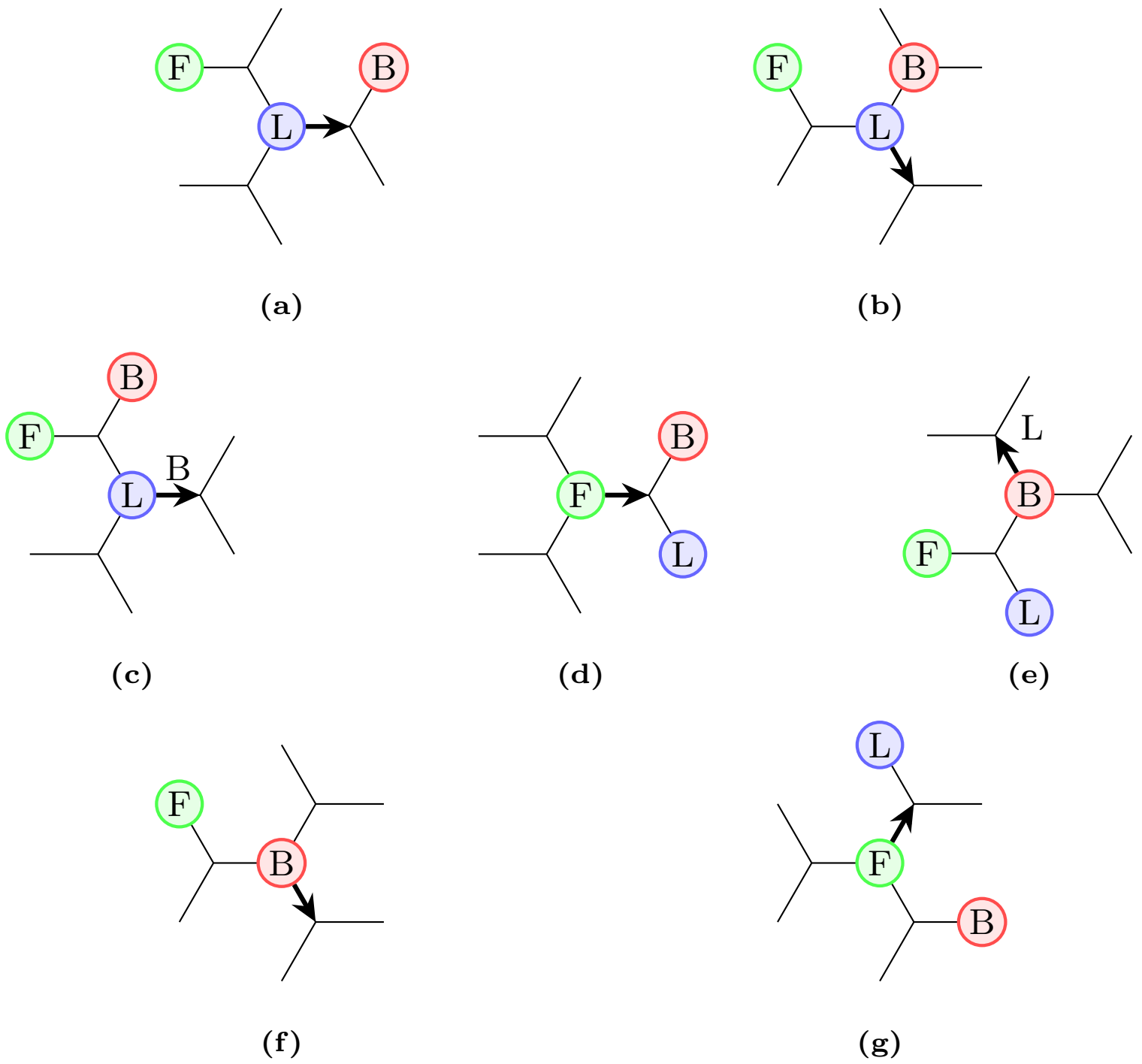

Just as in the previous algorithms, the moving group performs an adjustment when coming close to a beacon. The rules that allow for an adjustment to take place are shown in Figure~\ref{fig:APhexalg1adjustment}. The first two rules (Figures~\ref{fig:APhex1adj1} and \ref{fig:APhex1adj2}) allow the leader to move past the beacon for two rounds when first spotting it, as if it were not there. After this, the three robots are arranged in a triangle shape, all separated by a single unoccupied left-facing vertex. At this point, the leader and beacon robots change colors, switching roles. To be more precise, the leader robot continues moving in the same direction as before and changes its color to $B$ (Figure~\ref{fig:APhex1adj3}). The beacon robot moves away from the leader robot and changes its color to $L$ (Figure~\ref{fig:APhex1adj5}). The follower moves in the same direction as the leader, onto the free vertex in the center of the triangle (Figure~\ref{fig:APhex1adj4}). It does not change its color.

In the final round of the adjustment, the newly $L$-colored robot sees only the follower robot and acts according to the previously mentioned moving group rules. The unchanged follower robot sees both the new leader and beacon, and moves towards the leader (Figure~\ref{fig:APhex1adj7}). The newly $B$-colored robot moves away from the follower robot to correctly position itself for the next adjustment (Figure~\ref{fig:APhex1adj6}). The entire sequence of configurations during an adjustment can be seen in Figure~\ref{fig:APalgohex1adjustment}.

{
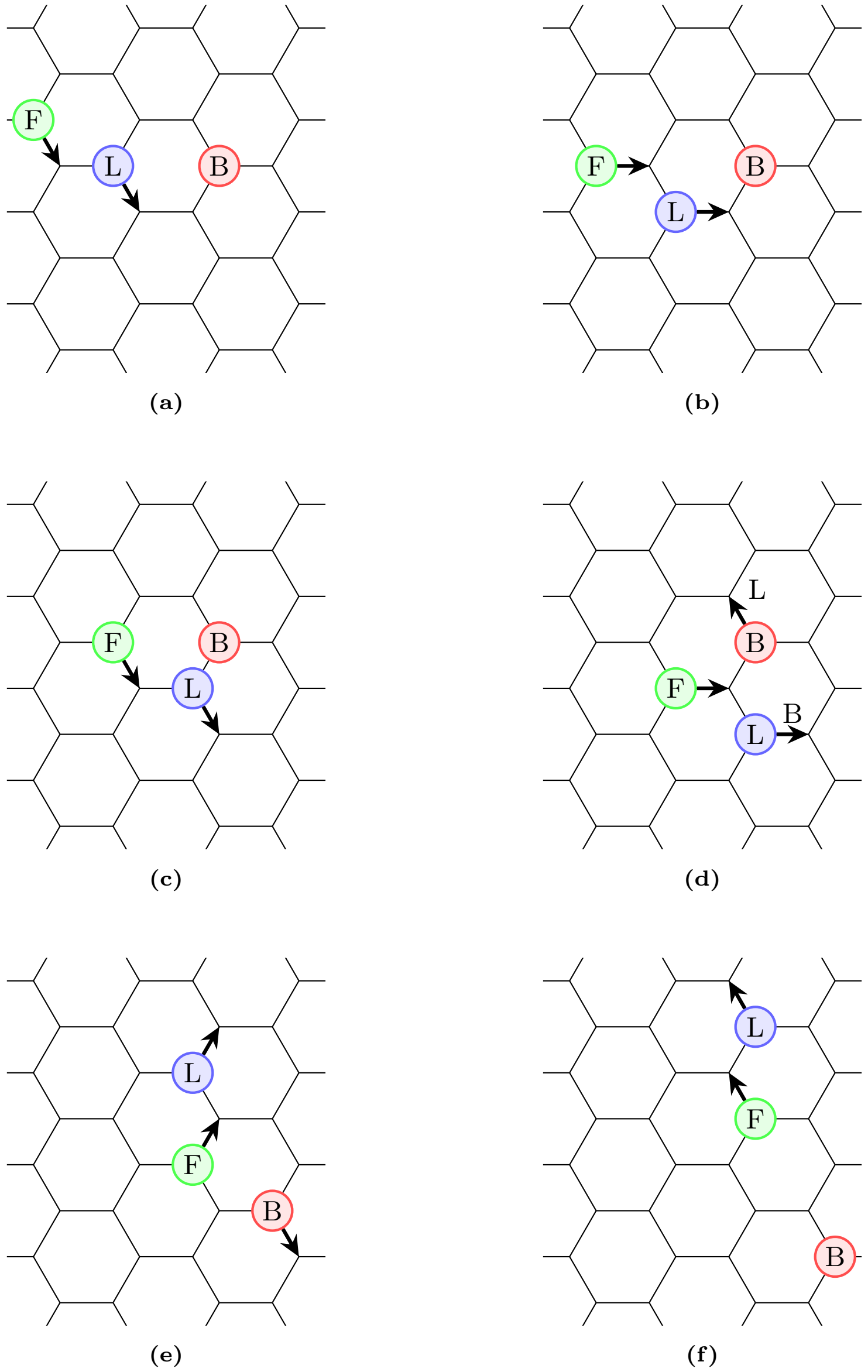
\begin{figure}
\centering
\hspace*{\fill}\begin{subfigure}{0.49\textwidth}
    \centering
    \captionsetup{justification=centering}
        \begin{tikzpicture}[triangular_slant,scale=\rulescale]
            \hexgrid{10}{9}
            \begin{scope}[shift={(5,7)}]
            \begin{scope}[every node/.style={robot}]
                \node (leader) at (-1, -2) [rblue] {L};
                \node (beacon) at (1, -2) [rred] {B};
                \node (follower) at (-2, -1) [rgreen] {F};
            \end{scope}
            \begin{scope} [moves]
                \draw (leader) -- +(0, -1);
                \draw (follower) -- +(0, -1);
            \end{scope}
            \end{scope}
        \end{tikzpicture}
        \caption{}
    \end{subfigure}\hfill \begin{subfigure}{0.49\textwidth}
    \centering
    \captionsetup{justification=centering}
        \begin{tikzpicture}[triangular_slant,scale=\rulescale]
            \hexgrid{10}{9}

            \begin{scope}[shift={(5,7)}]
            \begin{scope}[every node/.style={robot}]
                \node (leader) at (-1, -3) [rblue] {L};
                \node (beacon) at (1, -2) [rred] {B};
                \node (follower) at (-2, -2) [rgreen] {F};
            \end{scope}
            \begin{scope} [moves]
                \draw (leader) -- +(1, 0);
                \draw (follower) -- +(1, 0);
            \end{scope}
            \end{scope}
        \end{tikzpicture}
        \caption{}
    \end{subfigure}\hspace*{\fill}

    \par\vspace{0.70cm}

    \hspace*{\fill}\begin{subfigure}{0.49\textwidth}
    \centering
    \captionsetup{justification=centering}
        \begin{tikzpicture}[triangular_slant,scale=\rulescale]
            \hexgrid{10}{9}
            \begin{scope}[shift={(5,7)}]
            \begin{scope}[every node/.style={robot}]
                \node (leader) at (0, -3) [rblue] {L};
                \node (beacon) at (1, -2) [rred] {B};
                \node (follower) at (-1, -2) [rgreen] {F};
            \end{scope}
            \begin{scope} [moves]
                \draw (leader) -- +(0, -1);
                \draw (follower) -- +(0, -1);
            \end{scope}
            \end{scope}
        \end{tikzpicture}
        \caption{}
    \end{subfigure}\hfill \begin{subfigure}{0.49\textwidth}
    \centering
    \captionsetup{justification=centering}
        \begin{tikzpicture}[triangular_slant,scale=\rulescale]
            \hexgrid{10}{9}
            \begin{scope}[shift={(5,7)}]
            \begin{scope}[every node/.style={robot}]
                \node (leader) at (0, -4) [rblue] {L};
                \node (beacon) at (1, -2) [rred] {B};
                \node (follower) at (-1, -3) [rgreen] {F};
            \end{scope}
            \begin{scope} [moves]
                \draw (leader) edge node[above] {B} +(1, 0);
                \draw (follower) -- +(1, 0);
                \draw (beacon) edge node[above right] {L} +(0, 1);
            \end{scope}
            \end{scope}
        \end{tikzpicture}
        \caption{}
    \end{subfigure}\hspace*{\fill}

    \par\vspace{0.70cm}

    \hspace*{\fill}\begin{subfigure}{0.49\textwidth}
    \centering
    \captionsetup{justification=centering}
        \begin{tikzpicture}[triangular_slant,scale=\rulescale]
            \hexgrid{10}{9}
            \begin{scope}[shift={(5,7)}]
            \begin{scope}[every node/.style={robot}]
                \node (beacon) at (1, -4) [rred] {B};
                \node (leader) at (1, -1) [rblue] {L};
                \node (follower) at (0, -3) [rgreen] {F};
            \end{scope}
            \begin{scope} [moves]
                \draw (leader) -- +(1, 1);
                \draw (follower) -- +(1, 1);
                \draw (beacon) -- +(0, -1);
            \end{scope}
            \end{scope}
        \end{tikzpicture}
        \caption{}
    \end{subfigure}\hfill \begin{subfigure}{0.49\textwidth}
    \centering
    \captionsetup{justification=centering}
        \begin{tikzpicture}[triangular_slant,scale=\rulescale]
            \hexgrid{10}{9}
            \begin{scope}[shift={(5,7)}]
            \begin{scope}[every node/.style={robot}]
                \node (beacon) at (1, -5) [rred] {B};
                \node (leader) at (2, 0) [rblue] {L};
                \node (follower) at (1, -2) [rgreen] {F};
            \end{scope}
            \begin{scope} [moves]
                \draw (leader) -- +(0, 1);
                \draw (follower) -- +(0, 1);
            \end{scope}
            \end{scope}
        \end{tikzpicture}
        \caption{}
    \end{subfigure}\hspace*{\fill}

    \caption[The adjustment sequence in \hexalg{1}.]{The entire sequence of moves during an adjustment in the execution of~\hexalg{1}}
    \label{fig:APalgohex1adjustment}
\end{figure}
}

\paragraph{Algorithm \hexalg{1}.} To summarize, we formally define algorithm \hexalg{1} as the triplet $(\{L, F, B\}, I, T)$, where $I$ is the initial configuration as shown in Figure~\ref{fig:APalgohex1init} and $T$ is the transition function defined as the set of rules depicted in Figures~\ref{fig:APhexalg1initunique},~\ref{fig:APhexalg1movealong}~and~\ref{fig:APhexalg1adjustment}.

\begin{theorem}\label{the:APhex31}
Algorithm \hexalg{1} solves the IGE problem in the hexagonal grid using five robots of three colors under visibility range $\Phi=2$. It is well-defined.
\end{theorem}

\proof Like in previous proofs, we divide the execution into phases, show by induction that every phase is eventually reached and finally exhibit a specific set of nodes visited during each phase.

We assume the vertex marked by a circle in Figure~\ref{fig:APalgohex1init} to be the origin $\hex{0,0,0}$. We declare the configuration reached from $I$ after two rounds of execution as the beginning of phase 0, denoted by $C^0$. Configuration $C^0$ is then partitioned as
$$
C^0 = M^0\cup C^0_0 \cup C^0_1 \cup C^0_2,
$$
where
\begin{align*}
    M^0 &= \{(\hex{-1, 1, 1},F), (\hex{-2, 1, 1},L)\}\\
    C^0_0 &= \{(\hex{-1, 0, 1}, B)\},\\
    C^0_1 &= \{(\hex{1, -1, 1}, B)\},\\
    C^0_2 &= \{(\hex{0, 3, 1}, B)\}.
\end{align*}
Configuration $M^0$ contains the robots of the moving group and configurations $C^0_0$, $C^0_1$ and $C^0_2$ contain the beacons in the order in which they are visited during each phase -- $C^0_0$ contains the leftmost beacon, $C^0_1$ contains the bottom right beacon and $C^0_2$ contains the top right beacon.
The beginning of phase $i \in \mathbb{N}$ is then defined as the configuration
$$C^i = M^i \cup C^i_0 \cup C^i_1 \cup C^i_2,$$
where
\begin{align*}
    M^i &= \translate{\tri{-3i,0}}{M^0},\\
    C^i_0 &= \translate{\tri{-3i,0}}{C^0_0},\\
    C^i_1 &= \translate{\tri{0,-3i}}{C^0_1},\\
    C^i_2 &= \translate{\tri{3i,3i}}{C^0_2}.
\end{align*}
Note the use of the $\tri{\cdot}$ function in the translation notation. While the vector we translate by could be denoted using the $\hex{\cdot}$ function, we use the former for its descriptive nature.

Let us now prove that, starting at configuration $C^i$ in phase $i$, configuration $C^{i+1}$ is eventually reached. We begin at round $d$. The moving group and the beacon are in the triangle formation, all visible to each other. After two rounds, the first adjustment is complete and the configuration at time $d+2$ is
$$
\translate{\tri{-3i,0}}{
\{(\hex{-1, 0, 1},F), (\hex{0, -1, 1},L)\}
}
\cup C^{i+1}_0 \cup C^i_1 \cup C^i_2
.
$$

The moving group now travels towards the second beacon robot, alternatingly moving in the $\tri{1, 0}$ and $\tri{0, -1}$ directions (informally, it moves to the right and down). It reaches the beacon in $6i+2$ rounds. At time $d+6i+4$, just before the second adjustment takes place, the configuration is
$$
\translate{\tri{0,-3i}}{
\{(\hex{0, -1, 1},F), (\hex{1, -2, 1},L)\}
}
\cup C^{i+1}_0 \cup C^i_1 \cup C^i_2
,
$$
where the moving group and beacon are in the triangle formation. Note the change in the written coordinates of the moving group -- this is due to the movement in the last two rounds (as opposed to the $6i$ rounds, which modified the translation vector).

After two rounds, at time $d+6i+6$, the second adjustment is complete. The configuration becomes
$$
\translate{\tri{0,-3i}}{
\{(\hex{1, -1, 1},F), (\hex{1, 0, 1},L)\}
}
\cup C^{i+1}_0 \cup C^{i+1}_1 \cup C^i_2
.
$$
The moving group now travels upwards towards the third beacon, alternatingly moving in the $\tri{0,1}$ and $\tri{1,1}$ directions.

After $6i+4$ rounds, the third beacon is reached and the triangle formation is formed again. At time $d+12i+10$, the configuration is
$$
\translate{\tri{3i,3i}}{
\{(\hex{1, 1, 1},F), (\hex{1, 2, 1},L)\}
}
\cup C^{i+1}_0 \cup C^{i+1}_1 \cup C^i_2
.
$$
After another two rounds, the third adjustment is complete. The configuration at time $d+12i+12$ is
$$
\translate{\tri{3i,3i}}{
\{(\hex{0, 2, 1},F), (\hex{-1, 2, 1},L)\}
}
\cup C^{i+1}_0 \cup C^{i+1}_1 \cup C^{i+1}_2
.
$$

Finally, the group begins to travel back towards the first beacon, alternatingly taking steps in the $\tri{-1, -1}$ and $\tri{-1, 0}$ directions. It reaches the beacon in $6i+6$ rounds, at which point phase $i$ is over and phase $i+1$ begins. At time $d+18i+18$, the configuration is
$$
\translate{\tri{-3i,0}}{
\{(\hex{-3, 2, 1},F), (\hex{-4, 2, 1},L)\}
}
\cup C^{i+1}_0 \cup C^{i+1}_1 \cup C^{i+1}_2
.
$$
The configuration of the moving group can be written as
\begin{align*}
&\translate{\tri{-3i,0}}{
\{(\hex{-3, 2, 1},F), (\hex{-4, 2, 1},L)\}
} \\
&= \translate{\tri{-3i,0}}{
\{
(\hex{-1, 1, 1}+\tri{-3,0},F),
(\hex{-2, 1, 1}+\tri{-3,0},L)
\}
} \\
&= \translate{\tri{-3i-3,0}}{
\{
(\hex{-1, 1, 1},F),
(\hex{-2, 1, 1},L)
\}
} \\
&= M^{i+1}.
\end{align*}
Therefore, the entire configuration is
\begin{multline*}
\translate{\tri{-3i,0}}{
\{(\hex{-3, 2, 1},F), (\hex{-4, 2, 1},L)\}
}
\cup C^{i+1}_0 \cup C^{i+1}_1 \cup C^{i+1}_2 \\
= M^{i+1} \cup C^{i+1}_0 \cup C^{i+1}_1 \cup C^{i+1}_2 = C^{i+1}.
\end{multline*}

By induction, every phase eventually takes place -- configuration $C^0$ is reached after two rounds from configuration $I$ and thanks to the inductive step, $C^i$ is also reached within a finite amount of time.

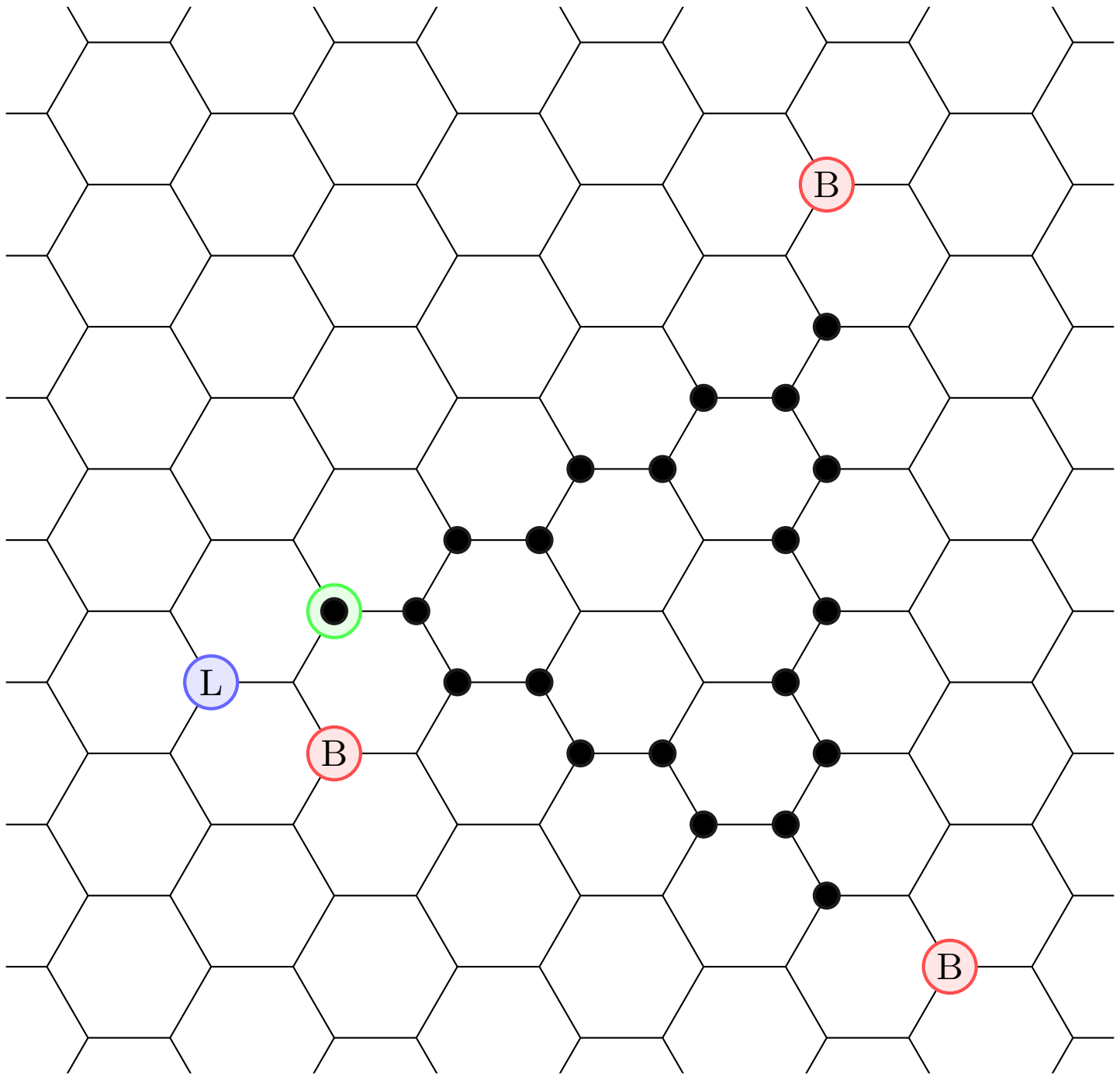
\begin{figure}[h!t]
\centering
    \begin{tikzpicture}[triangular_slant,scale=0.7]

        \hexgrid{21}{16}
        \begin{scope}[shift={(11,7)}]
        \begin{scope}[every node/.style={robot,rred}]
            \node at (-5, -2) {B};
            \node at (-4, 0) [rgreen] {};
            \node at (1, -5) {B};
            \node at (-6, -1) [rblue] {L};
            \node at (5, 6) {B};
        \end{scope}
        \foreach \pos in {(-4, 0), (-3, 0), (-3, -1), (-2, -1), (-2, -2), (-1, -2),
        (-1, -3), (0, -3), (0, -4),
        (1, -2), (1, -1), (2, 0), (2, 1), (3, 2), (3, 3), (4, 4),
        (2, 3), (1, 2), (0, 2), (-1, 1), (-2, 1)
        }
            \node at \pos [mark={6pt},fill=black] {};

        \end{scope}
    \end{tikzpicture}

    \caption[Nodes visited during the first phase of \hexalg{1}.]{Nodes visited during the first phase of \hexalg{1} marked by a dot}
    \label{fig:APhex1visited}
\end{figure}

When we follow the nodes occupied during each phase by the $L$ and $F$-colored robots, a shape best described as a wavy triangle is formed. An example of the shape can be seen in Figure~\ref{fig:APhex1visited}. This set of vertices can be defined by the corners of the triangle. Accordingly, we denote the set of vertices visited during phase $i$ as $S_i$ and define it as the nodes contained in such a triangle defined by the corner vertices $\hex{-3, 2, 1}+\tri{-3i,0}$, $\hex{1, -2, 1}+\tri{0,-3i}$ and $\hex{1, 2, 1}+\tri{3i,3i}$. Additionally, there are four vertices that are not contained in $S_0$, as seen in Figure~\ref{fig:APhex1visited}. The vertices $\hex{0, 0, 1}$, $\hex{0, 1, 1}$ and $\hex{-1, 1, 1}$ are all occupied in the initial configuration $I$ and the center vertex $\hex{0, 1, 0}$ is occupied after one round of execution from $I$. Subsequently, $\bigcup_{i\ge0}S_i \cup \{\hex{0, 0, 1}, \hex{0, 1, 1}, \hex{-1, 1, 1}, \hex{0, 1, 0}\} = V$, and therefore every vertex is visited and the algorithm \hexalg{1} solves the IGE problem.
\qed

\subsection{Nine robots, single color, visibility range three}
\label{sec:APhex1color}

\begin{figure}[h!t]
\centering
    \begin{tikzpicture}[triangular_slant,scale=0.6]

        \hexgrid{19.5}{13}
        \begin{scope}[shift={(11,7)}]
        \begin{scope}[every node/.style={robottiny,rblack}]
            \node (338) at (0, 3)[rblack] {R};
            \node (-1683) at (-6, -4)[rblack] {R};
            \node (2130) at (1, -2)[rblack] {R};
            \node (276) at (-2, -1)[rblack] {R};
            \node (-1621) at (-4, 0)[rblack] {R};
            \node (1206) at (-1, -2)[rblack] {R};
            \node (1144) at (-3, -6)[rblack] {R};
            \node (270) at (-1, 0)[rblack] {R};
            \node (2223) at (4, 4)[rblack] {R};
        \end{scope}
\begin{scope} [moves]
            \draw (338) -- +(1, 1);
            \draw (2130) -- +(1, 0);
            \draw (276) -- +(0, -1);
            \draw (-1621) -- +(0, 1);
            \draw (270) -- +(1, 0);
        \end{scope}
        \end{scope}
    \end{tikzpicture}

    \caption[Initial configuration of algorithm \hexalg{2}.]{Initial configuration of algorithm \hexalg{2}, including the actions taken in the first round}\label{fig:APalgohex2init}
\end{figure}
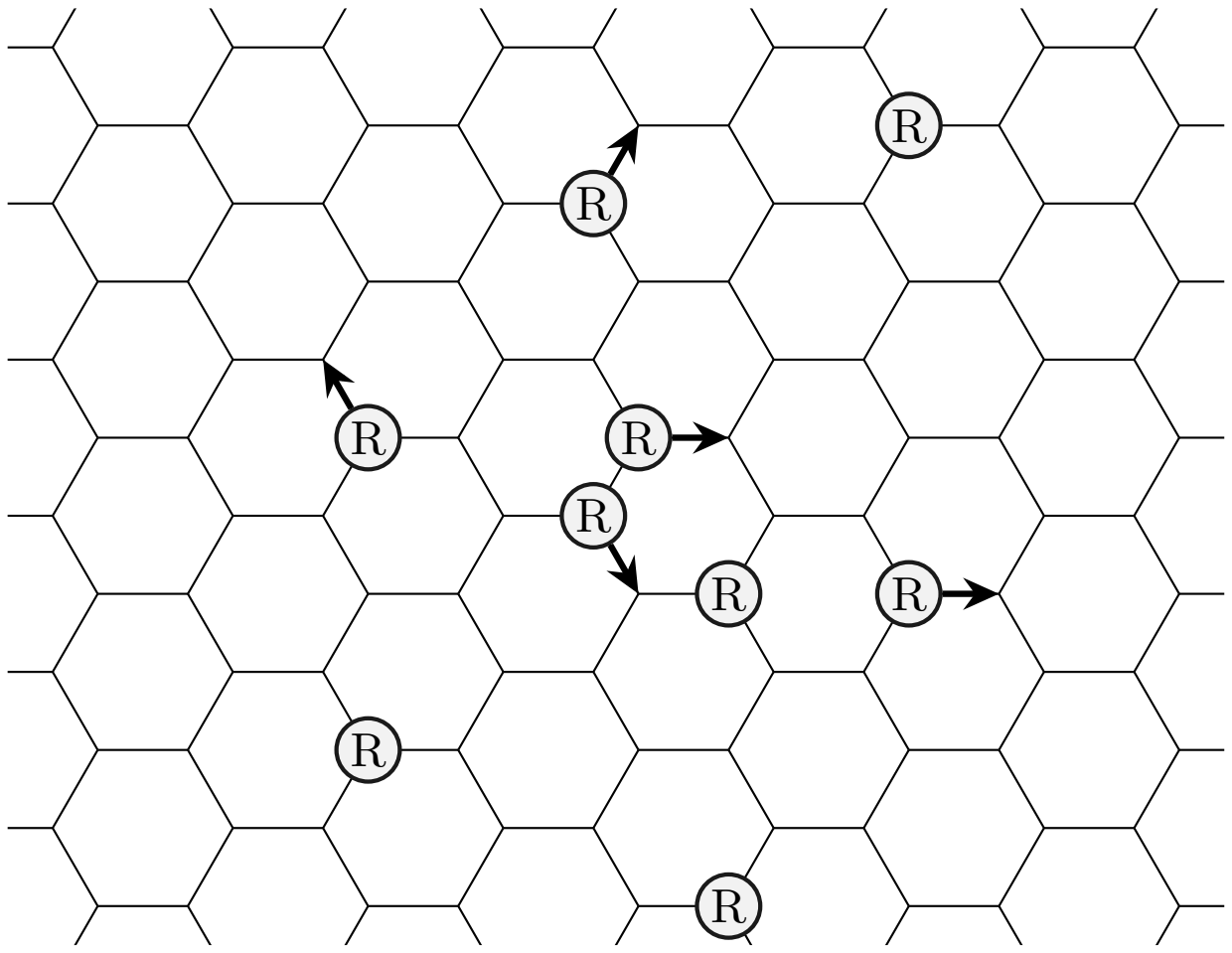

The second algorithm for the hexagonal grid, denoted as \hexalg{2}, solves the IGE problem using nine robots of a single color under visibility range three. It shows that infinite exploration is possible in the hexagonal grid using robots of a single color, while keeping visibility range small. The initial configuration is shown in Figure~\ref{fig:APalgohex2init}. It is similar to all previous algorithms in that it uses a moving group to explore and beacons to delimit the explored area, however, unlike the others, it uses six beacons instead of three. As such, the spiraling path taken by the moving group is hexagonal rather than triangular. The moving group is made up of three robots. The roles of the robots do not switch at any point.

The initial configuration is again constructed so that no vertices stay unvisited after the robots expand outwards. The inner three robots (who form the moving group) rearrange themselves through the help of the outer robots (beacons), which move to their required positions in the process. For this, a few single-use rules are needed. The moving group then begins to travel upward, during which the center-right beacon is once again nudged to its required position. On the fourth round, the final single-use rule is applied when the front robot of the moving group spots two beacons and performs two adjustments at the same time.

\begin{figure}[h!t]
\centering
    \begin{tikzpicture}[triangular_slant,scale=0.6]

        \hexgrid{10.5}{7}
        \begin{scope}[shift={(5,4)}]
        \begin{scope}[every node/.style={robottiny,rblack}]
            \node (1200) at (0, -1)[rblack] {R};;
            \node (1206) at (-1, -2)[rblack] {R};
            \node (1231) at (1, 1)[rblack] {R};
        \end{scope}
        \begin{scope} [moves]
            \draw (1200) -- +(0, 1);
            \draw (1206) -- (1200);
            \draw (1231) -- +(0, 1);
        \end{scope}
        \end{scope}
    \end{tikzpicture}

    \caption{The moving group of algorithm \hexalg{2}}
    \label{fig:APalgohex2movinggroup}
\end{figure}
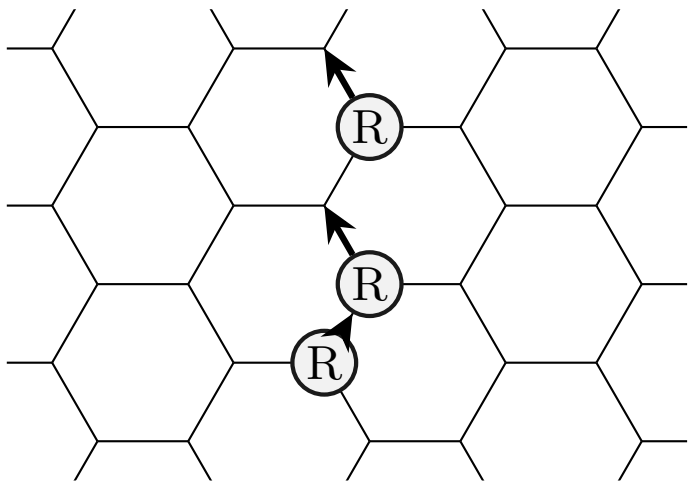

The moving group and its rules can be seen in Figure~\ref{fig:APalgohex2movinggroup}. Three robots are needed to somehow distinguish the front and back side. Each robot sees all other robots of the moving group thanks to the visibility range being three. However, a visibility range of two is sufficient for it to function.

The beacon robots stay separate from the moving group by only ever being seen at a distance of three. When the moving group first approaches a beacon during an adjustment, it is positioned directly across a hexagon, and the beacon moves away (Figure~\ref{fig:APhex2adj1}). In the next round, the beacon and the front robot of the moving group are still in view (Figure~\ref{fig:APhex2adj2}), which is distinct from the view of the last round. This allows them to take different actions. The beacon moves away from the front robot and assumes the correct position for the next adjustment. The front robot turns away, steering the moving group towards the next beacon. In the final round of the adjustment, the beacon is no longer in view and the moving group forms an L shape. The moving group then continues to travel in the direction of the front robot (Figure~\ref{fig:APhex2adj3}), regaining its regular shape.

{
\begin{figure}[h!t]
\centering
\hspace*{\fill}\begin{subfigure}{4cm}
\centering
\captionsetup{justification=centering}
    \begin{tikzpicture}[triangular_slant,scale=0.5]
        \hexgrid{12}{10}
        \begin{scope}[shift={(5,4)}]
        \begin{scope}[every node/.style={robottiny,rblack}]
            \node (1200) at (0, -1)[rblack] {R};
            \node (2229) at (3, 3)[rblack] {R};
            \node (1206) at (-1, -2)[rblack] {R};
            \node (1231) at (1, 1)[rblack] {R};
        \end{scope}
        \begin{scope} [moves]
            \draw (1200) -- +(0, 1);
            \draw (2229) -- +(1, 1);
            \draw (1206) -- (1200) [movestiny];
            \draw (1231) -- +(0, 1);
        \end{scope}
        \end{scope}
    \end{tikzpicture}
    \caption{}
    \label{fig:APhex2adj1}
\end{subfigure}\hfill \begin{subfigure}{4cm}
\centering
\captionsetup{justification=centering}
    \begin{tikzpicture}[triangular_slant,scale=0.5]
        \hexgrid{12}{10}
        \begin{scope}[shift={(5,4)}]
        \begin{scope}[every node/.style={robottiny,rblack}]
            \node (1200) at (0, -1)[rblack] {R};
            \node (1268) at (1, 2)[rblack] {R};
            \node (1237) at (0, 0)[rblack] {R};
            \node (2223) at (4, 4)[rblack] {R};
        \end{scope}
        \begin{scope} [moves]
            \draw (1200) -- (1237) [movestiny];
            \draw (1268) -- +(-1, 0);
            \draw (1237) -- +(1, 1);
            \draw (2223) -- +(1, 0);
        \end{scope}
        \end{scope}
    \end{tikzpicture}
    \caption{}
    \label{fig:APhex2adj2}
\end{subfigure}\hfill \begin{subfigure}{4cm}
\centering
\captionsetup{justification=centering}
    \begin{tikzpicture}[triangular_slant,scale=0.5]
        \hexgrid{12}{10}
        \begin{scope}[shift={(5,4)}]
        \begin{scope}[every node/.style={robottiny,rblack}]
            \node (1237) at (0, 0)[rblack] {R};
            \node (3190) at (5, 4)[rblack] {R};
            \node (301) at (0, 2)[rblack] {R};
            \node (1231) at (1, 1)[rblack] {R};
        \end{scope}
        \begin{scope} [moves]
            \draw (1237) -- (1231) [movestiny];
            \draw (301) -- +(0, 1);
            \draw (1231) -- +(0, 1);
        \end{scope}
        \end{scope}
    \end{tikzpicture}
    \caption{}
    \label{fig:APhex2adj3}
\end{subfigure}\hspace*{\fill}

    \caption{An adjustment in the execution of \hexalg{2}}
    \label{fig:APalgohex2adjustment}
\end{figure}
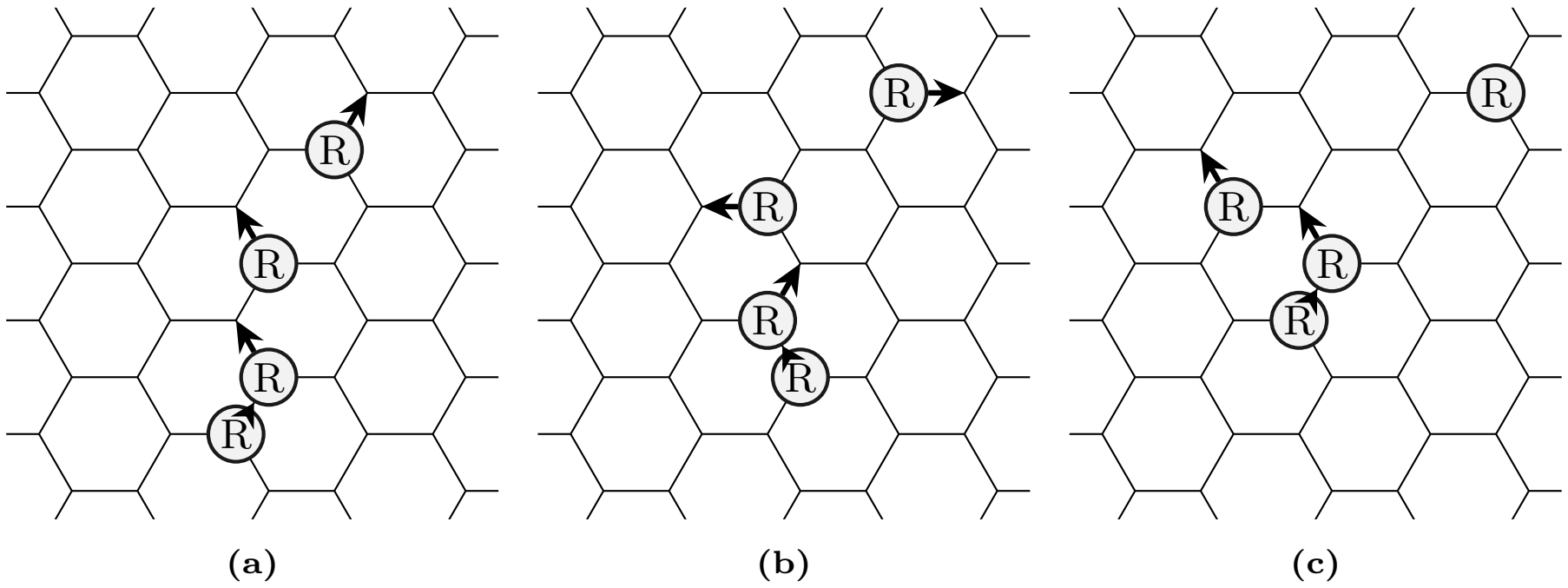
}

\paragraph{Algorithm \hexalg{2}.} To summarize, we formally define algorithm \hexalg{2} as the triplet $(\{R\}, I, T)$, where $I$ is the initial configuration as shown in Figure~\ref{fig:APalgohex2init} and $T$ is the transition function defined as the set of rules implicitly depicted in the figures of this subsection.

\begin{theorem}\label{the:APhex1color}
Algorithm \hexalg{2} solves the IGE problem in the hexagonal grid using nine robots of a single color under visibility range $\Phi=3$. It is well-defined.
\end{theorem}

\proof[Sketch] The proof is structured in the same way as for the previous algorithms.
We again divide the execution into phases, letting phase~$i$ begin in the round in which the moving group reaches the first of the six beacons for the $(i+1)$-st time, and we denote the configuration of this round by~$C^i$.
Starting from~$C^i$, the moving group travels along the six sides of a hexagon and repositions each of the beacons one step further away from the center as it passes it, until it reaches the first beacon again in the configuration~$C^{i+1}$.
An induction on~$i$ then shows that every phase is eventually reached, as the configuration~$C^0$ is reached from~$I$ in a finite number of rounds.
Denoting by $S_i$ the set of vertices occupied during phase~$i$, the sets $S_i$ form the boundaries of an increasing sequence of hexagons and hence, their union contains every vertex lying outside the hexagon traversed in phase~0.
It thus only remains to follow the execution from~$I$ and to check that the finitely many vertices in the center of the spiral are all visited before phase~0 begins.
Finally, no ambiguous rule is ever applied during the execution and therefore, the algorithm is well-defined.
\qed

\subsection{Exploration under visibility range one}
\label{sec:APhex1view}

We begin this section with a simple impossibility result, showing that any solution in the hexagonal grid using a visibility range of one will need to use at least one ambiguous rule.

\begin{lemma}\label{lem:APhexnowelldefined}
    There is no well-defined algorithm that solves the IGE problem in the hexagonal grid under a visibility range of one using robots without chirality.
\end{lemma}

\begin{figure}
\centering
    \begin{subfigure}[c]{0.58\textwidth}
    \centering
    \begin{tikzpicture}[triangular_slant,scale=0.6]

        \hexgrid{12.5}{8}
        \begin{scope}[shift={(7,1)}]
        \draw[dotted, thick] (-0.25,-0.5) -- (3.25, 6.5);
        \begin{scope}[every node/.style={robottiny,rviolet}]
            \node at (-3, 1) {};
            \node at (0, 4) {};
            \node at (-1, 5) {};
            \node at (-4, 2) {};
            \node at (-2, 5) {};
\end{scope}
        \foreach \i in {0, ..., 3}
            \node at (\i,2*\i) [mark={8pt}, rectangle, thick] {};
        \end{scope}
    \end{tikzpicture}

    \caption{The set $W$ marked by squares, with all robots in $I$ to the left}
    \label{fig:APhex1ambi}
    \end{subfigure}
    \hfill
    \begin{subfigure}[c]{0.38\textwidth}
    \centering
    \begin{tikzpicture}[triangular_slant,scale=0.8]
        \hexviewoneL

        \begin{scope}[every node/.style={robot,rviolet}]
            \node (r) at (0, 0) [rblack] {$r$};
            \node at (-1, 0) {};
        \end{scope}
        \node (up) at (1,1) [mark={8pt}, rectangle, thick] {};
        \node (down) at (0,-1) [mark={8pt}, rectangle, thick] {};
        \begin{scope}[moves]
            \draw[dotted] (r) edge node[above left] {?} (up);
            \draw[dotted] (r) edge node[below left] {?} (down);
        \end{scope}
    \end{tikzpicture}

    \caption[The view of robot $r$ as it is about to visit a vertex of $W$.]{The view of robot $r$ at round $d-1$ as it is about to visit a vertex of $W$}
    \label{fig:APhex1ambirobot}
    \end{subfigure}
\caption{Any algorithm for the hexagonal grid under visibility range one must use an ambiguous rule.}
\label{fig:APhex1ambiguity}
\end{figure}
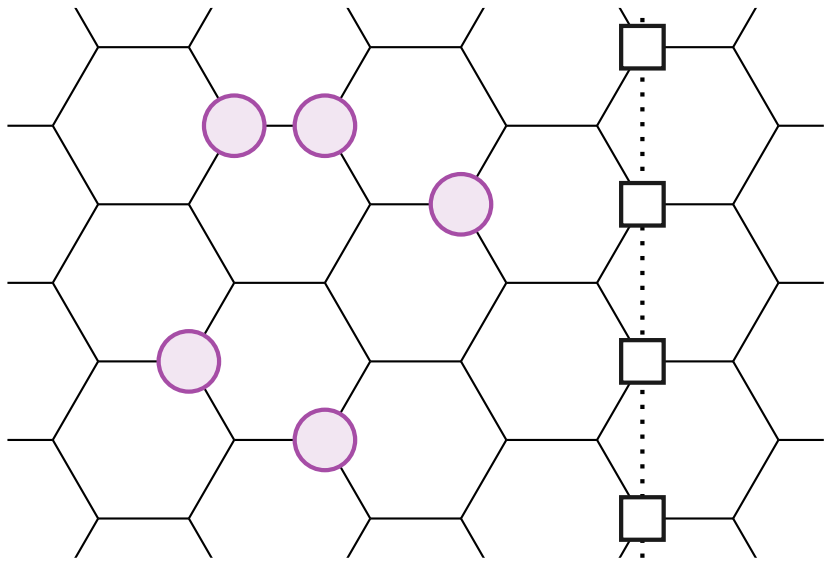
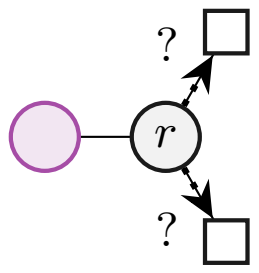

\proof We proceed by contradiction. Suppose there exists a well-defined algorithm $\mathcal{A}$ that solves the IGE problem in the hexagonal grid under a visibility range of one. Let us assume, without loss of generality, that the position of every robot in the initial configuration $I$ of $\mathcal{A}$ can be written as $\hex{x, y, z}$, where $x\in\mathbb{Z}^-, y\in\mathbb{Z}$ and $z\in\{0,1\}$. Therefore, the $x$ coordinate of every robot in the plane is less than 0. We also define a set of right-facing vertices $W = \{\hex{0,y,1}\ |\ y\in\mathbb{Z}\}$, depicted in Figure~\ref{fig:APhex1ambi}. Since the robots have to visit every vertex of the grid, they also have to eventually visit the vertices of $W$. We mark the first round in which a robot occupies a node in $W$ as round $d$.

We now analyze the configuration at round $d-1$ and the moves of the robots in this configuration. At round $d-1$, all robots still have to be to the left of $W$, as if there were a robot to the right, it would have had to have crossed and occupied a node of $W$ before round $d-1$. For a robot to occupy a vertex of $W$ at round $d$, there has to be a robot $r$ at position $\hex{0, y, 0}$ for some $y\in\mathbb{Z}$ in the previous round $d-1$. This robot cannot see any robots to the right, as they would be on the vertices of $W$. Therefore, the only place where $r$ can see a different robot is to its left, at position $\hex{-1, y, 1}$. If there were no robot at this position, the adversary could apply an indistinguishable transformation to make it move to the left. Now, no matter what colors are used for the two robots, the view of robot $r$ is symmetrical along the x-axis (see Figure~\ref{fig:APhex1ambirobot}), and therefore any action it takes to move onto a vertex of $W$ is ambiguous due to the lack of chirality.
\qed

\begin{figure}[h!t]
\centering
    \begin{tikzpicture}[triangular_slant,scale=0.65]

        \hexgrid[(-1.5,0)]{20.5}{12}
        \begin{scope}[shift={(12.5,7)}]
        \begin{scope}[every node/.style={robottiny,rblack}]
            \node (1200) at (0, -1)[rgreen] {F};
            \node (-1553) at (-3, 3)[rred] {B};
            \node (-753) at (-5, -5)[rred] {B};
            \node (-691) at (-3, -1)[rorange] {S};
            \node (2130) at (1, -2)[rorange] {S};
            \node (276) at (-2, -1)[rgreen] {F};
            \node (1237) at (0, 0)[rblue] {L};
            \node (-3512) at (-7, 0)[rred] {B};
            \node (2167) at (1, -1)[rred] {B};
            \node (301) at (0, 2)[rred] {B};
            \node (270) at (-1, 0)[rblue] {L};
            \node (-2607) at (-8, -4)[rred] {B};
        \end{scope}
        \begin{scope} [movestiny]
            \draw (1200) -- (1237);
            \draw[rgreen] (-691) -- (276);
            \draw[rred] (2130) -- (2167);
            \draw (276) -- (270);
            \draw (1237) -- +(1, 1);
            \draw[rgreen] (2167) -- (1200);
            \draw (270) -- +(0, 1);
        \end{scope}
\end{scope}
    \end{tikzpicture}

    \caption[Initial configuration of algorithm \hexalg{3}.]{Initial configuration of algorithm \hexalg{3}, including the actions taken in the first round}\label{fig:APalgohex3init}
\end{figure}
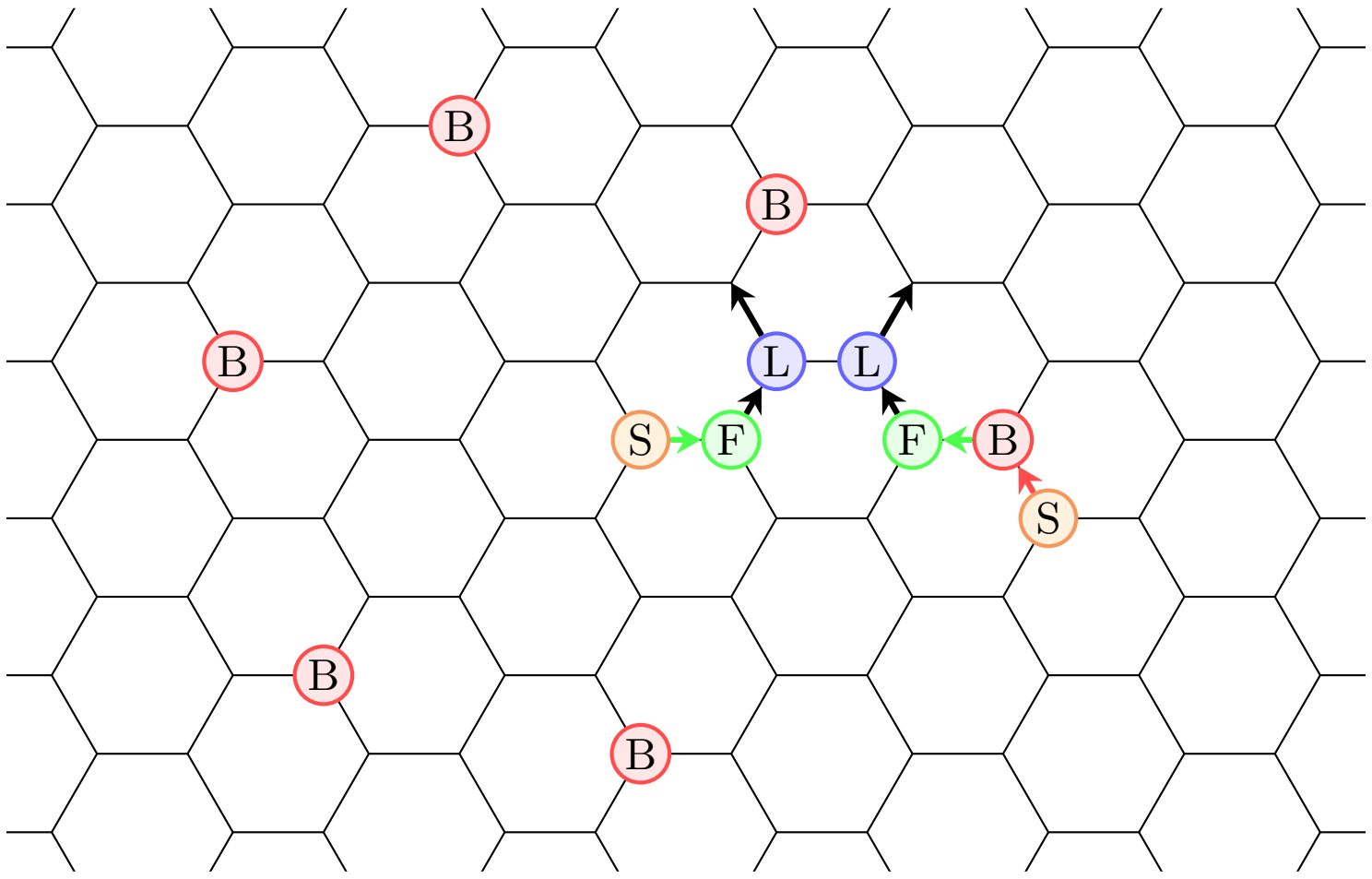

Although a well-defined algorithm is not able to solve the IGE problem under these parameters, we propose an algorithm that uses two ambiguous rules, denoted as \hexalg{3}. It uses twelve robots and four colors. Its existence shows that infinite grid exploration is possible in the hexagonal grid under visibility range one, albeit using some ambiguous rules. The initial configuration is shown in Figure~\ref{fig:APalgohex3init}.

In the first round, the moving group overlaps with the position at which the rightmost beacon is required. To circumvent this, the to-be beacon robot (the bottom right $S$-colored robot) is placed adjacent to the moving group and moves into position using an extra rule. One robot of the moving group is also recolored to allow this.

Just as the previous algorithms, algorithm \hexalg{3} uses the mechanism of a moving group and beacons to explore the grid. The moving group is made up of six robots and moves with a period equal to four, as shown in Figure~\ref{fig:APhex3movinggroup}. It is essentially made up of two symmetrical groups of three robots, which periodically diverge and meet again, thus overcoming the limitation of the small visibility range.

Here, one of the ambiguous rules is used to advance the leader of the moving group, shown in Figure~\ref{fig:APhex3move1} by dotted lines. The configuration in Figure~\ref{fig:APhex3move2} is one of four possible configurations. Nonetheless, they all lead to the configuration in Figure~\ref{fig:APhex3move3}. From there, the front $S$-colored robots can signal to the rest of the moving group about its orientation (Figure~\ref{fig:APhex3move4}), after which the cycle repeats.

Similar to algorithm \hexalg{2}, there are six beacons, and so the moving group explores in an outward hexagonal spiral.

{

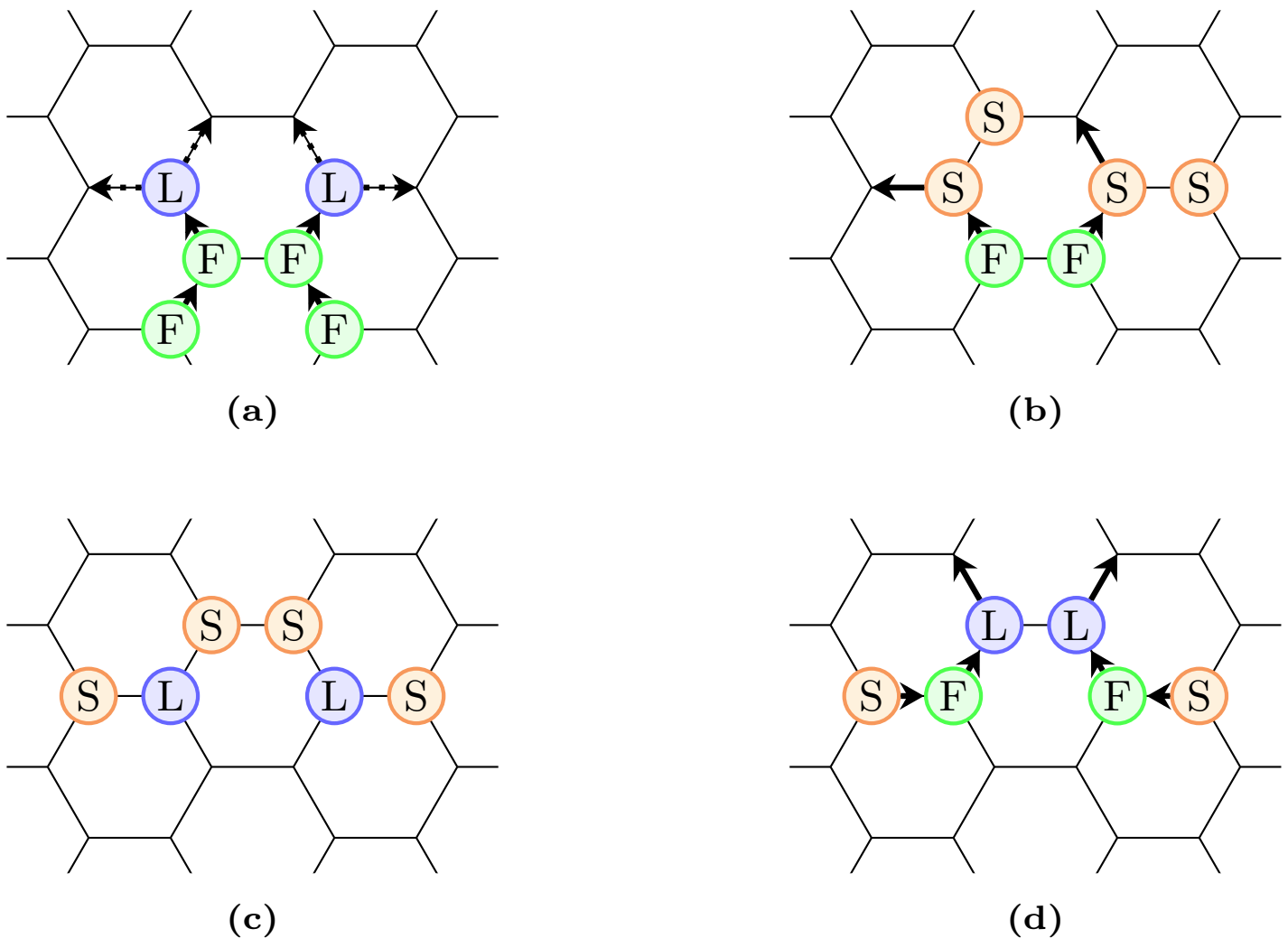
\begin{figure}
\centering
\hspace*{\fill}\begin{subfigure}{5cm}
\centering
\captionsetup{justification=centering}
    \begin{tikzpicture}[triangular_slant,scale=0.6]
    \hexgrid[(-1,-2)]{8.5}{6}
\begin{scope}[shift={(4,2)}]
    \begin{scope}[every node/.style={robottiny,rblack}]
        \node (1200) at (0, -1)[rgreen] {F};
        \node (307) at (-1, 1)[rblue] {L};
        \node (276) at (-2, -1)[rgreen] {F};
        \node (1237) at (0, 0)[rgreen] {F};
        \node (270) at (-1, 0)[rgreen] {F};
        \node (1231) at (1, 1)[rblue] {L};
    \end{scope}
    \begin{scope} [movestiny]
        \draw (1200) -- (1237);
        \draw[dotted] (307) -- +(1, 1);
        \draw[dotted] (307) -- +(-1, 0);
        \draw (276) -- (270);
        \draw (1237) -- (1231);
        \draw (270) -- (307);
        \draw[dotted] (1231) -- +(0, 1);
        \draw[dotted] (1231) -- +(1, 0);
    \end{scope}
    \end{scope}
\end{tikzpicture}
    \caption{}
    \label{fig:APhex3move1}
\end{subfigure}\hfill \begin{subfigure}{5cm}
\centering
\captionsetup{justification=centering}
    \begin{tikzpicture}[triangular_slant,scale=0.6]
    \hexgrid[(-1,-2)]{8.5}{6}
\begin{scope}[shift={(4,2)}]
    \begin{scope}[every node/.style={robottiny,rblack}]
        \node (307) at (-1, 1)[rorange] {S};
        \node (1237) at (0, 0)[rgreen] {F};
        \node (2198) at (2, 1)[rorange] {S};
        \node (301) at (0, 2)[rorange] {S};
        \node (270) at (-1, 0)[rgreen] {F};
        \node (1231) at (1, 1)[rorange] {S};
    \end{scope}
    \begin{scope} [movestiny]
        \draw (307) -- +(-1, 0);
        \draw (1237) -- (1231);
        \draw (270) -- (307);
        \draw (1231) -- +(0, 1);
    \end{scope}
    \end{scope}
\end{tikzpicture}
    \caption{}
    \label{fig:APhex3move2}
\end{subfigure}\hspace*{\fill}

\par\vspace{0.60cm}

\hspace*{\fill}\begin{subfigure}{5cm}
\centering
\captionsetup{justification=centering}
    \begin{tikzpicture}[triangular_slant,scale=0.6]
    \hexgrid[(-1,-2)]{8.5}{6}
\begin{scope}[shift={(4,2)}]
    \begin{scope}[every node/.style={robottiny,rblack}]
        \node (-660) at (-2, 1)[rorange] {S};
        \node (307) at (-1, 1)[rblue] {L};
        \node (1268) at (1, 2)[rorange] {S};
        \node (2198) at (2, 1)[rorange] {S};
        \node (301) at (0, 2)[rorange] {S};
        \node (1231) at (1, 1)[rblue] {L};
    \end{scope}
    \begin{scope} [movestiny]
\end{scope}
    \end{scope}
\end{tikzpicture}
    \caption{}
    \label{fig:APhex3move3}
\end{subfigure}\hfill \begin{subfigure}{5cm}
\centering
\captionsetup{justification=centering}
    \begin{tikzpicture}[triangular_slant,scale=0.6]
    \hexgrid[(-1,-2)]{8.5}{6}
\begin{scope}[shift={(4,2)}]
    \begin{scope}[every node/.style={robottiny,rblack}]
        \node (-660) at (-2, 1)[rorange] {S};
        \node (307) at (-1, 1)[rgreen] {F};
        \node (1268) at (1, 2)[rblue] {L};
        \node (2198) at (2, 1)[rorange] {S};
        \node (301) at (0, 2)[rblue] {L};
        \node (1231) at (1, 1)[rgreen] {F};
    \end{scope}
    \begin{scope} [movestiny]
        \draw (-660) -- (307);
        \draw (307) -- (301);
        \draw (1268) -- +(1, 1);
        \draw (2198) -- (1231);
        \draw (301) -- +(0, 1);
        \draw (1231) -- (1268);
    \end{scope}
    \end{scope}
\end{tikzpicture}
    \caption{}
    \label{fig:APhex3move4}
\end{subfigure}\hspace*{\fill}

\caption{The moving group of algorithm \hexalg{3}}
\label{fig:APhex3movinggroup}
\end{figure}
}

The moving group always encounters a beacon when in the H-shape (Figure~\ref{fig:APhex3adj1}). The outer side continues as if the beacon was not present, as it might not notice it until two rounds later. The inner side continues past the beacon robot, until all three robots have seen the beacon (Figures~\ref{fig:APhex3adj1}, \ref{fig:APhex3adj2} and \ref{fig:APhex3adj3}). Here, the second ambiguous rule is used, as the front inner-side robot is required to move away to prevent a collision. No matter the indistinguishable transformation used, the configuration in Figure~\ref{fig:APhex3adj4} is reached and no collision occurs. Also, the exploration does not depend on the outcome of this ambiguous rule, as both vertices are eventually visited in subsequent rounds.

In Figure~\ref{fig:APhex3adj4}, the inner side has moved into position and only changes its colors from now on. The outer side also moves into position, leaving behind a beacon robot in the correct position for the next adjustment (Figures~\ref{fig:APhex3adj4}, \ref{fig:APhex3adj5} and \ref{fig:APhex3adj6}). In the round following the configuration in Figure~\ref{fig:APhex3adj6}, the beacon robot changes its color to $L$ due to a rule required in the previous round. An additional rule is then required to change its color to $B$ again.

{

\begin{figure}
\centering
\hspace*{\fill}\begin{subfigure}{4cm}
\centering
\captionsetup{justification=centering}
    \begin{tikzpicture}[triangular_slant,scale=0.6]
    \hexgrid[(-1,-2)]{9}{7}
    \begin{scope}[shift={(4,2)}]
    \begin{scope}[every node/.style={robottiny,rblack}]
        \node (1200) at (0, -1)[rgreen] {F};
        \node (307) at (-1, 1)[rblue] {L};
        \node (276) at (-2, -1)[rgreen] {F};
        \node (1237) at (0, 0)[rgreen] {F};
        \node (301) at (0, 2)[rred] {B};
        \node (270) at (-1, 0)[rgreen] {F};
        \node (1231) at (1, 1)[rblue] {L};
    \end{scope}
    \begin{scope} [movestiny]
        \draw (1200) -- (1237);
        \draw (307) -- +(-1, 0);
        \draw (276) -- (270);
        \draw (1237) -- (1231);
        \draw (270) -- (307);
        \draw[dotted] (1231) -- +(0, 1);
        \draw[dotted] (1231) -- +(1, 0);
        \end{scope}
    \end{scope}
\end{tikzpicture}
    \caption{}
    \label{fig:APhex3adj1}
\end{subfigure}\hfill \begin{subfigure}{4cm}
\centering
\captionsetup{justification=centering}
    \begin{tikzpicture}[triangular_slant,scale=0.6]
    \hexgrid[(-1,-2)]{9}{7}
    \begin{scope}[shift={(4,2)}]
    \begin{scope}[every node/.style={robottiny,rblack}]
        \node (-660) at (-2, 1)[rblue] {L};
        \node (307) at (-1, 1)[rorange] {S};
        \node (1237) at (0, 0)[rgreen] {F};
        \node (2198) at (2, 1)[rorange] {S};
        \node (301) at (0, 2)[rred] {B};
        \node (270) at (-1, 0)[rgreen] {F};
        \node (1231) at (1, 1)[rorange] {S};
    \end{scope}
    \begin{scope} [movestiny]
        \draw[dotted] (-660) -- +(0, 1);
        \draw[dotted] (-660) -- +(-1, -1);
        \draw (307) -- (-660);
        \draw (1237) -- (1231);
        \draw (270) -- (307);
        \draw (1231) -- +(0, 1);
    \end{scope}
    \end{scope}
\end{tikzpicture}
    \caption{}
    \label{fig:APhex3adj2}
\end{subfigure}\hfill \begin{subfigure}{4cm}
\centering
\captionsetup{justification=centering}
    \begin{tikzpicture}[triangular_slant,scale=0.6]
    \hexgrid[(-1,-2)]{9}{7}
    \begin{scope}[shift={(4,2)}]
    \begin{scope}[every node/.style={robottiny,rblack}]
        \node (-660) at (-2, 1)[rgreen] {F};
        \node (307) at (-1, 1)[rblue] {L};
        \node (1268) at (1, 2)[rorange] {S};
        \node (2198) at (2, 1)[rorange] {S};
        \node (-654) at (-3, 0)[rgreen] {F};
        \node (301) at (0, 2)[rred] {B};
        \node (1231) at (1, 1)[rblue] {L};
    \end{scope}
    \begin{scope} [movestiny]
        \draw (-660) -- (307);
        \draw (307) -- +(0, -1);
\draw (-654) -- (-660);
\end{scope}
    \end{scope}
\end{tikzpicture}
    \caption{}
    \label{fig:APhex3adj3}
\end{subfigure}\hspace*{\fill}

\par\vspace{0.60cm}

\hspace*{\fill}\begin{subfigure}{4cm}
\centering
\captionsetup{justification=centering}
    \begin{tikzpicture}[triangular_slant,scale=0.6]
    \hexgrid[(-1,-2)]{9}{7}
    \begin{scope}[shift={(4,2)}]
    \begin{scope}[every node/.style={robottiny,rblack}]
\node (307) at (-1, 1)[rred] {B};
        \node (-660) at (-2, 1)[rgreen] {F};
        \node (1268) at (1, 2)[rred] {B};
        \node (2198) at (2, 1)[rorange] {S};
\node (301) at (0, 2)[rred] {B};
        \node (270) at (-1, 0)[rblue] {L};
\node (1231) at (1, 1)[rgreen] {F};
    \end{scope}
    \begin{scope} [movestiny]
\draw (1268) -- (301);
        \draw (2198) -- (1231);
        \draw (301) -- +(0, 1);
\draw (1231) -- (1268);
    \end{scope}
    \end{scope}
\end{tikzpicture}
    \caption{}
    \label{fig:APhex3adj4}
\end{subfigure}\hfill \begin{subfigure}{4cm}
\centering
\captionsetup{justification=centering}
    \begin{tikzpicture}[triangular_slant,scale=0.6]
    \hexgrid[(-1,-2)]{9}{7}
    \begin{scope}[shift={(4,2)}]
    \begin{scope}[every node/.style={robottiny,rblack}]
\node (338) at (0, 3)[rred] {B};
        \node (307) at (-1, 1)[rblue] {L};
        \node (-660) at (-2, 1)[rblue] {L};
        \node (1268) at (1, 2)[rorange] {S};
\node (301) at (0, 2)[rgreen] {F};
        \node (270) at (-1, 0)[rgreen] {F};
\node (1231) at (1, 1)[rgreen] {F};
    \end{scope}
    \begin{scope} [movestiny]
\draw (1268) -- +(1, 1);
        \draw (1231) -- (1268);
    \end{scope}
    \end{scope}
\end{tikzpicture}
    \caption{}
    \label{fig:APhex3adj5}
\end{subfigure}\hfill \begin{subfigure}{4cm}
\centering
\captionsetup{justification=centering}
    \begin{tikzpicture}[triangular_slant,scale=0.6]
    \hexgrid[(-1,-2)]{9}{7}
    \begin{scope}[shift={(4,2)}]
    \begin{scope}[every node/.style={robottiny,rblack}]
\node (338) at (0, 3)[rblue] {L};
        \node (307) at (-1, 1)[rgreen] {F};
        \node (-660) at (-2, 1)[rblue] {L};
        \node (1268) at (1, 2)[rgreen] {F};
\node (301) at (0, 2)[rgreen] {F};
        \node (1262) at (2, 3)[rred] {B};
        \node (270) at (-1, 0)[rgreen] {F};
\end{scope}
    \begin{scope} [movestiny]
        \draw[dotted] (338) -- +(1, 1);
        \draw[dotted] (338) -- +(-1, 0);
        \draw (307) -- (-660);
        \draw[dotted] (-660) -- +(0, 1);
        \draw[dotted] (-660) -- +(-1, -1);
        \draw (1268) -- (301);
        \draw (301) -- (338);
\draw (270) -- (307);
    \end{scope}
    \end{scope}
\end{tikzpicture}
    \caption{}
    \label{fig:APhex3adj6}
\end{subfigure}\hspace*{\fill}

\caption{An adjustment during the execution of \hexalg{3}}
\label{fig:APhex3adjustment}
\end{figure}
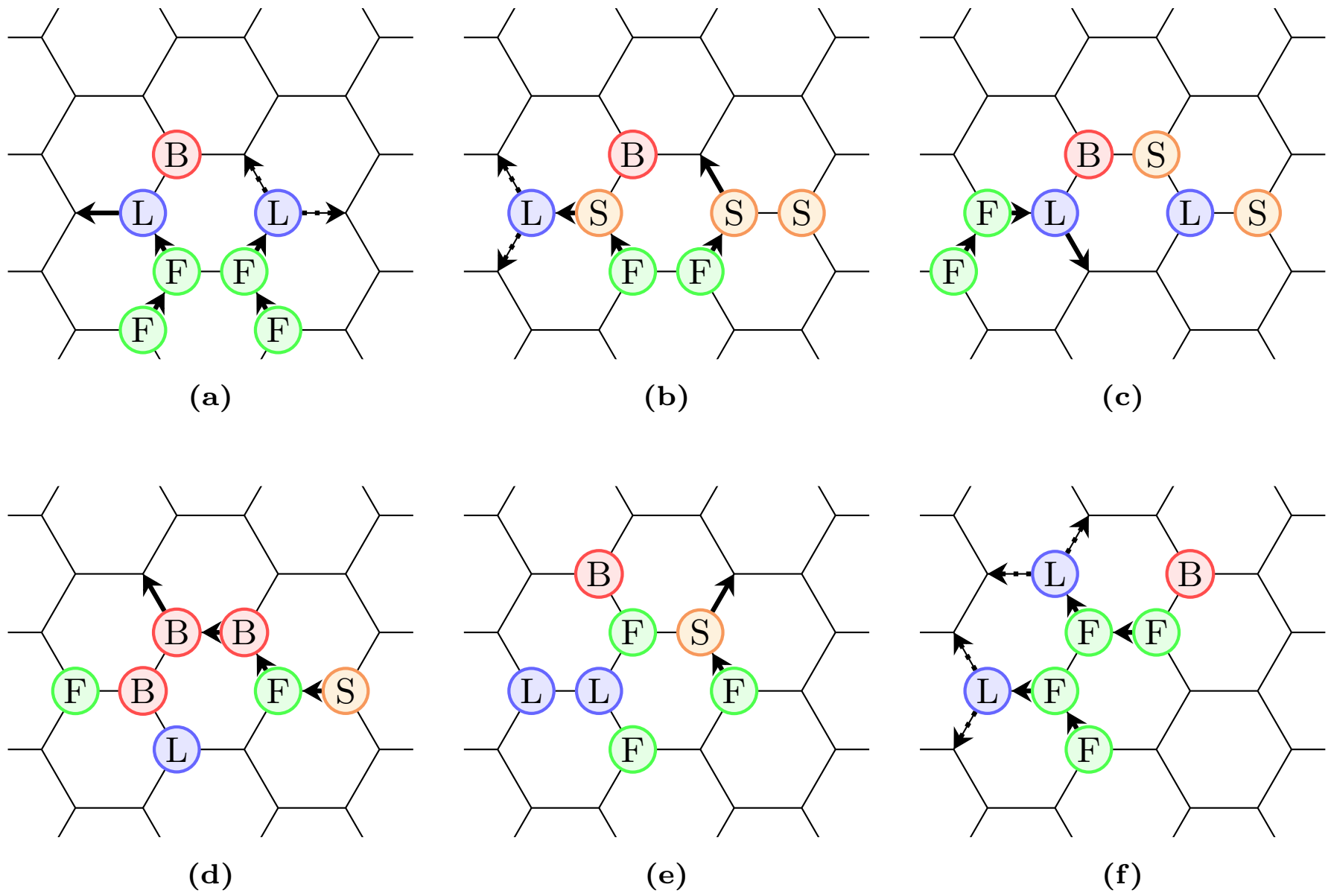

\paragraph{Algorithm \hexalg{3}.} To summarize, we formally define algorithm \hexalg{3} as the triplet $(\{L, F, B, S\}, I, T)$, where $I$ is the initial configuration as shown in Figure~\ref{fig:APalgohex3init} and $T$ is the transition function defined as the set of rules implicitly depicted in the figures of this subsection.

\begin{theorem}\label{the:APhex1view}
Algorithm \hexalg{3} solves the IGE problem in the hexagonal grid using twelve robots of four colors under visibility range $\Phi=1$.
\end{theorem}
}

\proof[Sketch] The proof follows the same scheme as the previous ones.
The execution is divided into phases, where phase~$i$ begins in the round in which the moving group reaches the first of the six beacons for the $(i+1)$-st time, and during a single phase the moving group travels once around the already explored area, repositioning each beacon further away from the center as it passes it.
Unlike the previous algorithms, \hexalg{3} is not well-defined and so, we have to argue about all of its executions at once.
This is made possible by the fact that the two ambiguous rules are applied only in the two situations described above, in which all the possible outcomes lead to a common configuration within a constant number of rounds (Figures~\ref{fig:APhex3move1}--\ref{fig:APhex3move3} and Figures~\ref{fig:APhex3adj1}--\ref{fig:APhex3adj3}).
Every execution therefore passes through the same sequence of phases and an induction on~$i$ shows that each of them is eventually reached.
As before, the vertices visited during phase~$i$ form a closed ring around the vertices visited in the earlier phases and thus, their union covers all but the finitely many vertices in the center of the spiral, which are checked to be visited during the first rounds of the execution.
\qed

\end{document}